\documentclass[structabstract, bibyear]{aa}
\usepackage{mathptmx}
\usepackage[T1]{fontenc}
\usepackage[utf8]{inputenc}
\usepackage{color}
\usepackage{amsmath}
\usepackage{amssymb}
\usepackage{graphicx}
\usepackage[authoryear]{natbib}

\makeatletter

\providecommand{\tabularnewline}{\\}

\usepackage[varg]{txfonts}
\usepackage{float}
\floatstyle{plain}
\restylefloat{figure}

\makeatother

\begin{document}

\title{Surface chemistry in the Interstellar Medium II. $\mathrm{H}_{2}$
formation on dust with random temperature fluctuations}

\author{Emeric Bron\inst{1,2}\and Jacques Le Bourlot\inst{1,2}\and Franck
Le Petit\inst{1}}

\institute{\inst{1}LERMA, Observatoire de Paris \& CNRS ; 5 place Jules Janssen
92190 Meudon, France\\
\inst{2}Université Paris Diderot ; 5 Rue Thomas Mann, 75205 Paris,
France\\\email{jacques.lebourlot@obspm.fr}}

\offprints{Jacques Le Bourlot}

\date{Received 18 June 2013; Accepted 3 June 2014}

\abstract{The $\mathrm{H_{2}}$ formation on grains is known to be sensitive
to dust temperature, which is also known to fluctuate for small grain
sizes due to photon absorption.} {We aim at exploring the consequences
of simultaneous fluctuations of the dust temperature and the adsorbed
$\mathrm{H}$-atom population on the $\mathrm{H}_{2}$ formation rate
under the full range of astrophysically relevant UV intensities and
gas conditions.} {The master equation approach is generalized to
coupled fluctuations in both the grain's temperature and its surface
population and solved numerically. The resolution can be simplified
in the case of the Eley-Rideal mechanism, allowing a fast computation.
For the Langmuir-Hinshelwood mechanism, it remains computationally
expensive, and accurate approximations are constructed. } {We find
the Langmuir-Hinshelwood mechanism to become an efficient formation
mechanism in unshielded photon dominated region (PDR) edge conditions
when taking those fluctuations into account, despite hot average dust
temperatures. It reaches an importance comparable to the Eley-Rideal
mechanism. However, we show that a simpler rate equation treatment
gives qualitatively correct observable results in full cloud simulations
under most astrophysically relevant conditions. Typical differences
are a factor of 2-3 on the intensities of the $\mathrm{H_{2}}$ $v=0$
lines. We also find that rare fluctuations in cloud cores are sufficient
to significantly reduce the formation efficiency.} {Our detailed
analysis confirms that the usual approximations used in numerical
models are adequate when interpreting observations, but a more sophisticated
statistical analysis is required if one is interested in the details
of surface processes. }


\maketitle

\section{Introduction}

$\mathrm{H}_{2}$ is the most abundant molecule in the interstellar
medium. It is found in a wide variety of astrophysical environments
in which it often plays a leading role in the physics and evolution
of objects (see the book by \citet{combes:01} for a review). As an
efficient coolant of hot gas, it controls for instance through thermal
balance the collapse of interstellar clouds leading ultimately to
star formation. Furthermore, its formation is the first step of a
long sequence of reactions leading to the great chemical complexity
found in dense clouds. In addition to its physical importance, $\mathrm{H}_{2}$
is also of great observational usefulness as a diagnostic probe for
many different processes in various environments.

The detailed mechanism of its formation is thus a key part of the
understanding and modeling of the interstellar medium. Direct gas
phase formation is very inefficient and cannot explain the observed
abundances. The ion-neutral reaction $\mathrm{H}+\mathrm{H}^{-}$
is more efficient, but the low $\mathrm{H}^{-}$ abundance does not
allow sufficient $\mathrm{H}_{2}$ formation this way. The $\mathrm{H_{2}}$
molecule is thus thought to form mainly on the surface of dust grains
\citep{gould:63,hollenbach:71}. Grains act as catalysts. They provide
a surface on which adsorbed $\mathrm{H}$ atoms can meet and react,
and they absorb that excess energy released by the formation that
prevented the gas phase formation of a stable molecule.

This process is usually modeled using the Langmuir-Hinshelwood (hereafter
LH) mechanism in which weakly bound adsorbed atoms migrate randomly
on the surface, and one $\mathrm{H}_{2}$ molecule is formed each
time two atoms meet. This mechanism has been studied in detail using
the master equation approach \citep{biham:02}, the moment equation
approach \citep{lipshtat:03,lepetit:09} and Monte Carlo simulations
\citep{Chang05,Cuppen05,chang:06}. Laboratory experiments have also
been conduced and their results modeled \citep{pirronello:97b,pirronello:97a,pirronello:99,katz:99}.
It is found to be efficient over a limited range of grain temperatures
of about $5-15\mathrm{K}$ for flat surfaces. However, observations
of dense photon dominated regions (PDRs) found efficient formation
despite higher grain temperatures \citep{habart:04,habart:11}. Various
modifications of the mechanism have been proposed to extend the range
of efficient formation. Some authors introduced sites of higher binding
energies due to surface irregularities \citep{Chang05,Cuppen05}.
Some introduced a reaction between chemisorbed atoms and atoms migrating
from a physisorption site (\citealp{cazaux:04,iqbal12}). Others proposed
that the Eley-Rideal (hereafter ER) mechanism for chemisorbed atoms
is a relevant formation process at the edge of PDRs \citep{duley:1996,habart:04,bachellerie:09,lebourlot:12}.
In this mechanism, chemisorbed $\mathrm{H}$ atoms, which are attached
to the surface by a chemical bond, stay fixed on the surface until
another atom of the gas falls into the same site, forming a $\mathrm{H}_{2}$
molecule.

The grain temperatures in illuminated environments are not only higher
but also strongly fluctuating. Small grains have small heat capacities,
and each UV-photon absorption makes their temperature fluctuate widely.
This effect has received much attention because some components of
dust emission are produced during these transient temperature spikes
\citep{desert86,li2001b,li2001a,compiegne:11}. Those fluctuations
are stronger in strong radiation field environments like PDRs. Moreover,
usual dust size distributions favor small grains so that they contribute
the most to the total dust surface, making the small grains dominant
for surface reactions. Those fluctuations can thus have an important
effect on $\mathrm{H}_{2}$ formation, especially in strongly illuminated
environments, as temperature spikes are likely to cause massive desorption.
This effect has so far only been studied by \citet{cuppen:06} using
Monte Carlo simulations for the LH mechanism, computing the $\mathrm{H}_{2}$
formation rate in one single specific diffuse cloud condition under
a standard interstellar radiation field. The dependance on the astrophysical
parameters was not investigated. Very recently, similar models using
the same Monte Carlo method were presented in \citet{Iqbal14} for
silicate surfaces. Those models also included the ER mechanism. The
authors remain limited by the prohibitive computation time (several
weeks) required by the kinetic Monte Carlo method when including temperature
fluctuations. We study here this problem for both ER and LH in a wider
range of environments, including PDRs and their high radiation fields,
and study the effect induced on cloud structure and observable line
intensities by introducing the computation of the fluctuation effect
into full cloud simulations.

We propose an analytical approach of the joint fluctuations problem
(temperature and surface coverage) based on the master equation, which
takes the form of an integral equation. Its numerical resolution incurs
a high computational cost, which grows with grain size. In the case
of the ER mechanism, the absence of non-linear term in the chemistry
allows a decomposition of the two-dimensional full master equation
into two one-dimensional equations : a marginal master equation for
the temperature fluctuations and an equation on the conditional average
population (conditional on $T$). This makes the numerical computation
much more tractable. In the case of the LH mechanism, such simplification
is not possible. The computationally heavy full resolution is used
to build and verify a fast and accurate approximation of the solution.
The final method proposed here allows computation of the total formation
rate with a computation time of the order of one minute.

In Sect.~\ref{sec:The-model}, we introduce the simple physical model
that we use and the exact resolution method. In Sect. \ref{sec:Approximations},
approximations of the solution are constructed for both mechanisms.
Section~\ref{sec:Results} presents the results of the numerical
computation of the $\mathrm{H}_{2}$ formation rate using the full
method for ER and the approximation for LH. Section~\ref{sec:PDR_code}
then shows how this computed effect affects the structure of the cloud
in PDR simulations. Section \ref{sec:Conclusion} is our conclusion.

\section{The model\label{sec:The-model}}

\subsection{Physical description}

We use a simple physical model of $\mathrm{H}_{2}$ formation on grains
to be able to solve the problem of the coupled fluctuations and to
obtain an estimate of the importance of dust temperature fluctuations
for $\mathrm{H_{2}}$ formation. 

Our system is a spherical grain of radius $a$, density $\rho_{\mathrm{gr}}$,
mass $m=\frac{4}{3}\,\pi\, a^{3}\,\rho$, and heat capacity $C(T)$.
Its thermal state can be equivalently described by its temperature
$T$ or by its thermal energy $E=\int_{0}^{T}dT'\, C(T')$%
\footnote{We neglect energy discretization for the lower states, and adopt the
normalization $E=0$ for $T=0$. See \citet{li2001a} for a detailed
state-by-state treatment.%
}. For polycyclic aromatic hydrocarbons (PAHs), we define the size
$a$ as the equivalent size of a sphere of the same mass (see \citealp{compiegne:11}).

We consider two groups of phenomena:
\begin{itemize}
\item interaction with the radiation field through photon absorption and
emission, which are both described as discrete events.
\item adsorption of $\mathrm{H}$ atoms of the gas onto the surface of the
grain and $\mathrm{H}_{2}$ formation on the surface (physisorption
and chemisorption are considered separately, as are the two corresponding
$\mathrm{H}_{2}$ formation mechanisms). 
\end{itemize}

\subsubsection{Photon absorption and emission\label{sub:Model-phtons}}

Under an ambient isotropic radiation field of specific intensity $I_{U}(U)$
in $\mathrm{W}\,\mathrm{m}^{-2}\,\mathrm{J}^{-1}\,\mathrm{sr}^{-1}$
(in this article, $U$ always denotes a photon energy), the power
received by the grain at a certain photon energy $U$ is $P_{\mathrm{abs}}(U)=4\pi^{2}\, a^{2}\, Q_{\mathrm{abs}}(U)\, I_{U}(U)$,
where $Q_{\mathrm{abs}}(U)$ is the absorption efficiency coefficient
of the grain at energy $U$. In later parts of the paper, we are interested
in transition rates between states of the grain. The rate of photon
absorptions at this energy $U$ is simply
\[
R_{\mathrm{abs}}(U)={\displaystyle \frac{P_{\mathrm{abs}}(U)}{U}}.
\]

We postulate that the grain emits according to a modified black body
law with a specific intensity $Q_{\mathrm{abs}}(U)\, B_{U}(U,T)$,
where $B_{U}(U,T)$ is the usual black body specific intensity. The
power emitted at energy $U$ is $P_{\mathrm{em}}(U,T)=4\pi^{2}\, a^{2}\, Q_{\mathrm{abs}}(U)\, B_{U}(U,T)$
and the photon emission rate is
\[
R_{\mathrm{em}}(U,T)=\frac{P_{\mathrm{em}}(U,T)}{U}.
\]

These events, which are supposed to occur as Poisson processes, cause
fluctuations of the temperature. We treat these fluctuations exactly
in Section \ref{sub:Method}.

When neglecting these fluctuations, the equilibrium temperature $T_{\mathrm{eq}}$
of the grain is defined as balancing the instantaneous emitted and
absorbed powers
\begin{equation}
\int_{0}^{+\infty}dU\, P_{\mathrm{abs}}(U)=\int_{0}^{E_{grain}}dU\, P_{\mathrm{em}}(U,T_{\mathrm{eq}}),\label{eq:T_eq}
\end{equation}
where the upper bound on the right hand side accounts for the finite
total energy of a grain. This upper cutoff of emission frequencies
can become very important for small cold grains, effectively stopping
their cooling.

In the following, we use a standard interstellar radiation field (\citealp{mathis:83})
and apply a scaling factor $\chi$ to the UV component of the field.
We measure the UV intensity of those fields using the usual $G_{0}=\frac{1}{u_{\mathrm{Habing}}}\,\int_{912\text{Å}}^{2400\text{Å}}d\lambda\, u_{\lambda}(\lambda)$,
where $u_{\mathrm{Habing}}=5.3\times10^{-15}\,\mathrm{J}\,\mathrm{m}^{-3}$.

The dust properties ($C(T)$, $Q_{\mathrm{abs}}(U)$ and $\rho$)
are taken from \citet{compiegne:11}. We consider PAHs, amorphous
carbon grains, and silicates dust populations and use the properties
used in this reference (see references therein, in their Appendix
A) for each of those populations.

\subsubsection{Surface chemistry via Langmuir-Hinshelwood mechanism\label{sub:LH_processes}}

We first consider the physisorption-based LH mechanism. In this mechanism,
atoms bind to the grain surface due to Van der Waals interactions,
leading to the so-called physisorption. These weakly bound atoms are
able to migrate from site to site. Formation can then occur when two
such migrating atoms meet.

Real grain surfaces are irregular, and binding properties vary from
site to site due to surface defects. To keep the model simple enough
to allow a detailed modeling of temperature fluctuations, we assume
identical binding sites uniformly distributed on the grain surface.
The number of sites is $N_{s}=\frac{4\pi\, a^{2}}{d_{\mathrm{s}}^{2}}$,
where $d_{\mathrm{s}}$ is the typical distance between sites. We
take $d_{\mathrm{s}}=0.26\,\mathrm{nm}$ as in \citet{lebourlot:12}.
We assume the physisorption sites to be simple potential wells of
depth $E_{\mathrm{phys}}$ without barrier.

Atoms from a gas at temperature $T_{\mathrm{gas}}$ collide with the
grain at a rate $k_{\mathrm{coll}}=\pi\, a^{2}\, n(H)\, v_{\mathrm{th}}$,
where $v_{\mathrm{th}}=\sqrt{\frac{8k\, T_{\mathrm{gas}}}{\pi\, m_{\mathrm{H}}}}$
is the thermal velocity. We call $s(T_{\mathrm{gas}})$ the sticking
probability and discuss later our choice for this function. The probability
to land on an empty site is $1-\frac{n}{N_{\mathrm{s}}}$ and the
physisorption rate is thus 
\[
R_{\mathrm{phys}}=k_{\mathrm{coll}}\, s(T_{\mathrm{gas}})\,\left(1-\frac{n}{N_{\mathrm{s}}}\right).
\]

We assume that gas atoms falling on an occupied site are never rejected
and react with the adsorbed atom to form a $\mathrm{H}_{2}$ molecule,
which is immediately desorbed by the formation energy. This assumption
of Eley-Rideal-like reaction for physisorbed atoms is similarly made
in the model of \citet{cuppen:06} with the difference that they do
not consider desorption at the formation of the new molecule. This
direct Eley-Rideal process is expected to have a large cross section
at temperatures relevant to the interstellar medium \citep{martinazzo:06},
and it is thus a reasonable assumption to assume that it dominates
over rejection. To estimate the influence of this assumption on our
results, we computed the contribution of this direct ER-like reaction
process to the total mean formation rate through physisorption and
found it to be a negligible fraction (always less than 1\% of the
total average formation rate). Despite the very small contribution
of this Eley-Rideal-like process to the physisorption-based formation
rate in our model, we keep calling the physisorption-based mechanism
LH mechanism to distinguish it from the chemisorption-based ER mechanism.

Surface atoms can then evaporate. The adsorbed atoms are assumed to
be thermalized at the grain temperature and must overcome the well
energy $E_{\mathrm{phys}}$. We take this desorption rate for one
atom to be of the form $k_{\mathrm{des}}(T)=\nu_{0}\,\exp\left(-\frac{T_{\mathrm{phys}}}{T}\right)$,
where $T$ is the grain temperature, $T_{\mathrm{phys}}=E_{\mathrm{phys}}/k$,
and $\nu_{0}$ is a typical vibration frequency taken as $\nu_{0}=\frac{1}{\pi}\,\sqrt{\frac{2\, E_{\mathrm{phys}}}{d_{0}^{2}\, m_{\mathrm{H}}}}$
with a typical size $d_{0}=0.1\,\textrm{nm}$ (\citealp{hasegawa:92}).
The total desorption rate on the grain is then 
\[
R_{\mathrm{des}}(T)=n\, k_{\mathrm{des}}(T).
\]

To migrate from site to site, an atom must cross a barrier of height
$E_{\mathrm{migr}}$. It can do so by thermal hopping or by tunneling
for which we use the formula from \citet{lebourlot:12}. The migration
rate for one physisorbed atom is thus 
\[
k_{\mathrm{migr}}(T)=\nu_{0}\,\exp\left(-\frac{T_{\mathrm{migr}}}{T}\right)+\frac{\nu_{0}}{1+\frac{T_{\mathrm{migr}}^{2}\sinh^{2}\left(d_{s}\sqrt{2m_{H}k(T_{\mathrm{migr}}-T)}/\hbar\right)}{4T(T_{\mathrm{migr}}-T)}}.
\]
 where $T_{\mathrm{migr}}=E_{\mathrm{migr}}/k$.

The main formation process is the meeting of two physisorbed atoms
during a migration event. In a simple approximation (see \citealp{lohmar:06,lohmar:09}
for a more detailed treatment of the sweeping rate), we take this
formation rate to be 
\[
R_{\mathrm{form}}^{(1)}(T)=k_{\mathrm{migr}}(T)\,\frac{n^{2}}{N_{\mathrm{s}}}.
\]

We also take direct reaction of a physisorbed atom with a gas atom
in an ER-like mechanism into account, with a rate
\[
R_{\mathrm{form}}^{(2)}(T)=k_{\mathrm{coll}}\frac{n}{N_{\mathrm{s}}}.
\]
This term is only significant for very low dust temperatures (a few
$\mathrm{K}$, depending on the collision rate with gas atoms).

We assume immediate desorption of the newly formed molecule because
of the high formation energy compared to the binding energy, thus
making a similar assumption to recent models \citep{iqbal12,lebourlot:12}.
Experimental results indicate that a small fraction of the formed
molecules may stay on the surface at formation \citep{katz:99,perets:07}.
Theoretical studies \citep{morisset:04,morisset:05} suggest that
even when the formed molecule does not desorb immediately at formation,
it desorbs after a few picoseconds by interacting with the surface
due to its high internal energy. It is thus equivalent to immediate
desorption compared to the timescales of the other events.

If the temperature fluctuations are neglected, an equilibrium surface
population can be computed using the equilibrium rate equation $R_{\mathrm{phys}}-R_{\mathrm{des}}(T)-2\, R_{\mathrm{form}}^{(1)}(T)-R_{\mathrm{form}}^{(2)}(T)=0$
(the factor of 2 comes from the fact that one formation through migration
event reduces the population by two). We find
\begin{multline}
n_{eq}(T)=\frac{\left(1+s(T_{\mathrm{gas}})\right)k_{\mathrm{coll}}}{4\, k_{\mathrm{mig}}(T)}\left(1+\frac{N_{\mathrm{s}}\, k_{\mathrm{des}}(T)}{k_{\mathrm{coll}}(1+s(T_{\mathrm{gas}}))}\right)\\
\left[\sqrt{1+\frac{8N_{\mathrm{s}}\, k_{\mathrm{mig}}(T)\, s(T_{\mathrm{gas}})}{k_{\mathrm{coll}}(1+s(T_{\mathrm{gas}}))^{2}}\frac{1}{\left(1+\frac{N_{\mathrm{s}}\, k_{\mathrm{des}}(T)}{k_{\mathrm{coll}}(1+s(T_{\mathrm{gas}}))}\right)^{2}}}-1\right],\label{eq:n_eq_LH}
\end{multline}

and the corresponding equilibrium formation rate on one grain at temperature
$T$ is
\begin{equation}
r_{\mathrm{H}_{2}}^{\mathrm{eq}}(T)=k_{\mathrm{migr}}(T)\frac{[n_{\mathrm{eq}}(T)]^{2}}{N_{\mathrm{s}}}+k_{\mathrm{coll}}\frac{n_{\mathrm{eq}}(T)}{N_{\mathrm{s}}}.\label{eq:rH2_eq_LH}
\end{equation}

Once again this expression is not valid for a fluctuating temperature
and the full calculation is described in the next Sect. \ref{sub:Method}.

\begin{table}
\caption{Model parameters for physisorption. $T_{\mathrm{phys}}$ is the physisorption
binding energy and $T_{\mathrm{migr}}$ the migration barrier. $d_{\mathrm{s}}$
is the distance between sites.\label{tab:H_binding}}

\centering%
\begin{tabular}{lcccc}
\hline 
\hline Dust Component & $T_{\mathrm{phys}}$ & $T_{\mathrm{migr}}$ & References & $d_{\mathrm{s}}$\tabularnewline
 & $(\mathrm{K})$ & $(\mathrm{K})$ &  & (nm)\tabularnewline
\hline 
Amorphous Carbon & $658.0$ & $510.6$ & 1 & $0.26$\tabularnewline
Amorphous Silicate & $510.0$ & $406.0$ & 2 & $0.26$\tabularnewline
Ices & $350.0$ & $100.0$ & 3 & $0.26$\tabularnewline
\hline 
\end{tabular}

$ $\tablebib{(1)~\citet{katz:99}; (2) \citet{perets:07}; (3) \citet{hasegawa:92}}
\end{table}

For the binding and migration energies, we use the values given in
Table \ref{tab:H_binding} associated to different substrates. 

Many different expressions of the sticking coefficient have been given
in the literature. For simplicity, we use the same sticking probability
for the physisorption-based LH mechanism and the chemisorption-based
ER mechanism, except for the additional presence of a barrier in the
ER case. We use the expression from \citet{lebourlot:12}:
\[
s(T_{\mathrm{gas}})=\frac{1}{1+\left(\frac{T_{\mathrm{gas}}}{T_{2}}\right)^{\beta}}
\]
using $T_{2}=464\,\mathrm{K}$ and $\beta=1.5$, which gives results
close to the expression of \citet{sternberg:95}.

We also simply assume an equipartition of the formation energy among
the kinetic energy of the newly formed molecule, its excitation energy,
and the heating of the grain.

The total volumic $\mathrm{H}_{2}$ formation rate is often parametrized
as $R_{\mathrm{H}_{2}}=R_{f}\, n_{\mathrm{H}}\, n(\mathrm{H})$ to
factor out the dependency to the atomic $\mathrm{H}$ abundance $n(\mathrm{H})$
and to the dust abundance, which is proportional to the total proton
density $n_{\mathrm{H}}$, coming from the collision rate between
gas $\mathrm{H}$ atoms and dust grains. The standard value of $R_{\mathrm{f}}$
is $3\times10^{-17}\mathrm{cm^{3}.s^{-1}}$. We also use the formation
efficiency defined for one grain as $\frac{2\, r_{\mathrm{H}_{2}}}{k_{\mathrm{coll}}}$,
which represents the fraction of the $\mathrm{H}$ atoms colliding
with the grain that are turned into $\mathrm{H}_{2}$.

\subsubsection{Surface chemistry via Eley-Rideal mechanism\label{sub:ER-processes}}

We also consider the chemisorption-based ER mechanism. The $\mathrm{H}$
atoms of the gas phase that hit the grain can bind to the surface
by a covalent bond after overcoming a repulsive barrier. This process
is called chemisorption (see for instance \citet{Jeloaica99}). Once
chemisorbed, the atoms either evaporate and return to the gas phase,
or react if another gas $\mathrm{H}$ atom lands on the same adsorption
site.

We assume the chemisorption sites to be similarly distributed with
the same inter-site distance $d_{\mathrm{s}}=0.26\,\mathrm{nm}$.
Each site is a potential well of depth $E_{\mathrm{chem}}$ and is
separated from the free state by a potential barrier of height $E_{\mathrm{barr}}$.

The collision rate $k_{\mathrm{coll}}$ is the same as in the LH case.
However, we have to take the presence of the chemisorption barrier
into acount in the sticking coefficient. Following \citet{lebourlot:12},
we use a sticking probability $s(T_{\mathrm{gas}})=\exp\left(-\frac{E_{\mathrm{barr}}}{k\, T_{\mathrm{gas}}}\right)\,\left(1+\left(\frac{T_{\mathrm{gas}}}{T_{2}}\right)^{\beta}\right)^{-1}$,
where the second term accounts for rebound of high energy atoms without
sticking with $T_{\mathrm{barr}}=E_{\mathrm{barr}}/k=300\,\mathrm{K}$,
$T_{2}=464\,\mathrm{K}$, and $\beta=1.5$. The energy barrier for
chemisorption on perfect graphitic surfaces has been found to be of
the order of $0.15\mbox{ eV}-0.2\mbox{ eV}$ (but adsorption in para-dimer
configuration has a much lower barrier of $0.04\mbox{ eV}$) \citep{sha02,kerwin:08}.
As discussed in \citet{lebourlot:12}, the choice made here of a lower
barrier for chemisorption is based on the idea that real grain surfaces
are not perfect flat surfaces like graphite, but present a large number
of defects. Surface defects are expected to induce a strongly reduced
barrier for chemisorption \citep{Ivanovskaya2010,casolo13}. Edge
sites on PAHs behave in a similar way \citep{bonfanti:11}. Moreover,
\citet{Mennella2008} found that chemisorption in aliphatic $\mathrm{CH_{2,3}}$
groups could also lead to efficient Eley-Rideal formation with a very
low chemisorption barrier.

Finally, only atoms arriving on an empty adsorption site stick, corresponding
to a probability $1-\frac{n}{N_{\mathrm{s}}}$, where $n$ is the
number of chemisorbed atoms.

\begin{table}
\caption{Model parameters for chemisorption. $T_{\mathrm{chem}}$ is the chemisorption
binding energy and $T_{\mathrm{barr}}$ the chemisorption barrier.
$d_{\mathrm{s}}$ is the distance between sites. See references in
the text.\label{tab:H_binding-1}}

\centering%
\begin{tabular}{ccc}
\hline 
\hline $T_{\mathrm{chem}}$ & $T_{\mathrm{barr}}$ & $d_{\mathrm{s}}$\tabularnewline
$(\mathrm{K})$ & $(\mathrm{K})$ & (nm)\tabularnewline
\hline 
$7000$ & $300$ & $0.26$\tabularnewline
\hline 
\end{tabular}
\end{table}

The chemisorption rate is thus 
\[
R_{\mathrm{chem}}=k_{\mathrm{coll}}\, s(T_{\mathrm{gas}})\,\left(1-\frac{n}{N_{\mathrm{s}}}\right).
\]

Atoms hitting an occupied site react with the adsorbed atom to form
a $\mathrm{H}_{2}$ molecule. Some theoretical studies show a very
small activation barrier of $9.2\,\mathrm{meV}$ \citep{morisset:04b}
for this reaction, but the existence of this barrier is under
debate \citep{rougeau:11}. To stay consistent with the model of \citet{lebourlot:12}
to which we compare our results in Sect. \ref{sec:PDR_code}, we neglect
this barrier compared to the stronger adsorption barrier. We found
that including it in the model reduces the formation efficiency by
less than $10\%$ in the range of gas temperatures where the ER mechanism
is important compared to the physisorption-based LH mechanism. 

As the formation reaction releases $4.5\,\mathrm{eV}$, which is much
more than the binding energy, we assume that the new molecule is immediately
released in the gas. The formation rate is then 
\[
R_{\mathrm{form}}=k_{\mathrm{coll}}\,\frac{n}{N_{\mathrm{s}}}.
\]

Last, chemisorbed atoms can evaporate. This effect is usually negligible
at typical grain temperatures, but the fluctuating temperatures of
small grains can easily make excursions into the domain where evaporation
is significant. The adsorbed atoms are supposed to be thermalized
at the grain temperature, and must overcome the well energy $E_{\mathrm{chem}}$.
The desorption process is similar to the LH case, and we take again
$k_{\mathrm{des}}(T)=\nu_{0}\,\exp\left(-\frac{T_{\mathrm{chem}}}{T}\right)$,
where $T_{\mathrm{chem}}=E_{\mathrm{chem}}/k$, and $\nu_{0}=\frac{1}{\pi}\,\sqrt{\frac{2\, E_{\mathrm{chem}}}{d_{0}^{2}\, m_{\mathrm{H}}}}$
(with $d_{0}=0.1\,\textrm{nm}$). The total desorption rate on the
grain is then 
\[
R_{\mathrm{des}}(T)=n\, k_{\mathrm{des}}(T).
\]

Theoretical studies find the chemisorption energy on graphene to be
in the range $0.67\,\mathrm{eV}-0.9\,\mathrm{eV}$ \citep{sha02,lehtinen04,casolo09}.
On small PAHs, \citet{bonfanti:11} find binding energies in the range
$1\,\mathrm{eV}-3\,\mathrm{eV}$ for edge sites and in the range $0.5\,\mathrm{eV}-1\,\mathrm{eV}$
for non edge sites. In this overall range $5800\,\mathrm{K}-35000\,\mathrm{K}$,
we choose a low value of $7000\,\mathrm{K}$ to estimate the maximum
effect of temperature fluctuations on the ER mechanism.

We sum up the model parameters in Table \ref{tab:H_binding-1}.

When neglecting the fluctuations of the grain temperature, we can
calculate the equilibrium surface population $n_{\mathrm{eq}}$ using
the equilibrium rate equation $R_{\mathrm{chem}}-R_{\mathrm{form}}-R_{\mathrm{des}}(T)=0$,
which gives
\begin{equation}
n_{\mathrm{eq}}(T)=N_{\mathrm{s}}\,\frac{s(T_{\mathrm{gas}})}{1+s(T_{\mathrm{gas}})+\frac{N_{\mathrm{s}}\, k_{\mathrm{des}}(T)}{k_{\mathrm{coll}}}},\label{eq:n_eq_ER}
\end{equation}
and the corresponding equilibrium formation rate on one grain at temperature
$T$ is
\begin{equation}
r_{\mathrm{H}_{2}}^{\mathrm{eq}}(T)=k_{\mathrm{coll}}\frac{n_{\mathrm{eq}}(T)}{N_{\mathrm{s}}}.\label{eq:rh2_eq_ER}
\end{equation}

However, when temperature fluctuations are taken into account, this
equilibrium situation is not valid. The method we use to calculate
this effect is described in Section \ref{sub:Method}.

The energy liberated from the formation reaction is split between
the excitation and kinetic energy of the molecule and the heating
of the grain. As in \citet{lebourlot:12}, we take the results of
\citet{sizun:10} for the ER mechanism : $2.7\,\mathrm{eV}$ goes
into the excitation energy of the molecule, $0.6\,\mathrm{eV}$ into
its kinetic energy, and $1\,\mathrm{eV}$ is given to the grain.

We must note that the part of the formation energy given to the grain
creates another coupling between surface population and grain temperature.
This fraction of the formation energy ($1\,\mathrm{eV}$) is not completely
negligible compared to photon energies. This additional retro-coupling
between surface chemistry and temperature is ignored in the statistical
calculation presented in this article. However, its effect on the
equilibrium situation when neglecting fluctuations is evaluated in
Appendix \ref{app:form_energy}. It is found to be negligible as the
power given to the grain is usually much smaller that the radiative
power it receives.

We can again define the formation parameter $R_{f}$ as $R_{\mathrm{H}_{2}}=R_{f}\, n_{\mathrm{H}}\, n(\mathrm{H})$
and the formation efficiency for one grain as $\frac{2\, r_{\mathrm{H}_{2}}}{k_{\mathrm{coll}}}$.

\subsection{Method\label{sub:Method}}

To take into account temperature fluctuations, we adopt a statistical
point of view, and we describe the statistical properties of the fluctuating
variables.

The necessary statistical information on our system is contained in
the probability density function (PDF) $f(X)$, giving the probability
to find the system in the state $X$ that is defined by the two variables
$E$, the thermal energy of the grain (equivalent to its temperature
$T$), and $n$, the number of adsorbed H atoms on its surface. The
function $f$ is thus the joint PDF in the two variables. As we treat
our two formation mechanisms independently, $n$ counts chemisorbed
\emph{or }physisorbed atoms depending on which mechanism we are discussing.

The time evolution of this PDF is governed by the master equation:
\[
\frac{df(X)}{dt}=\int_{\mathrm{states}}dY\, p_{Y\rightarrow X}\, f(Y)-\int_{\mathrm{states}}dY\, p_{X\rightarrow Y}\, f(X),
\]
where $p_{X\rightarrow Y}$ is the transition rate from state $X$
to $Y$. This equation expresses that the probability of being in
a given state varies in time due to the imbalance between the rate
of arrivals and the rate of departures. In a stationary situation,
the two rates, which we call as the gain and loss rates respectively,
must compensate each other. 

We are thus looking for a solution to the equation
\[
\int_{\mathrm{states}}dY\, p_{Y\rightarrow X}\, f(Y)-f(X)\int_{\mathrm{states}}dY\, p_{X\rightarrow Y}=0.
\]

Two different kinds of transitions occur: the emission or absorption
of a photon, which changes only the thermal energy $E$ of the grain,
and the adsorption, desorption, or reaction of hydrogen atoms, which
changes only the surface population $n$. The two variables are only
coupled by the transition rates for the chemical events, which depend
on the grain temperature. The population fluctuations are thus entirely
driven by the temperature fluctuations with no retroactions (as mentioned
before we neglect grain heating by the formation process).

As the transition mechanisms modify only one variable at a time, we
split each of the gain and loss terms into two parts, respectively
for photon processes (affecting only $E$) and for surface chemical
processes (affecting only $n$),
\begin{equation}
G_{E}(E,n)+G_{n}(E,n)-L_{E}(E,n)-L_{n}(E,n)=0,\label{eq:main_master}
\end{equation}
with 
\[
G_{E}(E,n)=\int_{0}^{+\infty}dE'\, p_{E'\rightarrow E}\, f(E',n)
\]
\[
G_{n}(E,n)=\sum_{n'=0}^{N_{s}}p_{n'\rightarrow n}(E)\, f(E,n')
\]
\[
L_{E}(E,n)=f(E,n)\int_{0}^{+\infty}dE'\, p_{E\rightarrow E'}
\]
\[
L_{n}(E,n)=f(E,n)\sum_{n'=0}^{N_{s}}p_{n\rightarrow n'}(E),
\]
where we have simplified the notations for the transition rates as
transitions only affect one variable at a time and the transitions
affecting $E$ (photon absorptions or emissions) have rates that are
independent of $n$.

We now first show how the thermal energy marginal PDF $f_{E}(E)$
(or the equivalent temperature PDF $f_{T}(T)$) can be derived from
our formalism, which reduces to equations similar to previous works
\citep{desert86,li2001b}. We then show how the formation rates in
the LH case and in the ER case can be derived.

\subsubsection{Thermal energy PDF\label{sub:Thermal-energy-distribution}}

Summing Eq. \ref{eq:main_master} over all $n$ values, we obtain
\begin{equation}
f_{E}(E)\int_{-\infty}^{+\infty}dE'\, p_{E\rightarrow E'}=\int_{-\infty}^{+\infty}dE'\, p_{E'\rightarrow E}\, f_{E}(E'),\label{eq:pre-thermal}
\end{equation}
where $f_{E}(E)=\sum_{n}f(E,n)$ is the marginal PDF for the single
variable $E$. The marginal PDF $f_{E}$ can easily be converted into
a grain temperature PDF $f_{T}$ through the relationship
\[
f_{T}(T)=f_{E}(E)\, C(T).
\]

We thus obtain an independent master equation for the temperature
fluctuations. 

Detailing the transition rates in Eq. \ref{eq:pre-thermal}, we can
rewrite it as
\begin{multline}
f_{E}(E)=\\
\frac{\int_{0}^{E}dE'\, R_{\mathrm{abs}}(E-E')\, f_{E}(E')+\int_{E}^{+\infty}dE'\, R_{\mathrm{em}}(E'-E,T(E'))\, f_{E}(E')}{\int_{E}^{+\infty}dE'\, R_{\mathrm{abs}}(E'-E)+\int_{0}^{E}dE'\, R_{\mathrm{em}}(E-E',T(E))}.\label{eq:thermal}
\end{multline}

This equation is an eigenvector equation for the linear integral operator
defined by the right-hand side: $f_{E}(E)=\mathcal{L}[f_{E}](E)$.
A stationary PDF for the grain thermal energy is a positive and normalized
eigenvector of this operator for the eigenvalue $1$. The existence
and unicity of this eigenvector is proven in Appendix \ref{app:Eigenvalues-f_E}.
Moreover, the other eigenvalues all have real parts that are lower
than $1$ (see Appendix \ref{app:Eigenvalues-f_E}). We can thus converge
toward the solution by building a sequence $\mathcal{L}^{n}[f_{E}]$
(the exponent refers to operator composition) from an initial guess.
We solve this equation numerically by choosing an energy grid $\left\{ E_{i}\right\} $
and solving for piecewise linear functions on this grid.

\subsubsection{Eley-Rideal mechanism\label{sub:Eley-Rideal-mechanism}}

We first present the method for the ER mechanism as the linearity
of the chemical rates in $n$ simplifies the problem significantly.
We can avoid directly solving the full master equation for the joint
PDF in two variables (Eq. \ref{eq:main_master}).

We are interested in the average ER formation rate 
\[
\left\langle R_{\mathrm{form}}\right\rangle =\int_{0}^{+\infty}dE\sum_{n=0}^{N_{s}}f(E,n)\, k_{\mathrm{coll}}\,\frac{n}{N_{\mathrm{s}}}.
\]
Knowing $f_{E}(E)$, we can rewrite it as :
\[
\left\langle R_{\mathrm{form}}\right\rangle =\int_{0}^{+\infty}dE\, f_{E}(E)\, k_{\mathrm{coll}}\,\frac{\left\langle n\mid E\right\rangle }{N_{s}},
\]
 where $\left\langle n\mid E\right\rangle =\sum_{n=0}^{N_{s}}n\frac{f(E,n)}{f_{E}(E)}$
is the conditional expectation of the surface population at thermal
energy $E$. It is by definition the expected value (or ensemble average)
of the surface population of a grain, \emph{knowing that }this grain
has thermal energy $E$. Multiplying Eq~(\ref{eq:main_master}) by
$n$ and summing over all $n$ values gives the equation governing
this quantity $\left\langle n\mid E\right\rangle $,
\begin{equation}
\frac{k_{\mathrm{coll}}\, s}{M(E)}=\left\langle n\mid E\right\rangle -\int_{0}^{+\infty}dE'\,\left\langle n\mid E'\right\rangle \, K(E,E'),\label{eq:ER_equation_first_form}
\end{equation}
where
\begin{multline*}
M(E)=\int_{E}^{+\infty}dE'\, R_{\mathrm{abs}}(E'-E)+\int_{0}^{E}dE'\, R_{\mathrm{em}}(E-E')\\
+\frac{k_{\mathrm{coll}}(1+s)}{N_{s}}+k_{\mathrm{des}}(E)
\end{multline*}
 and
\[
K(E,E')=\begin{cases}
{\displaystyle \frac{f_{E}(E')\, R_{\mathrm{abs}}(E-E')}{f_{E}(E)\, M(E)}} & \mathrm{if}\qquad E'\in[0,E]\\
\\
{\displaystyle \frac{f_{E}(E')\, R_{\mathrm{em}}(E'-E,T(E'))}{f_{E}(E)\, M(E)}} & \mathrm{if}\qquad E'>E
\end{cases}.
\]

After a similar discretization, this is a linear system of equations.
However, while being nonsingular, it converges exponentially fast
toward a singular system as the grain size $a$ grows, making standard
numerical procedures unusable.

To avoid this problem, we rewrite this equation. Multiplying by $M(E)f_{E}(E)$
and integrating it over $E$ yields 
\begin{equation}
k_{\mathrm{coll}}\, s=\int_{0}^{+\infty}dE'\left[\frac{k_{\mathrm{coll}}(1+s)}{N_{s}}+k_{\mathrm{des}}(E')\right]f_{E}(E')\,\left\langle n\mid E'\right\rangle .\label{eq:norm_ER}
\end{equation}
Dividing again by $M(E)$ and subtracting it from the initial equation
gives
\begin{equation}
0=\left\langle n\mid E\right\rangle -\int_{0}^{+\infty}dE'\,\left\langle n\mid E'\right\rangle \, K'(E,E'),\label{eq:ER-nbar}
\end{equation}
 where $K'(E,E')=K(E,E')+\frac{k_{\mathrm{coll}}(1+s)+N_{\mathrm{s}}\, k_{\mathrm{des}}(E')}{N_{\mathrm{s}}\, M(E)}\, f_{E}(E')$.
This is again an eigenvector equation for the eigenvalue $1$ of a
linear integral operator. After discretization, we numerically compute
this eigenvector and use Equation \ref{eq:norm_ER} as a normalization
condition. This procedure proved to work on the entire range of relevant
grain sizes.

\subsubsection{Langmuir-Hinshelwood mechanism\label{sub:LH-method}}

For the LH mechanism, the formation rate contains a quadratic term.
When trying to apply a similar method to the LH mechanism, the equivalent
of Eq.~(\ref{eq:ER-nbar}) is then an infinite system of equations
on the conditional moments of the population $\left\langle n\mid E\right\rangle $,
$\left\langle n^{2}\mid E\right\rangle $, $\left\langle n^{3}\mid E\right\rangle $...
We thus directly solve the main master equation (Eq. \ref{eq:main_master}),
despite the computational burden.

Writing explicitly the transition rates in Eq. \ref{eq:main_master},
we get
\begin{multline}
0=\int_{0}^{E}dE'\, R_{abs}(E-E')\, f(E',n)\\
+\int_{E}^{+\infty}dE'\, R_{em}(E'-E,T(E'))\, f(E',n)\\
+f(E,n-1)k_{coll}s(T_{gas})\left(1-\frac{n-1}{N_{s}}\right)\\
+f(E,n+1)\left[(n+1)\, k_{des}(T)+k_{coll}\frac{n+1}{N_{s}}\right]\\
+f(E,n+2)k_{mig}(T)\frac{(n+2)(n+1)}{N_{s}}\\
-\left[\int_{E}^{+\infty}dE'\, R_{\mathrm{abs}}(E'-E)+\int_{0}^{E}dE'\, R_{\mathrm{em}}(E-E')\right]\, f(E,n)\\
-f(E,n)\left[k_{coll}s(T_{gas})\left(1-\frac{n}{N_{s}}\right)+n\, k_{des}(T)+k_{coll}\frac{n}{N_{s}}\right.\\
\left.+k_{mig}(T)\frac{n(n-1)}{N_{s}}\right],\label{eq:LH-master-raw}
\end{multline}
where, as boundary conditions, all expressions $n-1$, $n+1$ and
$n+2$ are implicitly taken to be 0 if they become negative or greater
than $N_{s}$.

We define the total loss rate as
\begin{multline*}
Q(E,n)=\int_{E}^{+\infty}dE'\, R_{\mathrm{abs}}(E'-E)+\int_{0}^{E}dE'\, R_{\mathrm{em}}(E-E')\\
+k_{coll}s(T_{gas})\left(1-\frac{n}{N_{s}}\right)+n\, k_{des}(T)+k_{coll}\frac{n}{N_{s}}+k_{mig}(T)\frac{n(n-1)}{N_{s}},
\end{multline*}
the integral operator (for photon induced transitions) as
\begin{multline*}
\mathcal{G}[f](E,n)=\int_{0}^{E}dE'\, R_{abs}(E-E')\, f(E',n)\\
+\int_{E}^{+\infty}dE'\, R_{em}(E'-E,T(E'))\, f(E',n),
\end{multline*}
and the discrete jump operator (for chemical transitions) as
\begin{multline*}
\mathcal{J}[f](E,n)=\\
f(E,n-1)k_{coll}s(T_{gas})\left(1-\frac{n-1}{N_{s}}\right)\\
+f(E,n+1)\left[(n+1)\, k_{des}(T)+k_{coll}\frac{n+1}{N_{s}}\right]\\
+f(E,n+2)k_{mig}(T)\frac{(n+2)(n+1)}{N_{s}}.
\end{multline*}

We can then rewrite Eq. \ref{eq:LH-master-raw} in a simplified form
as 
\begin{equation}
f(E,n)=\frac{1}{Q(E,n)}\left[\mathcal{G}[f](E,n)+\mathcal{J}[f](E,n)\right].\label{eq:LH-master-simple}
\end{equation}

This is an eigenvector equation for the eigenvalue $1$, which is
formally similar to Eq. \ref{eq:thermal}. For the same reasons as
previously, we expect the operator defined by the right hand side
to have all its eigenvalues with real parts that are strictly lower
than 1, except the eigenvalue 1, which is simple. We can thus find
the solution by iterating the application of the operator from an
initial guess.

The equation is solved numerically using this iterative procedure.
The computation time is found to explode when increasing the grain
size $a$. In addition to the number of possible values of $n$ increasing
as $a^{2}$, the number of iterations necessary to converge toward
the stationary solution is also found to quickly increase with $a$.
Grouping the values of $n$ in bins to reduce the matrix size does
not solve the problem, as the reduction of the computation time due
to the smaller matrices is found to be balanced by a roughly equivalent
increase in the number of iterations necessary to converge to a given
stationarity threshold.

We can thus only use this method up to moderate grain sizes ($\sim20\mbox{ nm}$
with 2 days of computation). In later sections, we see, however, that
this is sufficient to observe the range of sizes for which the temperature
fluctuations significantly impact the chemistry. In addition, a much
quicker yet sufficiently accurate approximation is presented in Sect.
\ref{sec:Approximations}.

\section{Approximations\label{sec:Approximations}}

In this section, we construct fast approximations of the formation
rate based on simple physical arguments and compare their results
with those of the exact method presented in Sect. \ref{sub:Method}.
An approximation is especially necessary for the LH mechanism to avoid
the prohibitive computational cost of the exact method. A similar
approximation is given for completeness for the ER mechanism. Those
methods assume the temperature PDF $f_{T}(T)$ to be known. It can
be computed, for instance, by the method described in Sect. \ref{sub:Thermal-energy-distribution}.

\subsection{Approximation of the Langmuir-Hinshelwood formation rate\label{sub:LH-Approx}}

We build our method around an approximation of $\left\langle \left.n\right|T\right\rangle $,
the average population on the grains that are at a given temperature
$T$. We first write a balance equation for this average population,
taking into account both chemical processes (changing only the grain's
population), and grains leaving or reaching this temperature $T$.

We first consider the chemical processes. Those are described in Sect.
\ref{sub:LH_processes}:
\begin{itemize}
\item The grain can gain atoms through adsorption (average rate $k_{ads}=k_{coll}s(T_{gas})\left(1-\frac{\left\langle \left.n\right|T\right\rangle }{N_{sites}}\right)$).
\item The grain can lose atoms through desorption (rate $\left\langle \left.n\right|T\right\rangle k_{des}(T)$).
\item The grain can lose atoms through LH reaction, a term we express later.
For now, let us write it as $\left\langle \left.r_{H_{2}}^{LH}\right|T\right\rangle $.
We discuss this term in depth below.
\item The grain can lose atoms through direct ER reaction (rate $k_{coll}\frac{\left\langle \left.n\right|T\right\rangle }{N_{sites}}$).
\end{itemize}
On the other hand, the grain can also leave the temperature $T$.
At equilibrium, the rate at which grains leave the state $T$ is exactly
balanced by the rate of arrivals from other states. The net effect
on the conditional average $\left\langle \left.n\right|T\right\rangle $
can be a loss or a gain depending on whether the average population
of the grains arriving from other states is higher or lower than $\left\langle \left.n\right|T\right\rangle $.
In general, noting $k_{\mathrm{leave}}(T)$, the rate at which grains
leave the state $T$, and $\left\langle \left.n_{\mathrm{arrivals}}\right|T\right\rangle $,
the average population on grains arriving in state $T$, the net gain
rate is $k_{\mathrm{leave}}(T)\left(\left\langle \left.n_{\mathrm{arrivals}}\right|T\right\rangle -\left\langle \left.n\right|T\right\rangle \right)$.

Up to now, no approximation has been made.

As we expected high temperatures states, where desorption dominates
all other processes, to have negligible formation and extremely low
average population, we focus our approximation on a low temperature
regime.

We want to estimate what fraction of the grains leaving state $T$
will come back bare (or with a negligible population compared to $\left\langle \left.n\right|T\right\rangle $).
We make the following approximation : Grains leaving $T$ to reach
the regime where desorption dominates, come back with no population.
Fluctuations that do not reach the desorption regime leave the surface
population unchanged. We define the temperature $T_{lim}$ that delimits
these regimes as the temperature at which the desorption rate for
one atom becomes equal to the adsorption rate for the grain (we call
$E_{\mathrm{lim}}$ the corresponding thermal energy):
\[
T_{\mathrm{lim}}=\frac{T_{\mathrm{phys}}}{\ln\left(\frac{\nu_{0}}{k_{\mathrm{coll}}\, s(T_{\mathrm{gas}})}\right)}.
\]

We thus define
\begin{equation}
k_{phot}(T)=\int_{\max(E_{\mathrm{lim}}-E(T),0)}^{+\infty}dU\, R_{abs}(U),\label{eq:k_phot}
\end{equation}
the rate at which grains leave state $T$ to reach a temperature above
the limit. The net loss rate due to fluctuations is then $k_{phot}(T)\left\langle \left.n\right|T\right\rangle $.

We have thus obtained a rate equation for the conditional average
population at temperature $T$:
\begin{multline}
k_{coll}s(T_{gas})\left(1-\frac{\left\langle \left.n\right|T\right\rangle }{N_{sites}}\right)=\\
\left(k_{des}(T)+k_{phot}(T)\right)\left\langle \left.n\right|T\right\rangle +k_{coll}\frac{\left\langle \left.n\right|T\right\rangle }{N_{sites}}+2\left\langle \left.r_{H_{2}}^{LH}\right|T\right\rangle .\label{eq:approxLH_rate_eq}
\end{multline}

This equation is in closed form if we can express $\left\langle \left.r_{H_{2}}^{LH}\right|T\right\rangle $
as a function of $\left\langle \left.n\right|T\right\rangle $ only.
We distinguish the two regimes $\left\langle \left.n\right|T\right\rangle >1$
and $\left\langle \left.n\right|T\right\rangle <1$. From now on,
we simplify the notations by noting $s=s(T_{\mathrm{gas}})$.

\subsubsection{Regime with $\left\langle \left.n\right|T\right\rangle >1$}

In this regime, we assume that the discrete nature of the number of
surface atoms is negligible and treat this variable as continuous.
We then have $\left\langle \left.r_{H_{2}}^{LH}\right|T\right\rangle =k_{mig}(T)\frac{\left\langle \left.n^{2}\right|T\right\rangle }{N_{sites}}$.
We also know that when $\left\langle \left.n\right|T\right\rangle \gg1$,
direct ER reaction is likely to dominate so that a precise determination
of the LH reaction rate is not necessary. We thus make the simple
approximation $ $$\left\langle \left.n^{2}\right|T\right\rangle =\left\langle \left.n\right|T\right\rangle ^{2}$.
Using this approximation in Eq. \ref{eq:approxLH_rate_eq}, we then
get
\begin{multline}
\left\langle \left.n\right|T\right\rangle =\frac{1+s}{4}\frac{k_{coll}}{k_{mig}(T)}\left(1+\frac{N_{sites}(k_{des}(T)+k_{phot}(T))}{k_{coll}(1+s)}\right)\\
\times\left[\sqrt{1+\frac{8s}{(1+s)^{2}}\frac{k_{mig}(T)}{k_{coll}}\frac{N_{sites}}{\left(1+\frac{N_{sites}(k_{des}(T)+k_{phot}(T))}{k_{coll}(1+s)}\right)^{2}}}-1\right].
\end{multline}

If this results becomes smaller than $1$, we switch to the other
regime.

\subsubsection{Regime with $\left\langle \left.n\right|T\right\rangle <1$}

In this regime, discretisation effects are very important. Formation
is dominated by LH reaction, which can only happen if two atoms are
present on the grain at the same time. We use reasoning that is similar
to the modified rate equation approach developed by \citet{garrod08}.

As the average population is low, we can make the approximation $\left\langle \left.n\right|T\right\rangle \simeq p(n=1\,|\, T)$,
where the right hand side is the conditional probability of having
one surface atom knowing that the temperature is $T$. The LH formation
rate can then be computed the following way. The formation rate is
the rate at which gas atoms fall on a grain that already had one adsorbed
atom AND reaction occurs before any other process removes one of the
atoms. The first part of the sentence gives a rate 
\[
k_{coll}\, s(1-\frac{1}{N_{sites}})\, p(n=1\,|\, T)\simeq k_{coll}\, s(1-\frac{1}{N_{sites}})\,\left\langle \left.n\right|T\right\rangle 
\]
 using the previous approximation. Once we have two atoms on the grain,
the probability that they react before anything else removes one atom
is
\[
P=\frac{\frac{2k_{mig}(T)}{N_{sites}}}{\frac{2k_{mig}(T)}{N_{sites}}+\frac{2k_{coll}}{N_{sites}}+2(k_{des}(T)+k_{phot}(T))}.
\]
This gives
\[
\left\langle \left.r_{H_{2}}^{LH}\right|T\right\rangle =\frac{k_{coll}\, s(1-\frac{1}{N_{sites}})}{1+\frac{k_{coll}}{k_{mig}(T)}+\frac{N_{sites}(k_{des}(T)+k_{phot}(T))}{k_{mig}(T)}}\,\left\langle \left.n\right|T\right\rangle .
\]

We inject this expression in Eq. \ref{eq:approxLH_rate_eq} and finally
find the solution for this regime:
\begin{equation}
\left\langle \left.n\right|T\right\rangle =\frac{N_{sites}}{N_{sites}\frac{(k_{des}(T)+k_{phot}(T))}{k_{coll}s}+\frac{1+s}{s}+\frac{2(N_{sites}-1)}{1+\frac{k_{coll}}{k_{mig}(T)}+N_{sites}\frac{k_{des}(T)+k_{phot}(T)}{k_{mig(T)}}}}.
\end{equation}

\subsubsection{Total formation rate}

The total formation rate is then computed by integrating over the
temperature PDF $f(T)$ and distinguishing the two regimes:
\begin{multline}
\left\langle r_{H_{2}}\right\rangle =\int_{0}^{T_{switch}}dT\, f(T)\,\left(\frac{k_{coll}}{N_{sites}}\left\langle \left.n\right|T\right\rangle +\frac{k_{mig}(T)}{N_{sites}}\left\langle \left.n\right|T\right\rangle ^{2}\right)+\\
\int_{T_{switch}}^{+\infty}dT\, f(T)\,\times\\
\left(\frac{k_{coll}}{N_{sites}}\left\langle \left.n\right|T\right\rangle +\frac{k_{coll}\, s(1-\frac{1}{N_{sites}})}{1+\frac{k_{coll}}{k_{mig}(T)}+\frac{N_{sites}(k_{des}(T)+k_{phot}(T))}{k_{mig}(T)}}\,\left\langle \left.n\right|T\right\rangle \right),
\end{multline}
where $T_{switch}$ is the temperature at which $\left\langle \left.n\right|T\right\rangle $
becomes $<1$ ($\left\langle \left.n\right|T\right\rangle >1$ for
$T<T_{switch}$, and $\left\langle \left.n\right|T\right\rangle <1$
for $T>T_{switch}$).

This approximation is compared to the exact method of Sect. \ref{sub:LH-method}
in Appendix \ref{sec:Approx_comparison}. It is found to give a very
accurate estimate of the total average formation rate (within at most
6\%). This approximation is used in all results shown in the following
sections as the exact method is not practically usable.

\subsection{Approximation of the Eley-Rideal formation rate \label{sub:ER-Appprox}}

A similar approximation can be constructed in the case of the ER mechanism.
We again write a rate equation for the conditional average $\left\langle \left.n\right|T\right\rangle $,
including a fluctuation loss term. We use the same approximation for
this loss term with the limiting temperature being
\[
T_{\mathrm{lim}}=\frac{T_{\mathrm{chim}}}{\ln\left(\frac{\nu_{0}}{k_{\mathrm{coll}}\, s(T_{\mathrm{gas}})}\right)},
\]
and the loss rate being $k_{\mathrm{phot}}(T)\left\langle \left.n\right|T\right\rangle $
with $k_{\mathrm{phot}}(T)$ as defined as previously by Eq. \ref{eq:k_phot}.
The resulting rate equation is then directly obtained in closed form:
\[
k_{coll}s(T_{gas})\left(1-\frac{\left\langle \left.n\right|T\right\rangle }{N_{sites}}\right)=\left(k_{des}(T)+k_{phot}(T)\right)\left\langle \left.n\right|T\right\rangle +k_{coll}\frac{\left\langle \left.n\right|T\right\rangle }{N_{sites}},
\]
and no further approximation is needed. The solution is
\[
\left\langle \left.n\right|T\right\rangle =\frac{s}{1+s}N_{sites}\frac{1}{1+\frac{N_{sites}}{1+s}\left(\frac{k_{des}(T)+k_{phot}(T)}{k_{coll}}\right)},
\]
and the average $\mathrm{H}_{2}$ formation rate can then be computed
as
\[
\left\langle r_{H_{2}}\right\rangle =\int_{0}^{+\infty}dT\, f(T)\,\frac{k_{coll}}{N_{sites}}\left\langle \left.n\right|T\right\rangle .
\]

A comparison of this method with the exact result is also performed
in Appendix \ref{sec:Approx_comparison}. However, as the exact method
is easily tractable, the approximation is not used in the results
presented in the rest of this article.

\section{Results\label{sec:Results} }

The method described in the previous two sections was implemented
as a stand alone code called \texttt{Fredholm}. The temperature PDF
is computed using the exact method of Sect. \ref{sub:Thermal-energy-distribution}
and the ER formation rate using the exact method of Sect. \ref{sub:Eley-Rideal-mechanism},
while we use the approximation presented in Sect. \ref{sec:Approximations}
for the LH formation rate. This code is now used to study the effect
of the temperature fluctuations on the formation rate and the influence
of the external conditions (gas conditions and radiation field). We
first present a few results on the temperature PDFs before presenting
the results on $\mathrm{H}_{2}$ formation by the ER and LH mechanisms.

\subsection{Dust temperature probability density functions and grain emission\label{sub:Temperature_distributions_results}}

\begin{figure}
\centering\includegraphics[width=1\columnwidth]{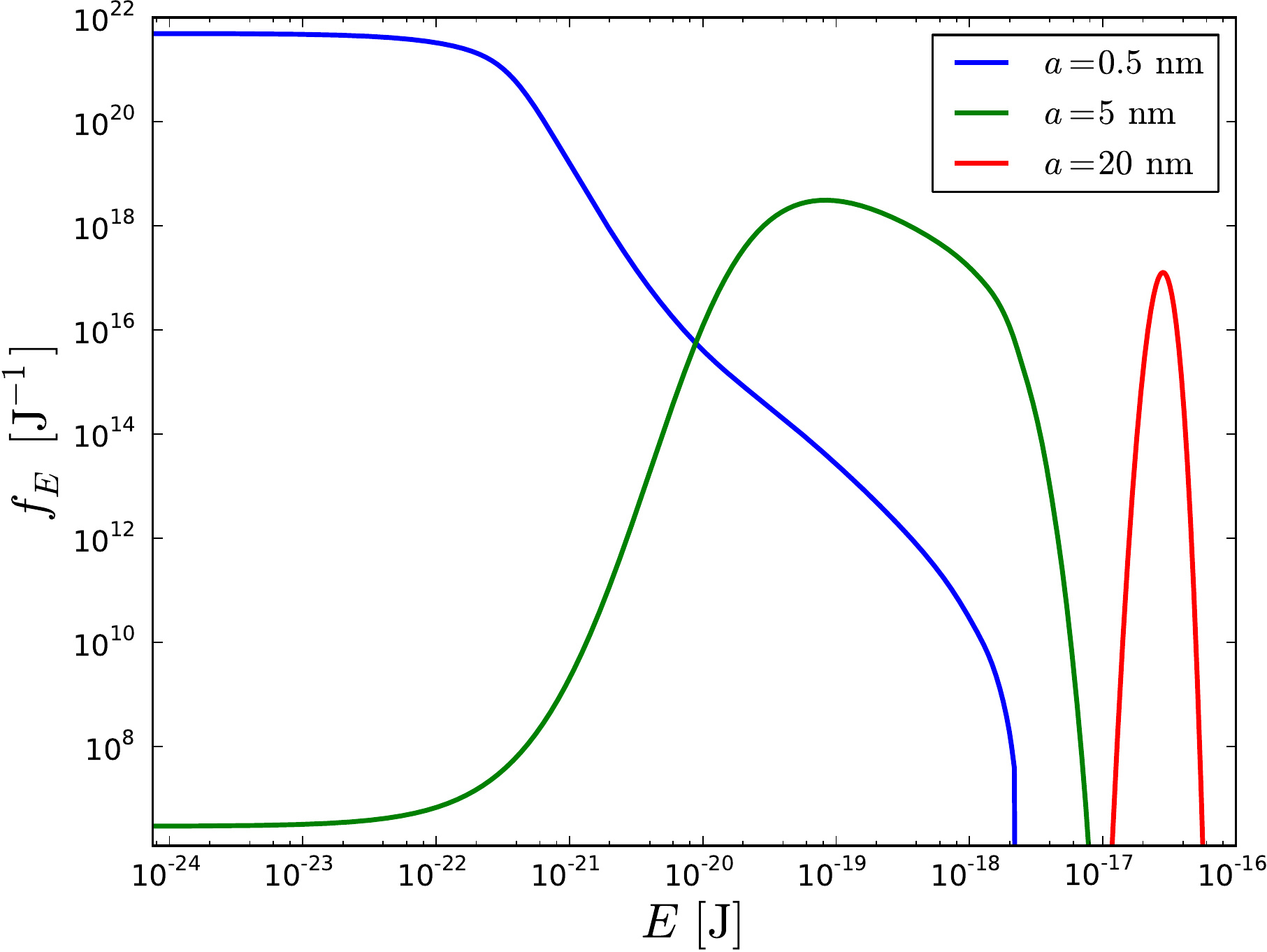}

\caption{Dust thermal energy PDFs for three grain sizes (amorphous carbon grains
under a radiation field with $G_{0}=120$).\label{fig:Temperature_PDFs}}
\end{figure}
\begin{figure}
\centering\includegraphics[width=1\columnwidth]{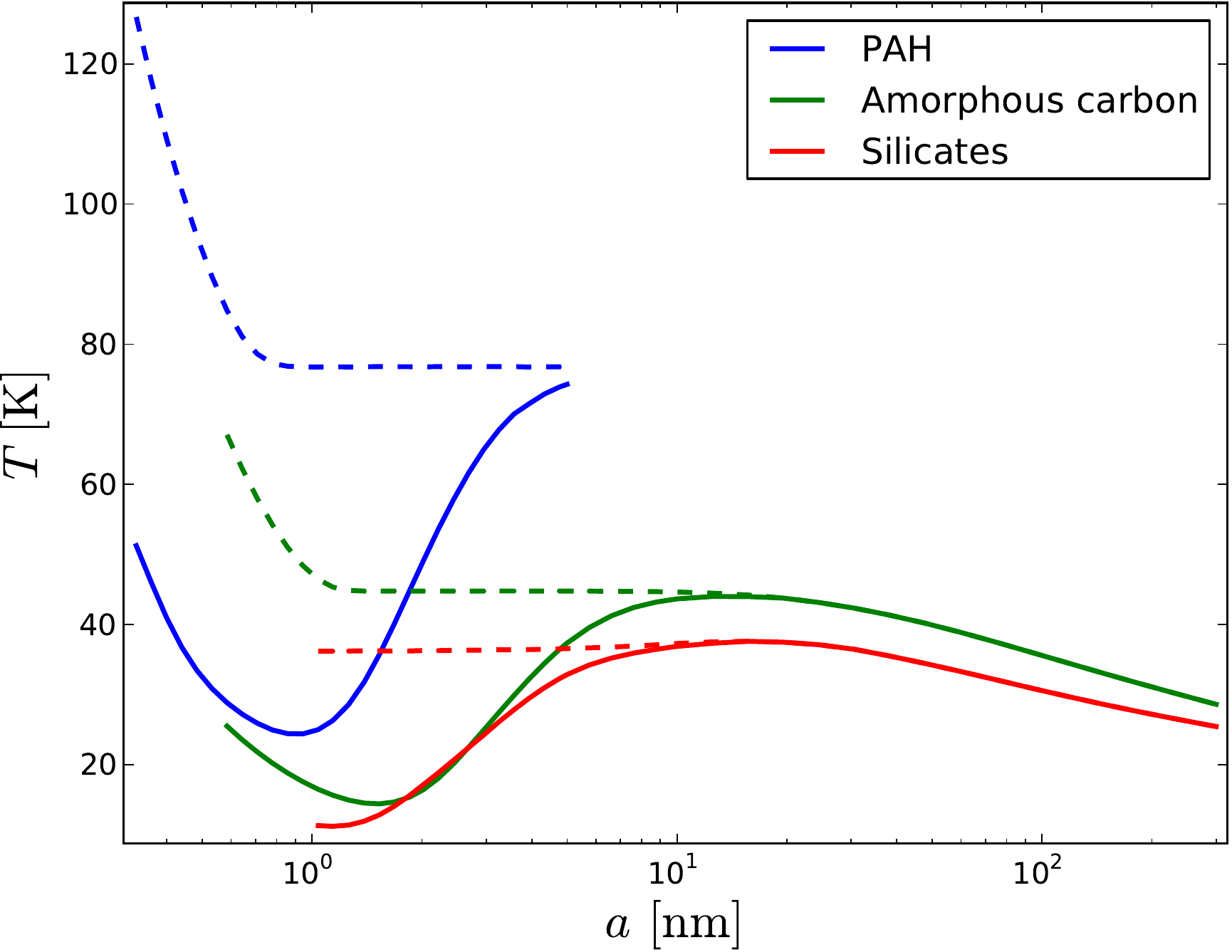}

\caption{Equilibrium temperature (dashed line) and actual average temperature
(solid line), as a function of the grain size for the three types
of grains (radiation field $G_{0}=120$).\label{fig:Mean_temperature}}
\end{figure}

The thermal energy PDFs that we compute are shown on Fig.~\ref{fig:Temperature_PDFs},
which highlights the change of shape as we go from small grains to
bigger grains. For small sizes, most of the grains are in the lowest
energy states (the non-zero limit at zero energy can be easily derived
from the master equation), while the high energy tail extends very
far (the abrupt cut around $2\times10^{-18}\,\mathrm{J}$ is due to
the Lyman cutoff present in the radiation field). As we go to bigger
grains, the fraction of grains in the low energy plateau decreases,
and the PDF takes a more peaked form but keeps a clear asymmetry.
For the bigger sizes, the PDF becomes very sharp around its maximum
and its relative width decreases towards zero, but the shape remains
asymmetrical and could be asymptotically log-normal.

As expected, the average temperature converges toward the usual equilibrium
temperature at large grain sizes but is significantly lower at small
sizes, as shown on Fig.~\ref{fig:Mean_temperature}. High temperature
grains have a much higher contribution to the average energy balance
through their high emissivity. Thus, a very low fraction of high temperature
grains is sufficient to lower the average temperature of the PDF compared
to the equilibrium temperature. The temperature increase at very low
grain sizes, which we can see on both the equilibrium and average
temperatures, is due to the property that the smallest sizes have
decreased cooling efficiency (as they cannot emit photons with higher
energies than their own internal energy, their emission spectrum is
reduced).

\begin{figure}
\centering\includegraphics[width=1\columnwidth]{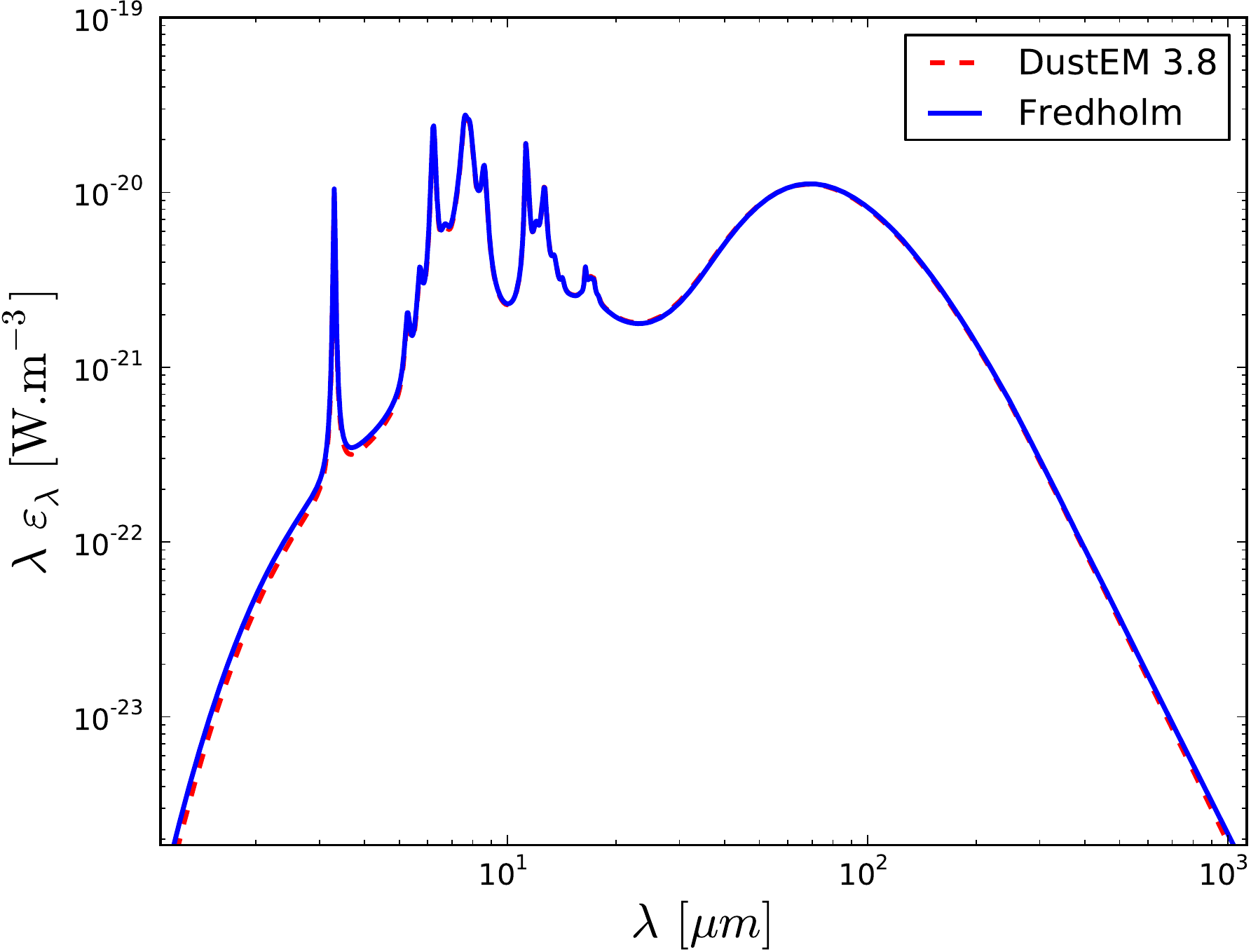}

\caption{Local dust emissivity (emitted power per unit of volume) as a function
of wavelength for a dust population corresponding to \citet{compiegne:11}
under a radiation field with $G_{0}=100$. We compare our code, \textsf{Fredholm},
with \textsf{DustEM}.\label{fig:compa_dustem}}
\end{figure}

This part of the computation has been done before \citep{desert86,li2001b}
and especially in the \texttt{DustEM} code \citep[see][]{compiegne:11}
to compute the infrared emission of dust. We checked the results of
our code, which is named \texttt{Fredholm}, by comparing both the
temperature PDFs, and the derived grain emissivity to the public \texttt{DustEM}
code. We found a good agreement between the results, and despite small
differences in the temperature PDFs (due mainly to the continuous
cooling approximation in \texttt{DustEM}), the final emissivities
are extremely close. Figure~\ref{fig:compa_dustem} shows the final
comparison on a dust population corresponding to \citet{compiegne:11}
with PAHs, small carbonaceous grains, large carbonaceous grains, and
large silicate grains. 

\subsection{$\mathrm{H}_{2}$ formation rate : LH mechanism\label{sub:Results-LH}}

We apply the approximation method described in Sect. \ref{sec:Approximations}
to compute the $\mathrm{H}_{2}$ formation rate on physisorption sites.
The results presented here concern unshielded gas in which the equilibrium
rate equation method gives very inefficient LH formation rate due
to high dust temperatures (i.e., PDR edges). For results in a more
shielded gas, see Sect. \ref{sec:PDR_code} describing the coupling
with full cloud simulations.

The results shown in this section use binding and barrier energies
corresponding to either ices or amorphous carbon surfaces (see Table
\ref{tab:H_binding}) as specified for each figure. Silicates have
intermediate values and are thus not shown here. On the other hand,
the optical and thermal properties of the grains are those of amorphous
carbon grains.

\begin{figure}
\includegraphics[width=1\columnwidth]{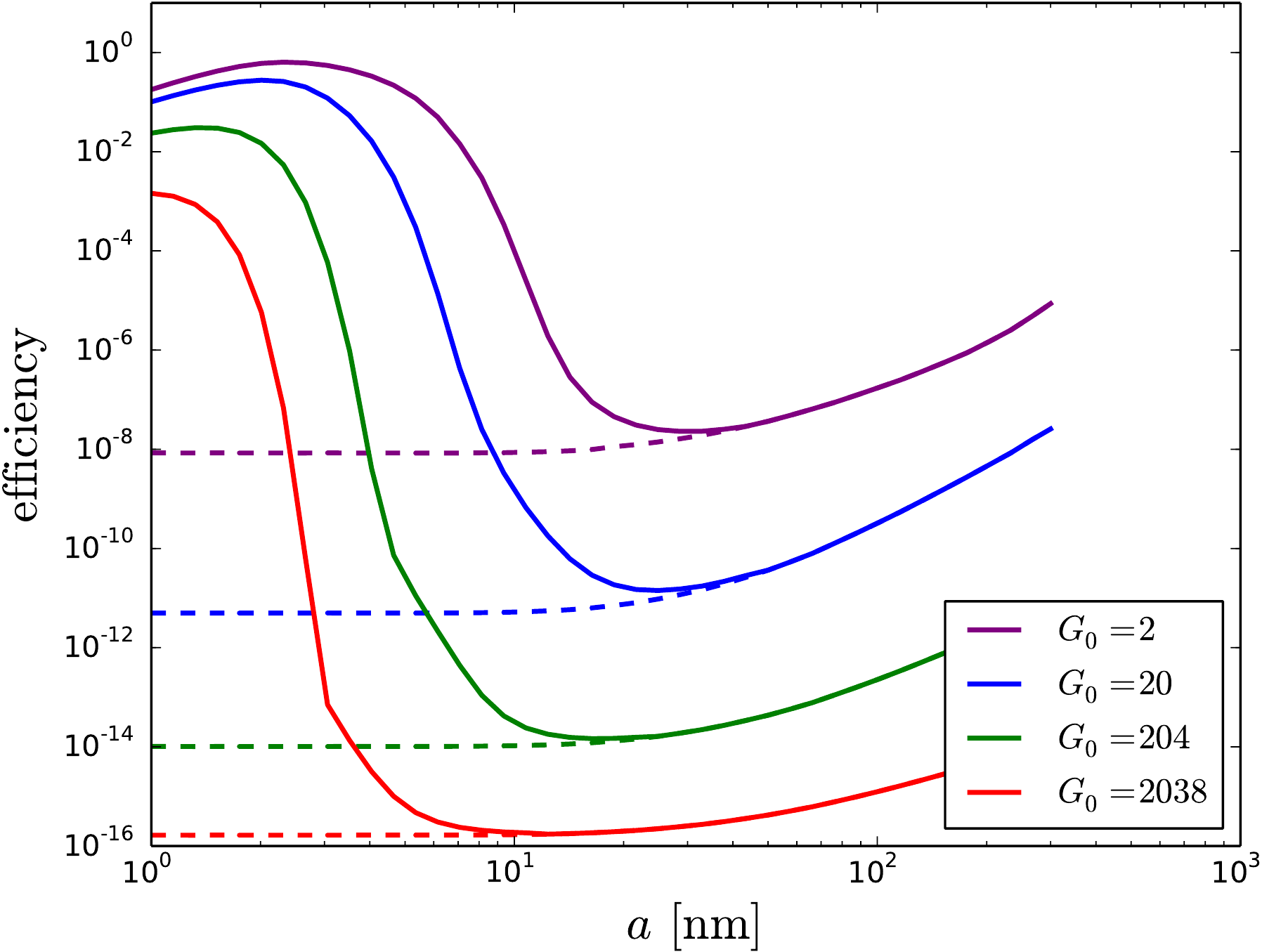}

\caption{Formation efficiency (LH mechanism) on one grain (carbonaceous) as
a function of size for different radiation field intensities (in a
gas at $n_{\mathrm{H}}=10^{4}\,\mathrm{cm}^{-3}$, $T=100\,\mathrm{K}$).
Solid lines: full computation. Dashed line: equilibrium rate equation
results. The binding and barrier energies correspond to ices.\label{fig:LH-rh2-vs-a}}
\end{figure}

We first show results for ice surfaces. Figure~\ref{fig:LH-rh2-vs-a}
shows the formation efficiency on one carbonaceous grain as a function
of the grain size under various radiation field intensities and compares
the equilibrium rate equation method with our estimation of the fluctuation
effects. The results for small grains are different by several orders
of magnitude; the method presented here giving a much more efficient
formation. For large grains, the two methods converge as expected.

\begin{figure}
\includegraphics[width=1\columnwidth]{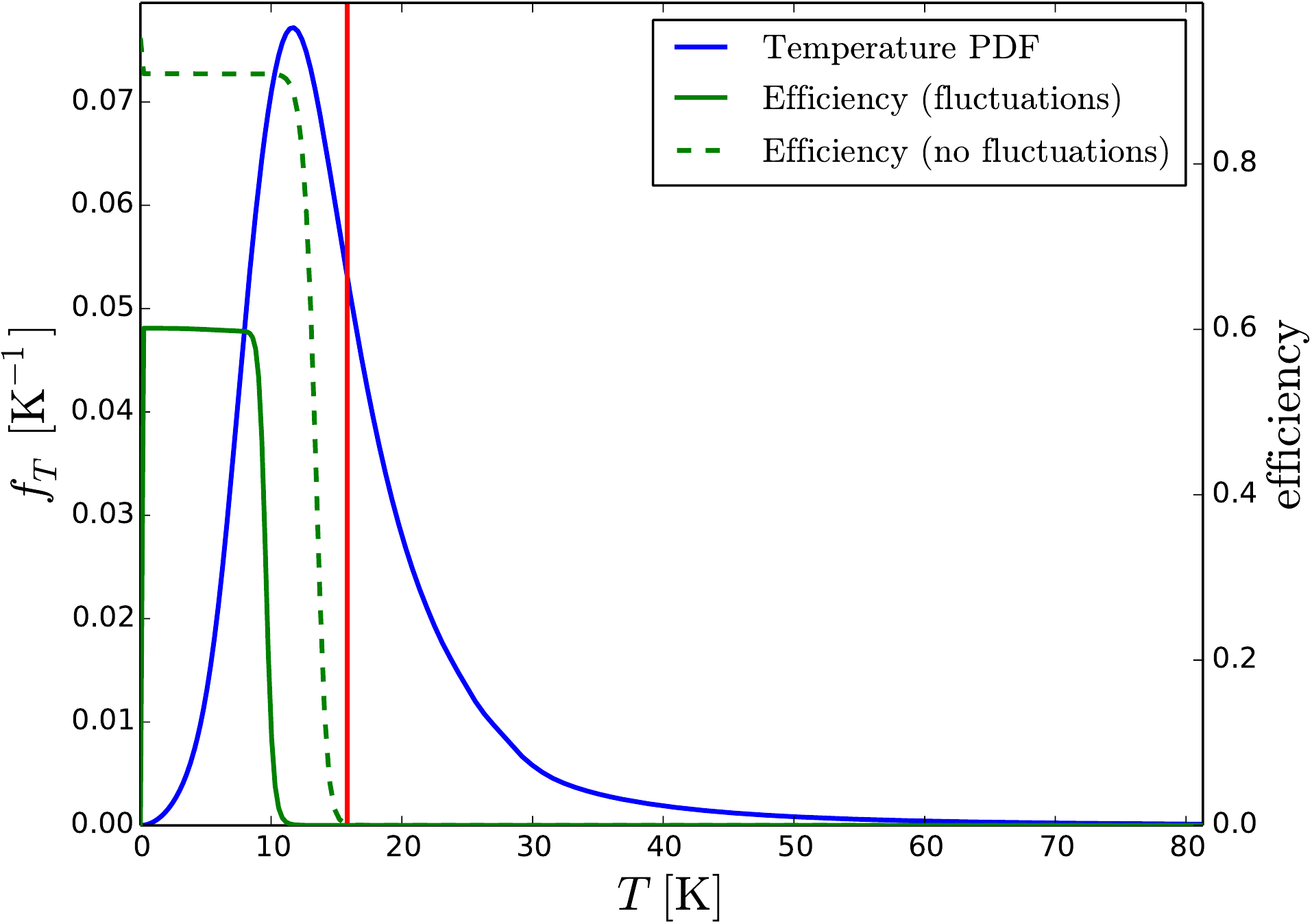}

\caption{Temperature PDF (blue line, left axis) for a $3\,\mathrm{nm}$ carbonaceous
dust grain, as compared to the formation efficiency (LH mechanism)
as a function of temperature (green lines, right axis). The red vertical
line marks the mean temperature of the blue PDF. The grain receives
a $G_{0}=20$ radiation field and is surrounded by a $100\,\mathrm{K}$
gas with $n_{H}=10^{4}\,\mathrm{cm}^{-3}$. The energy values correspond
to ices.\label{fig:overplot_fT_efficiency}}
\end{figure}

To understand those differences, we show how the temperature of a
$3\,\mathrm{nm}$ dust grain is distributed compared to the formation
efficiency domain on Fig.~\ref{fig:overplot_fT_efficiency}. We show
both the efficiency curve that would be obtained from an equilibrium
computation without fluctuations (dashed line) and the true conditional
mean efficiency $\frac{2\left\langle \left.r_{\mathrm{H}_{2}}\right|T\right\rangle }{k_{\mathrm{coll}}}$(solid
line), for fluctuating grains when they are at the given temperature
$T$. We see two different effects caused by the fluctuations. First,
the efficiency domain is reduced in both range and maximum value.
This occurs because grains do not stay in the low temperature regime
but undergo frequent fluctuations going to the high temperature domain
where their surface population evaporates quickly before cooling down.
Their average coverage when they are at low temperatures is thus decreased.
The amplitude of this effect depends on the competition between the
fluctuation timescale and the adsorption timescale. 

The other effect comes from the spread of the temperature PDF. While
the average temperature of the PDF falls in the low efficiency domain
(and the equilibrium temperature, being higher, is even worse), a
significant part of the temperature PDF actually falls in the high
efficiency domain, resulting in a quite efficient formation as seen
on Fig.~\ref{fig:LH-rh2-vs-a}. The grain spends most of its time
in the cold states and makes quick excursions into the high temperatures
that shift the average toward high temperatures. The resulting formation
rate is thus determined by the fraction of the time spent in the high
efficiency domain and not by the equilibrium or average temperature
of the grain. As described in Sect. \ref{sub:Temperature_distributions_results},
the smaller the grain, the more asymmetric the temperature PDF with
most of the PDF below its average and a long high temperature tail.
The smallest grains are thus the most affected, and we see (Fig.~\ref{fig:LH-rh2-vs-a})
that formation on the very small grains is significantly improved
even for the highest radiation field intensities. As the radiation
field intensity is decreased, larger grains are affected and become
efficient. Grains up to $20\,\mathrm{nm}$ are affected. We note here
that \citet{cuppen:06} do not consider grains smaller than $5\,\mathrm{nm}$
and thus do not include part of the affected range; they, however,
find a similar effect from $5\,\mathrm{nm}$ to $\sim10\,\mathrm{nm}$
on their flat surface model.

\begin{figure}
\includegraphics[width=1\columnwidth]{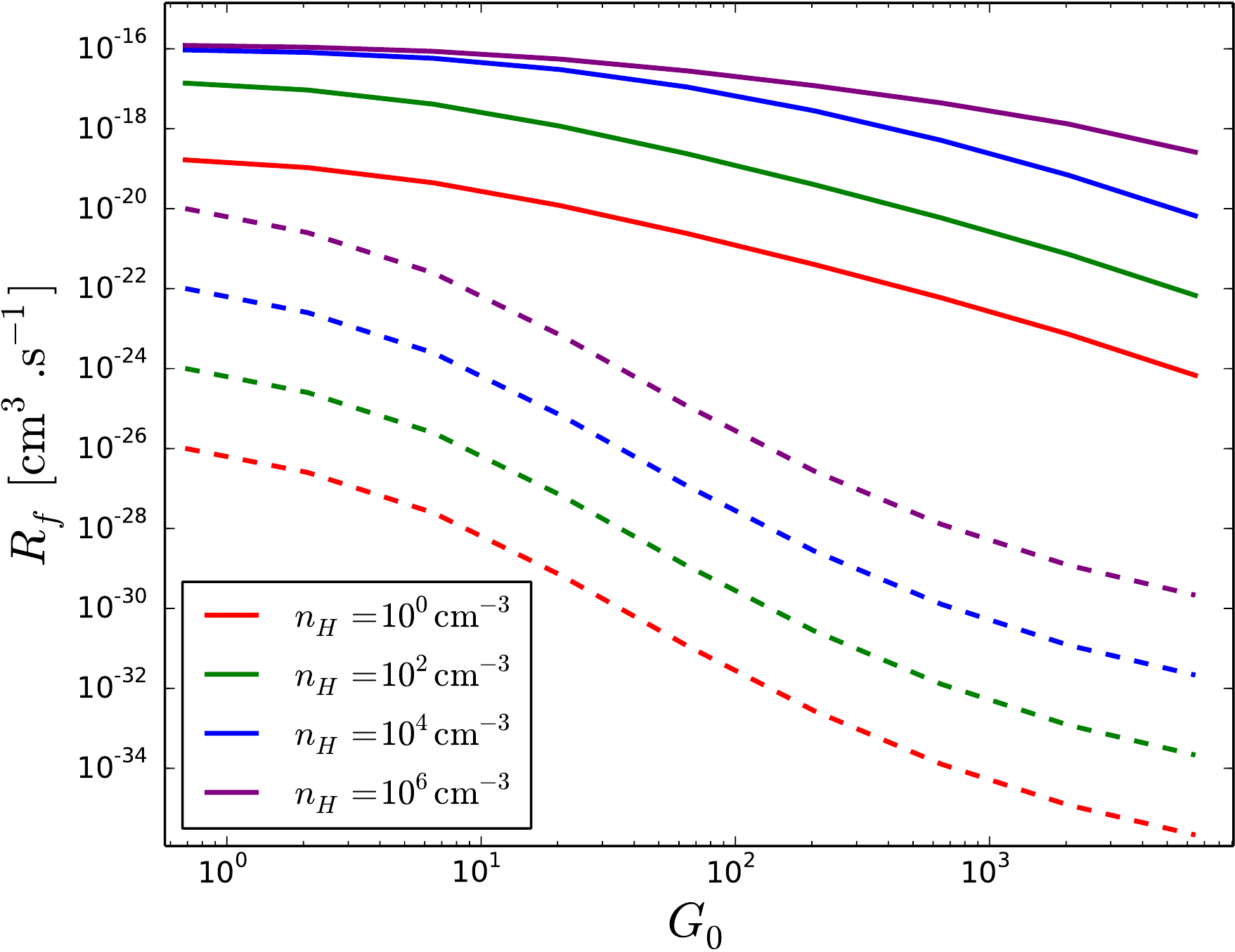}

\caption{Integrated formation rate parameter $R_{\mathrm{f}}$ (LH mechanism)
for a carbonaceous dust population (see Table \ref{tab:dust-dist})
as a function of the radiation field intensity $G_{0}$ for different
gas densities (the gas temperature is fixed at $100\,\mathrm{K}$).
Solid lines: full computation. Dashed lines: equilibrium rate equation
results. Binding and barrier energies corresponding to ices.\label{fig:LH-RH2-vs-G0_ices}}
\end{figure}

\begin{figure}
\includegraphics[width=1\columnwidth]{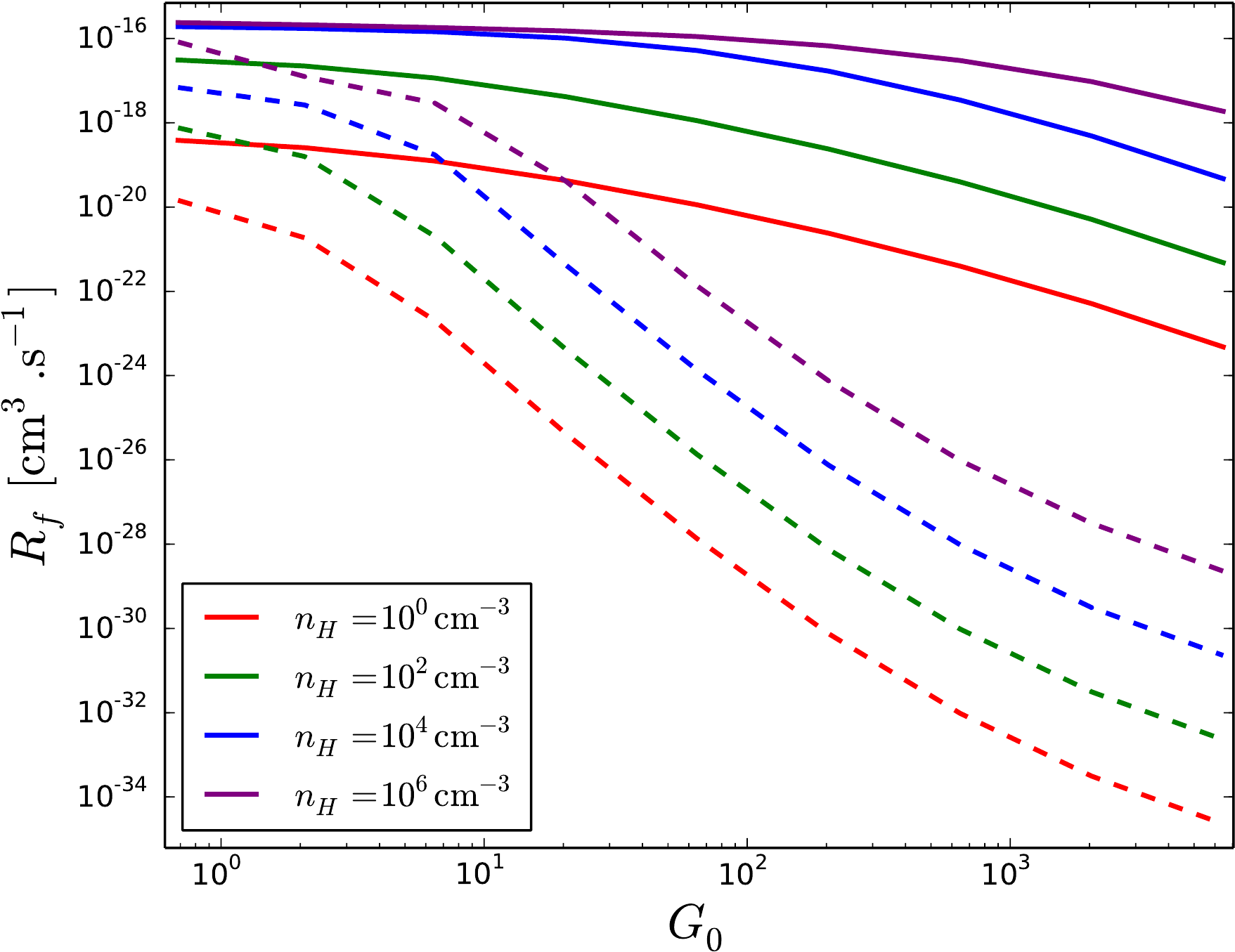}

\caption{Same as Fig. \ref{fig:LH-RH2-vs-G0_ices} with binding and barrier
energies corresponding to amorphous carbon surfaces.\label{fig:LH-RH2-vs-G0_amC}}
\end{figure}

To evaluate the importance of this effect on the global formation
rate, we integrate our formation rate on a MRN-like power law distribution
(\citealp{mathis:77}) of carbonaceous grains with an extended size
range that goes from $1\:\mathrm{nm}$ to $0.3\:\mathrm{\mu m}$ (see
Table \ref{tab:dust-dist}) and obtain a volumic formation rate $R_{H_{2}}$.

\begin{table}
\caption{Parameters of the dust population. The distribution is a MRN-like
power law with exponent $\beta$, with grain sizes going from $a_{\mathrm{min}}$
to $a_{\mathrm{max}}$.\label{tab:dust-dist}}

\centering%
\begin{tabular}{cccc}
\hline 
\hline $a_{\mathrm{min}}$ & $a_{\mathrm{max}}$ & $\beta$ & dust type\tabularnewline
$(\mathrm{nm})$ & $(\mathrm{nm})$ &  & \tabularnewline
\hline 
$1$ & $300$ & $-3.5$ & amorphous carbon\tabularnewline
\hline 
\end{tabular}
\end{table}

As this distribution strongly favors small grains in term of surface,
this integrated rate is strongly affected by the effect on small grains.
Figure~\ref{fig:LH-RH2-vs-G0_ices} shows this total formation rate
as a function of the radiation field intensity for various gas densities
(still using ice energy values). We obtain very efficient formation
from the LH mechanism in unshielded PDR edge conditions for high gas
densities. In all cases, the formation rate is increased by several
orders of magnitude compared to the equilibrium computation. As the
smallest grains remain efficient even for strong radiation fields,
the total formation rate decreases more slowly when increasing the
radiation field. 

Moreover, we note a difference in the dependency to the gas density.
The gas density dependency that comes from the collision rate is already
factored out in the definition of the $R_{f}$ parameter shown on
this graph. The equilibrium rate equation result is in the low efficiency
regime, where the efficiency is determined by the competition between
the adsorption rate (proportional to $n_{H}$) and the desorption
rate, and is thus proportional to the gas density. The result from
the method presented here is mainly determined by the fraction of
the time spend in the high efficiency regime. This fraction is fully
determined by the photon absorption and emission processes and is
independent of the gas density. The final result for high gas densities
($n_{H}>10^{3}\mbox{ cm}^{-3}$) is thus largely independent of the
gas density. For low densities, the reduction of the conditional average
efficiency curve (noted on Fig. \ref{fig:overplot_fT_efficiency})
becomes dominant. As this effect results from the competition between
the fluctuation rate (due to photons) and the adsorption rate (proportional
to $n_{\mathrm{H}}$), the resulting formation parameter $R_{f}$
becomes proportional again to the gas density for low densities ($n_{H}<10^{3}\mbox{ cm}^{-3}$).
This change of behavior corresponds to the point where the fluctuation
timescale becomes significantly smaller than the adsorption timescale.

Figure \ref{fig:LH-RH2-vs-G0_amC} shows the same result when using
binding and barrier energies corresponding to amorphous carbon surfaces.
As a consequence of the higher binding energy for amorphous carbon
surface, the equilibrium result yields higher formation rates. The
increase caused by the fluctuations is thus less dramatic but remains
very large for strong radiation fields.

We can already note here that the formation rates obtained in those
unshielded PDR edge conditions are comparable to the typical value
of $3\times10^{-17}\mathrm{cm}^{3}.\mathrm{s}^{-1}$. The LH and the
ER mechanisms thus become of comparable importance in PDR edges for
high densities, as seen in the results of full cloud simulations shown
in Sect. \ref{sec:PDR_code}.

\subsection{$\mathrm{H}_{2}$ formation rate : ER mechanism\label{sub:Results-ER}}

\begin{figure}
\centering\includegraphics[width=1\columnwidth]{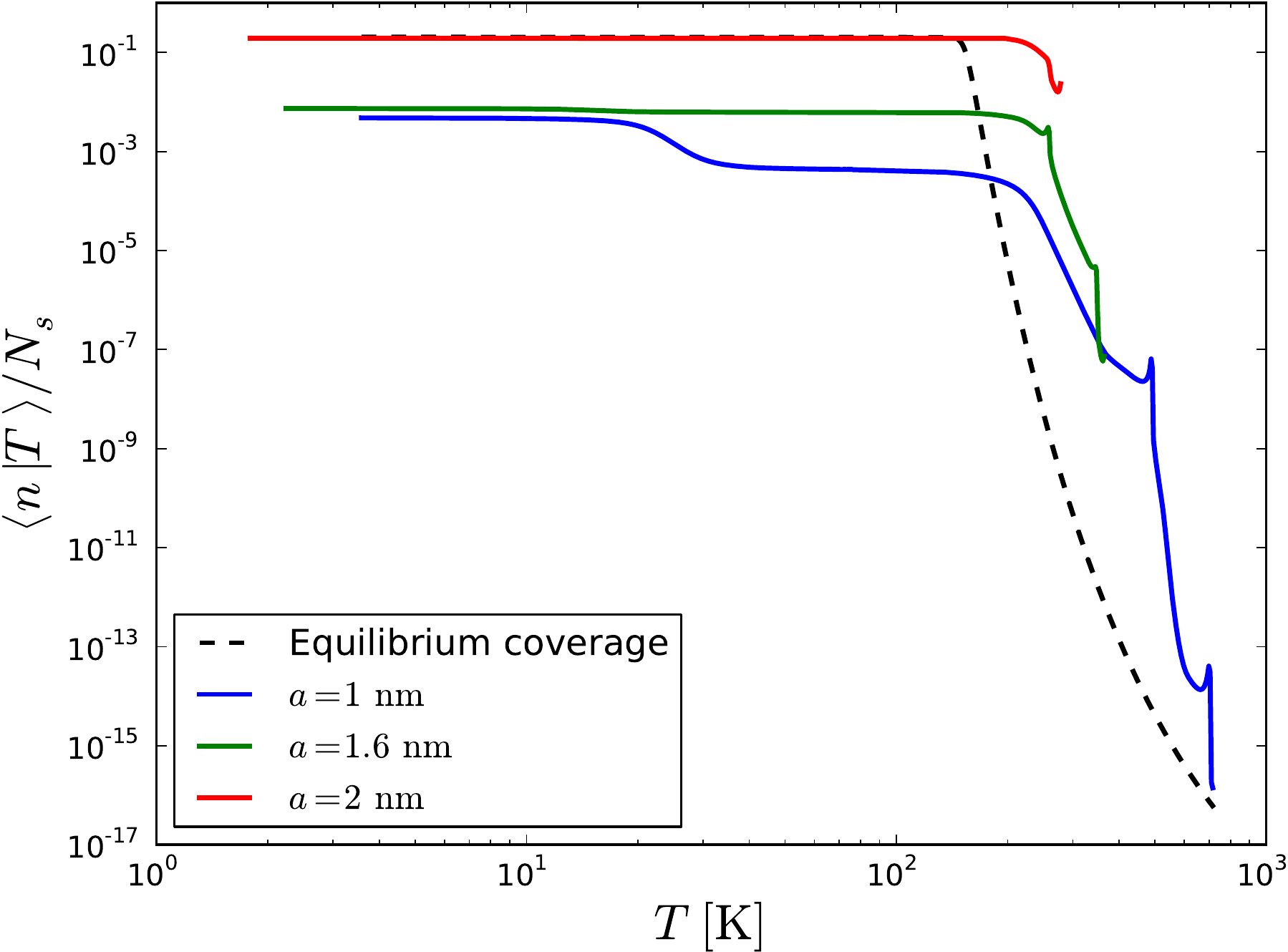}

\caption{Conditional mean of the surface coverage (chemisorption only) as a
function of grain temperature for three sizes of amorphous carbon
grains (solid lines) and equilibrium coverage (dashed line). The grain
is surrounded by gas at density $n_{\mathrm{H}}=10^{4}\,\mathrm{cm}^{-3}$
and temperature $T=350\,\mathrm{K}$ under a radiation field with
$G_{0}=120$.\label{fig:n_Es}}
\end{figure}

The first step in our computation of the formation rate is the conditional
mean of the surface population knowing the grain temperature. We show
this quantity (as a fraction of the total number of sites) on Fig.~\ref{fig:n_Es},
as compared to the coverage fraction for a grain at equilibrium at
constant temperature (this equilibrium coverage fraction is independent
of the grain size). As expected, the equilibrium coverage shows a
plateau as long as the desorption rate is negligible compared to the
adsorption rate. This plateau is lower than unity as the sticking
is less than unity. Then the equilibrium coverage decreases as desorption
becomes dominant. The dynamical effect of the temperature fluctuations
has a consequence that the actual average populations are not at equilibrium.
The grains undergoing a fluctuation to high temperature arrive in
their new state with their previous population and then go through
a transient desorption phase, which lasts a certain time, thus increasing
the average population at the high temperatures compared to the equilibrium.
When cooling down back to the low temperature states, re-accretion
of their population takes some time, and this effect lowers the average
population at low temperatures. Because this lowering effect is important,
as seen on Fig.~\ref{fig:n_Es}, it means that the characteristic
time between two temperature spikes is sufficiently short compared
to the adsorption timescale. The mechanism at play is thus that grains
that go to high temperatures lose their population but then cool down
faster than they re-adsorb their surface population and do not spend
enough time in the low temperature state before the next fluctuation
to have regained their equilibrium population. Of course, as the fluctuations
are less dramatic for bigger grains (as a given photon energy makes
a smaller temperature change: a Lyman photon at $912\,\AA$
brings a $1\,\mathrm{nm}$ grain to $\sim500\,\mathrm{K}$ but a $2\,\mathrm{nm}$
grain to only $\sim200\,\mathrm{K}$), the effect disappears for larger
grains, as seen on Fig.~\ref{fig:n_Es}.

\begin{figure}
\centering\includegraphics[width=1\columnwidth]{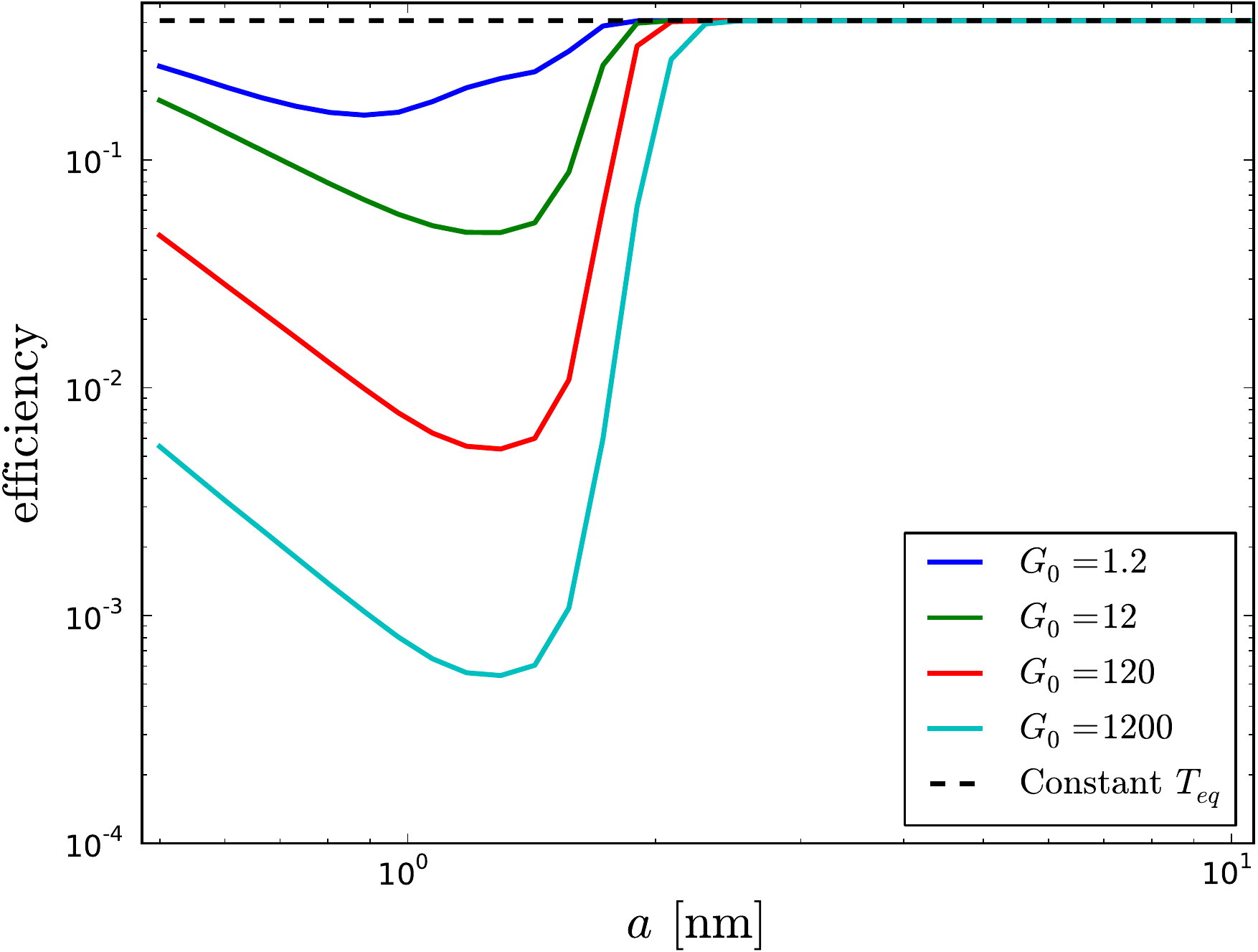}

\caption{$\mathrm{H}_{2}$ formation efficiency (ER mechanism) as a function
of grain size for various radiation fields (carbonaceous grains in
a gas at $n_{\mathrm{H}}=10^{4}\,\mathrm{cm}^{-3}$ and $T=350\,\mathrm{K}$).\label{fig:Formation-rate}\label{fig:RH2-vs-a_ISRFs}}
\end{figure}

The formation rate on one grain is then simply $r_{\mathrm{H}_{2}}=\int_{0}^{+\infty}dT\, f_{T}(T)\, k_{\mathrm{coll}}\frac{\left\langle n\mid T\right\rangle }{N_{\mathrm{s}}}$.
The resulting formation efficiency on one grain is shown on Fig.~\ref{fig:RH2-vs-a_ISRFs}
as a function of grain size.

As explained above, the smaller grains have a lower coverage than
equilibrium due to temperature fluctuations and, thus, a decreased
formation rate. Only grains below $2\:\mathrm{nm}$ are affected,
as the same photon absorption causes a smaller temperature fluctuation
for bigger grains. For stronger radiation fields, fluctuations occur
with a shorter timescale, leaving less time for the surface population
to reach equilibrium. The effect is then stronger and extends to higher
sizes.

The affected range seems very small, but we have to remember that
usual dust size distributions are such that the small grains represent
most of the surface (e.g., the MRN distribution). Thus, the contribution
from the smallest grain sizes dominates the global formation rate.
By integrating over the dust size distribution (same as in Sect. \ref{sub:Results-LH}),
we obtain the volumic formation rate in the gas, which we
can express as $R_{\mathrm{H}_{2}}=R_{f}\, n(H)\, n_{\mathrm{H}}$
defining $R_{f}$. In the following, we compare the integrated formation
rate that we obtain to the one obtained with the rate equation at
constant temperature as used in \citet{lebourlot:12} and plot our
results as a fraction of this reference formation rate.

The two main parameters are the radiation field intensity and the
H atom collision rate (and the sticking coefficient, which modifies
the effective collision rate). We use standard interstellar radiation
fields with a scaling factor as explained in Sect. \ref{sub:Model-phtons}.

\begin{figure}
\centering\includegraphics[width=1\columnwidth]{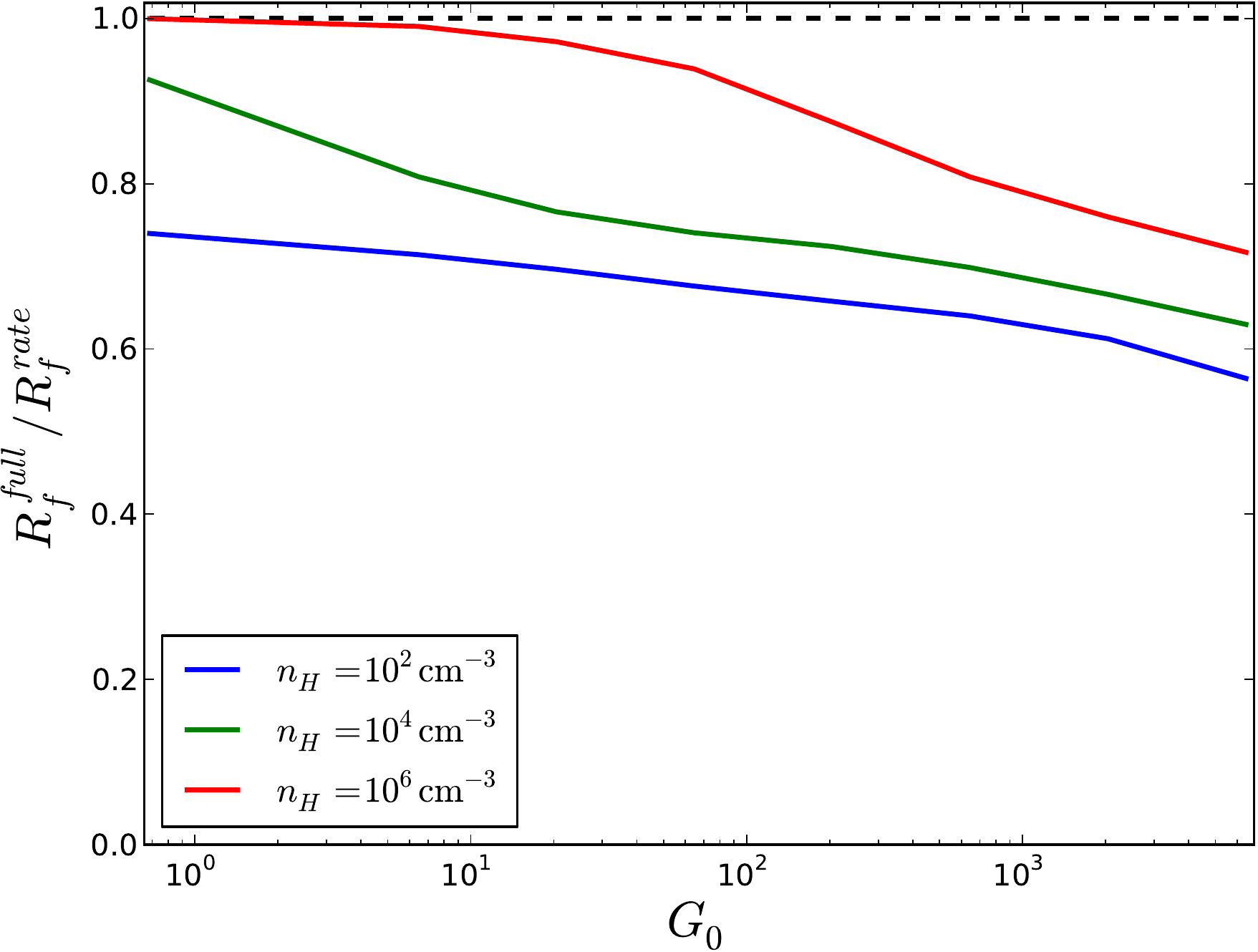}

\caption{Integrated formation rate $R_{f}$ (ER mechanism) over a full dust
size distribution (see Table \ref{tab:dust-dist}) as a fraction of
the equilibrium rate equation result. The result is shown for three
different gas densities (thus changing the collision rate) as a function
of the radiation field intensity.\label{fig:RH2_nH}}
\end{figure}

Figure~\ref{fig:RH2_nH} shows the effect of these two parameters.
We show the formation rate as a function of the radiation field intensity
for different gas densities to probe the effect of the collision rate.
Note that the reference rate equation result at constant temperature
does not depend on either of these parameters: the density dependency
(from the collision rate) is eliminated in the definition of $R_{f}$,
and the radiation field has no effect when the fluctuations are neglected
(the equilibrium temperature is too low to trigger desorption). The
effect we show is thus fully the effect of grain temperature fluctuations.
As expected, the formation rate decreases when either the radiation
intensity is increased or the collision rate is decreased, as the
effect results from a competition between the fluctuation rate and
the collision rate. The observed reduction remains limited and reaches
a $\sim40\%$ reduction or lower in the range of parameters that we
explore. The result that an effect affecting only a tiny fraction
of the range of grain sizes can have a significant effect on the integrated
formation rate illustrates how the smallest sizes dominate the chemistry
in the usual MRN size distribution.

\begin{figure}
\centering\includegraphics[width=1\columnwidth]{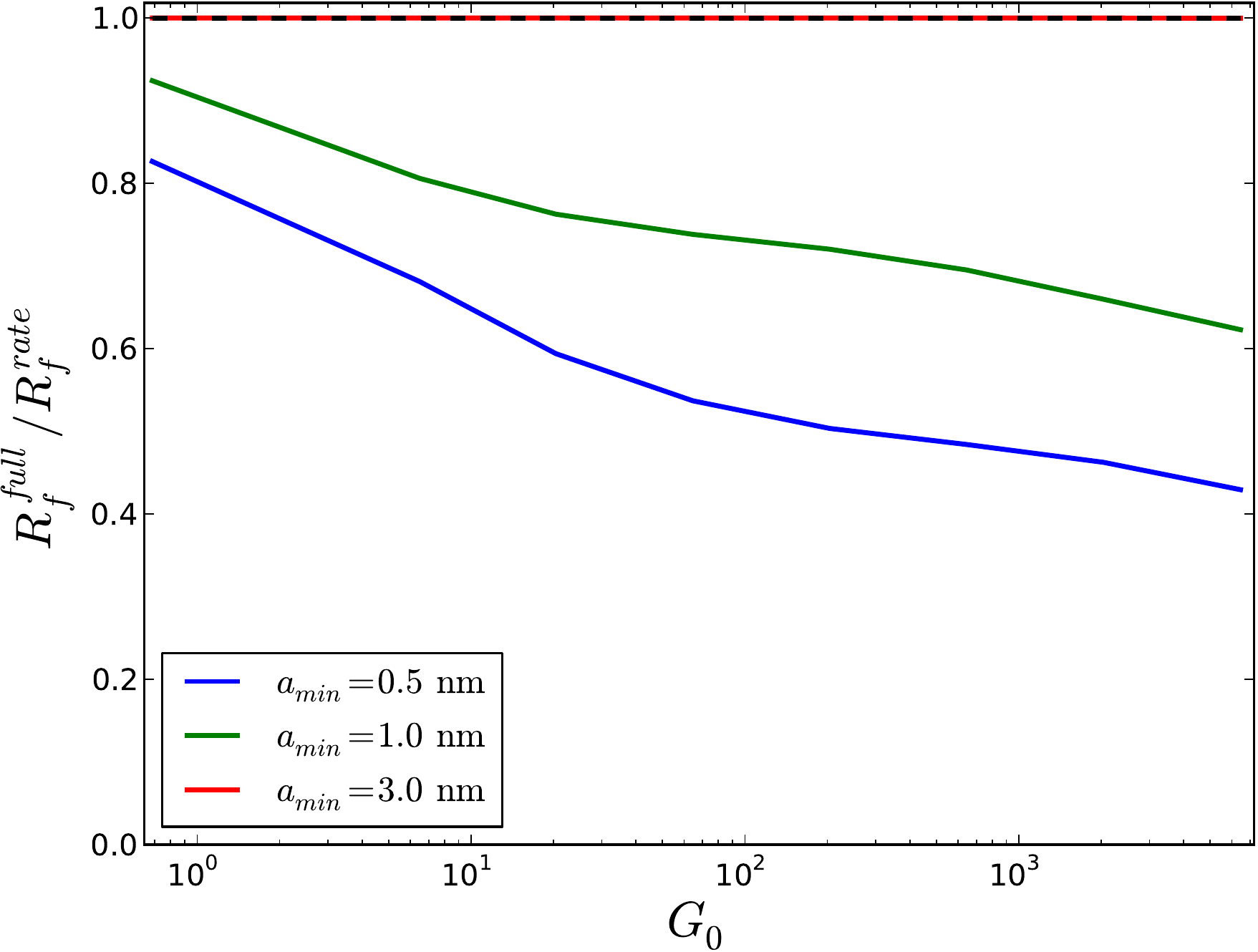}

\caption{Integrated formation rate ratio (ER mechanism) over a full dust size
distribution, as a function of the radiation field intensity when
changing the inferior limit of the grain size distribution ($n_{\mathrm{H}}=10^{4}\,\mathrm{cm}^{-3}$).\label{fig:RH2_amins}}
\end{figure}

The choice of the lower limit of this size distribution is thus of
crucial importance. The lower this minimal size, the larger the usual
equilibrium rate equation formation rate will be, as the total dust
surface is increased for a constant total dust mass. However, as we
lower this limit, we extend the range affected by the fluctuations,
and thus obtain a stronger decrease compared to the rate equation
result, as shown on Fig.~\ref{fig:RH2_amins}. We observe no effect
if the size distribution starts above the affected size (e.g., for
$a_{\mathrm{min}}=3\,\mathrm{nm}$), while a lower limit causes a
stronger effect (up to $\sim60\%$ with $a_{\mathrm{min}}=0.5\,\mathrm{nm}$).

\begin{figure}
\centering\includegraphics[width=1\columnwidth]{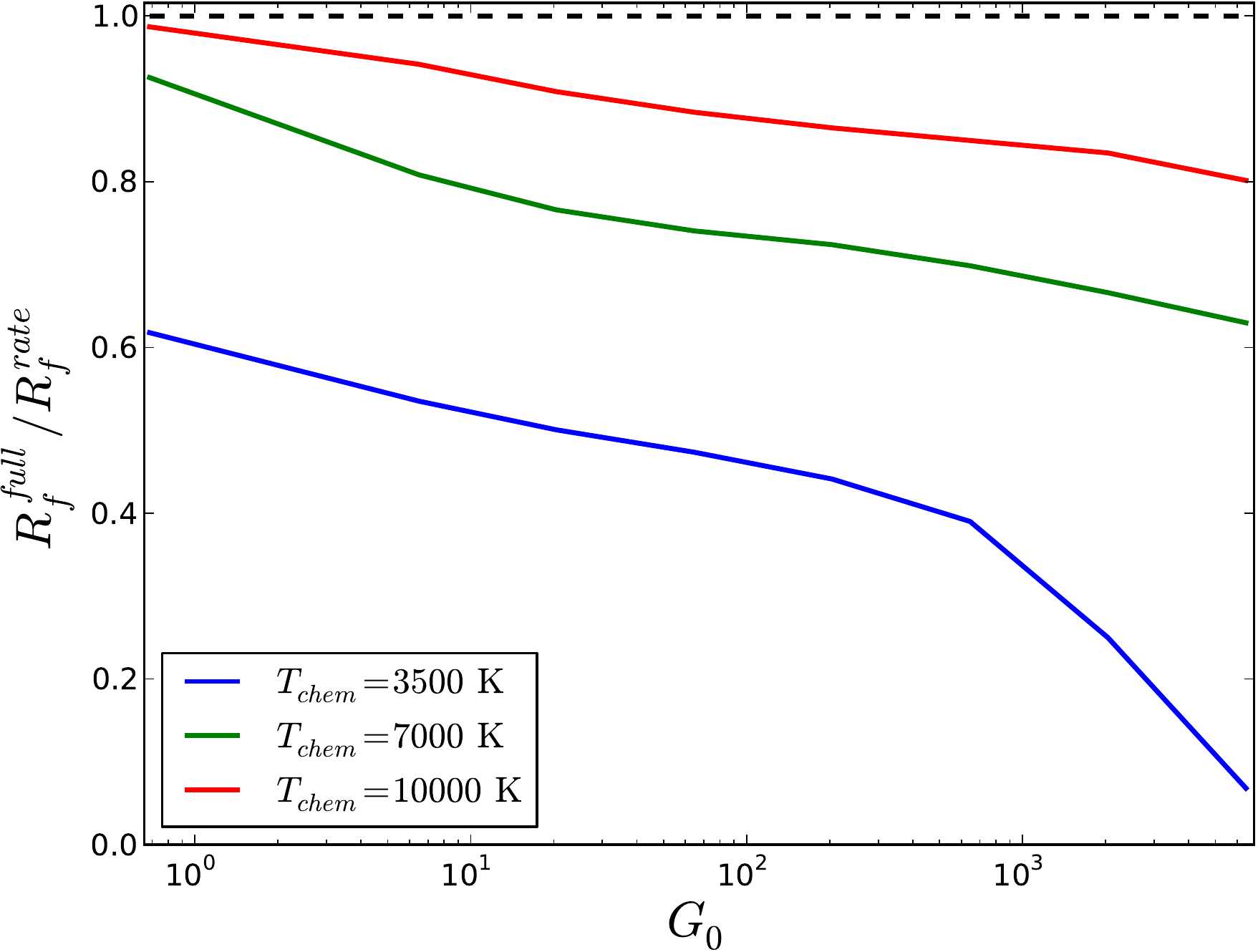}

\caption{Integrated formation rate ratio (ER mechanism) over a full dust size
distribution, as a function of the radiation field intensity for different
chemisorption binding energies ($n_{\mathrm{H}}=10^{4}\,\mathrm{cm}^{-3}$).\label{fig:RH2_Tchims}}
\end{figure}

Last, as noted in Sect.~\ref{sub:ER-processes}, the microphysical
parameters of the adsorption process are not well known. The adsorption
barrier affects the sticking and changes the effective collision rate
(see Fig.~\ref{fig:RH2_nH} for the effect of the collision rate).
The vibration frequency $\nu_{0}$ has a negligible effect in the
range of reasonable values. The most important microphysical parameter
is the chemisorption energy $T_{\mathrm{chem}}$. Figure~\ref{fig:RH2_Tchims}
shows the strong influence of this parameter on the result. A lower
chemisorption energy allows for smaller fluctuations to trigger desorption.
Bigger grains are thus affected, resulting in a much lower formation
rate (e.g., a reduction of more than $90\%$ for $T_{\mathrm{chem}}=3500\,\mathrm{K}$).

\section{Integration into the Meudon PDR code\label{sec:PDR_code}}

We now investigate the effects of the changes in the formation rate
found in the previous section on the overall structure and chemistry
of an interstellar cloud with special attention to the observable
line intensities.

Our code \texttt{Fredholm}, which computes the $\mathrm{H}_{2}$ formation
rate from the LH and ER mechanisms as described in Sect. \ref{sec:The-model}
and \ref{sec:Approximations}, has been coupled to the Meudon PDR
code described in \citet{lepetit:06}. 

The Meudon PDR code models stationary interstellar clouds in a 1D
geometry with a full treatment of the radiative transfer (including
absorption and emission by both gas and dust), thermal balance, and
chemical equilibrium done in a self-consistent way. In its standard
settings, the formation of $\mathrm{H}_{2}$ on dust grains is treated
using rate equations and including both the LH and ER mechanism, as
described in \citet{lebourlot:12}. We have now enabled the possibility
to compute the $\mathrm{H}_{2}$ formation rate using our code \texttt{Fredholm}
from the gas conditions and the radiation field sent by the PDR code
at each position in the cloud, thus taking temperature fluctuations
into account. We compare the results of those coupled runs with the
simpler rate equation treatment. The dust population in the PDR models
presented here is a single component following a MRN-like power-law
distribution from $1\,\mathrm{nm}$ to $0.3\,\mathrm{\mu m}$ with
an exponent of $-3.5$, using properties corresponding to a 70\%-30\%
mix of graphites and silicates above $3\,\mbox{nm}$ and a progressive
transition to PAH properties from $3\,\mathrm{nm}$ to $1\,\mathrm{nm}$.

A direct comparison between the rate equation approximation and a
full account of stochastic effects is thus possible. In addition to
the radiation field illuminating the front side of the cloud, whose
strength is indicated for each model, the back side of the cloud is
always illuminated by a standard interstellar radiation field ($G_{0}=1$)
in the following models.

\subsection{Detailed analysis}

\subsubsection{Effects on PDR edges}

We focus first on the effects on PDR edges, as the main effects of
a change in formation rate come from the shift of the $\mathrm{H}/\mathrm{H}_{2}$
transition that is induced. We have seen in Sect. \ref{sec:Results}
that the full treatment of fluctuations reduces the efficiency of
the ER mechanism while strongly increasing the efficiency of the LH
mechanism (compared to a rate equation treatment). The resulting effect
in full cloud models depends on the relative importance of the two
mechanisms. 

\begin{figure}
\includegraphics[width=1\columnwidth]{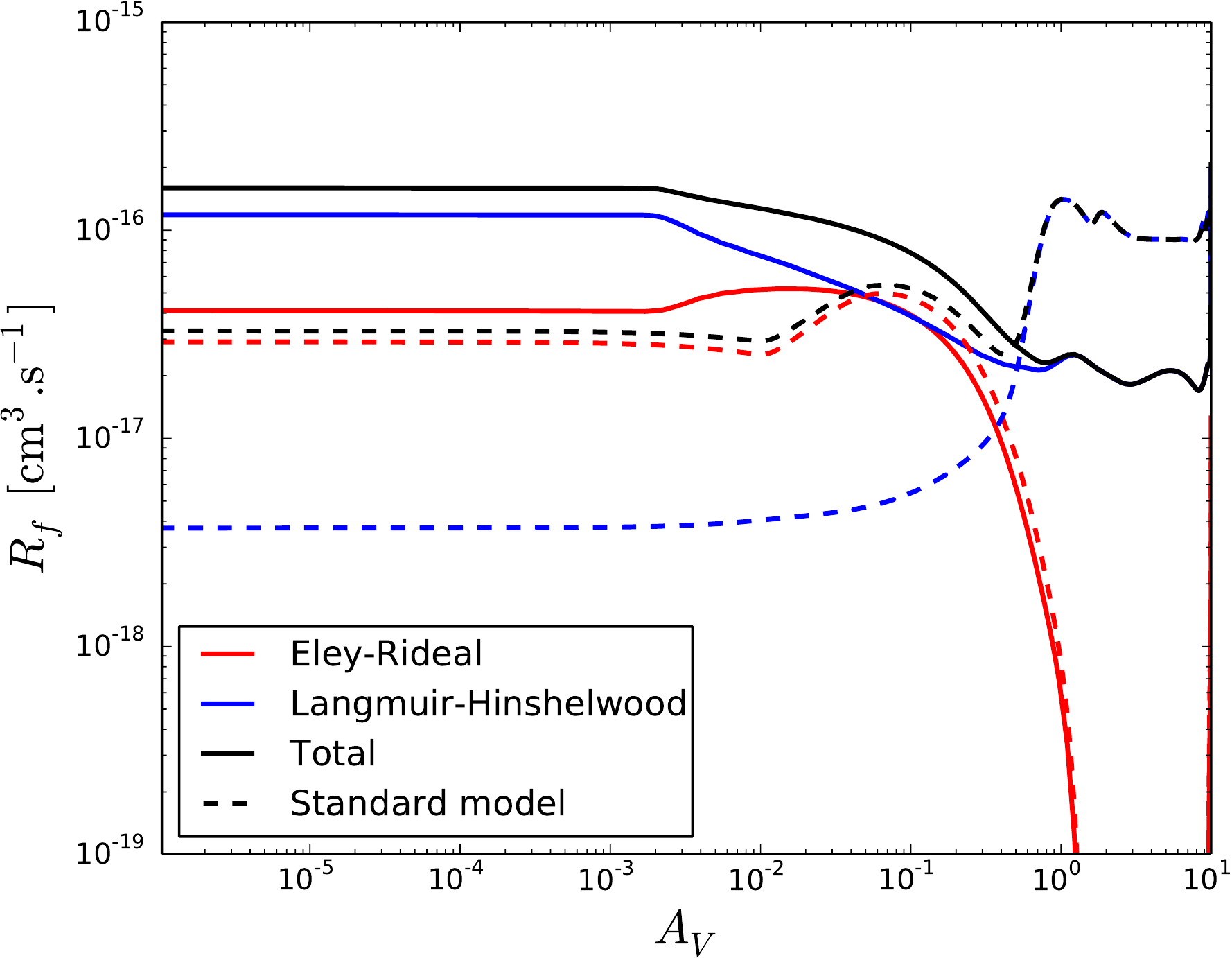}

\caption{$\mathrm{H_{2}}$ formation parameter $R_{f}$ (full treatment:
solid lines, standard treatment: dashed lines) for the two processes
in a cloud with $n_{\mathrm{H}}=10^{3}\,\mathrm{cm}^{-3}$ and $G_{0}=10$.\label{fig:n1E3khi10_form}}
\end{figure}

\begin{figure}
\includegraphics[width=1\columnwidth]{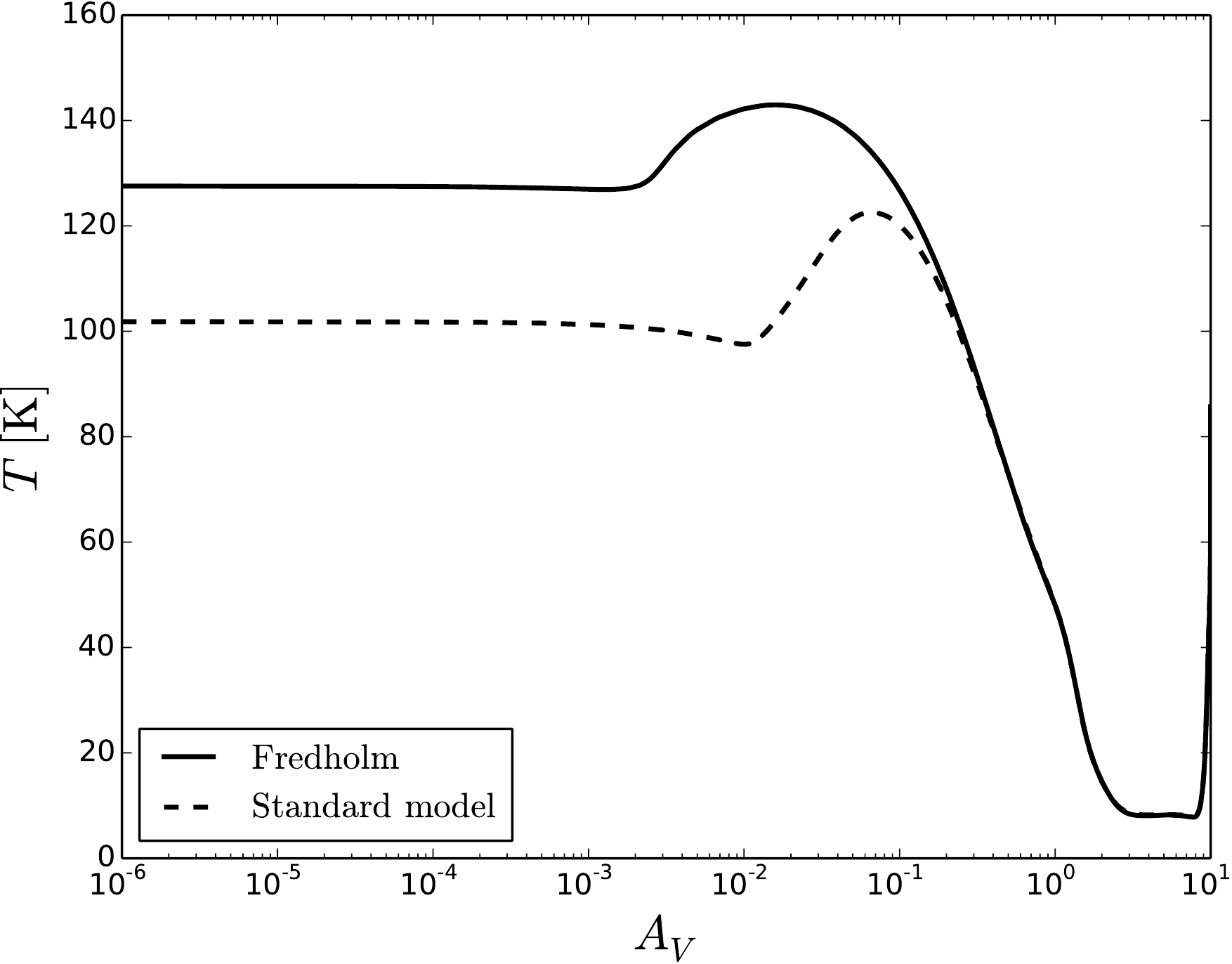}

\caption{Gas temperature (same cloud as Fig.~\ref{fig:n1E3khi10_form}).\label{fig:n1E3khi10_temp}}

\end{figure}

For low radiation field models, the full treatment increases the LH
mechanism at the edge to a level comparable or even higher than the
ER mechanism, as shown on Fig.~\ref{fig:n1E3khi10_form}. The LH
formation rate is increased by more than one order of magnitude and
becomes higher than the ER rate, while it was one order of magnitude
below in the standard rate equation model. The net result is an increase
of the total formation rate by a factor of 4. While we were expecting
the ER mechanism to be slightly decreased by the full treatment, we
can note that it is slightly increased here. This is due to the gas
temperature sensitivity of the ER mechanism. In this low radiation
field model, the heating of the gas due to $\mathrm{H}_{2}$ formation
is not negligible compared to the usually dominant photoelectric effect.
A higher total $\mathrm{H}_{2}$ formation rate increases the heating
of the gas resulting in a higher temperature at the edge (see Fig.~\ref{fig:n1E3khi10_temp}).
The ER mechanism is thus made slightly more efficient.

\begin{figure}
\includegraphics[width=1\columnwidth]{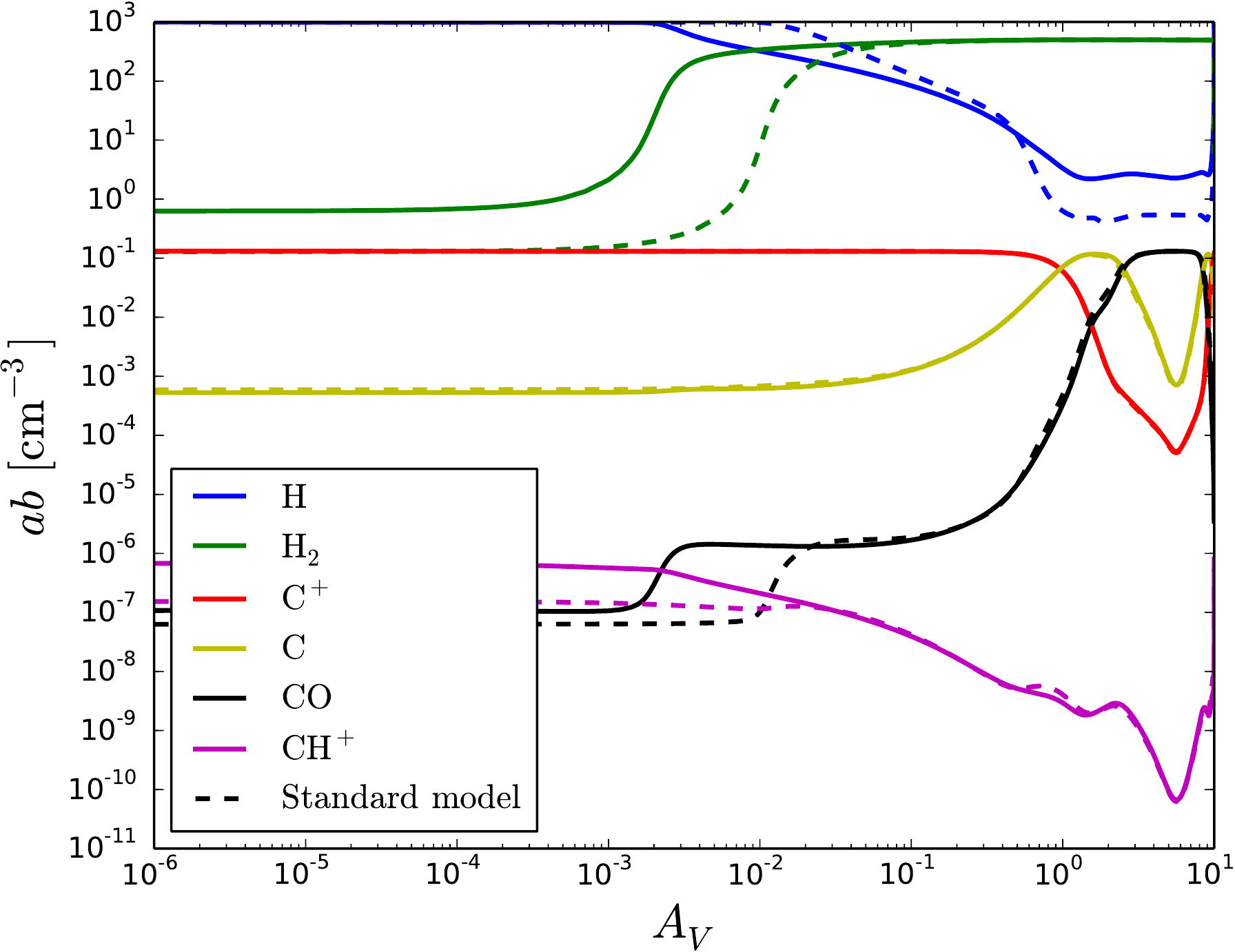}

\caption{Abundances of the species of interest in the same cloud as Fig.~\ref{fig:n1E3khi10_form}\label{fig:n1E3khi10_struct}}
\end{figure}

The main effect is a shift of the $\mathrm{H}/\mathrm{H}_{2}$ transition
from $A_{V}=4\times10^{-2}$ to $A_{V}=8\times10^{-3}$, as shown
of Fig.~\ref{fig:n1E3khi10_struct}. As a result, the column density
of excited $\mathrm{H}_{2}$ is increased, as shown in Table \ref{tab:n1E3khi10}.
As we are in low excitation conditions, the effect is mainly visible
on the levels $v=0,\, J=2$ and $v=0,\, J=3$.

\begin{table}
\centering{}\caption{$\mathrm{H}_{2}$ intensities in the same cloud as Fig.~\ref{fig:n1E3khi10_form}.\label{tab:n1E3khi10}}
\begin{tabular}{lcc}
\hline 
\hline  & \multicolumn{2}{c}{Integrated intensity}\tabularnewline
 & \multicolumn{2}{c}{$\mathrm{W}\,\mathrm{m}^{-2}\,\mathrm{sr}^{-1}$}\tabularnewline
$\mathrm{H}_{2}$ line & Standard model & Full treatment\tabularnewline
\hline 
$(0-0)\, S(0)$ & $8.14\times10^{-10}$ & $1.06\times10^{-9}$\tabularnewline
$(0-0)\, S(1)$ & $6.40\times10^{-11}$ & $1.56\times10^{-10}$\tabularnewline
$(0-0)\, S(2)$ & $3.74\times10^{-11}$ & $4.37\times10^{-11}$\tabularnewline
$(0-0)\, S(3)$ & $3.53\times10^{-11}$ & $4.21\times10^{-11}$\tabularnewline
$(0-0)\, S(4)$ & $1.88\times10^{-11}$ & $2.15\times10^{-11}$\tabularnewline
$(0-0)\, S(5)$ & $3.28\times10^{-11}$ & $3.72\times10^{-11}$\tabularnewline
$(1-0)\, S(1)$ & $4.40\times10^{-11}$ & $5.72\times10^{-11}$\tabularnewline
\hline 
\end{tabular} 
\end{table}

We also note a significant increase of the $\mathrm{H}_{2}$ abundance
before the transition, which results in increased $\mathrm{CH}^{+}$
formation at the edge. This effect and the higher temperature at the
edge result in stronger excited lines of $\mathrm{CH}^{+}$$ $$ $with
differences of a factor of $2$ on line $(4-3)$. The excitation temperature
computed from the observed intensities above $(4-3)$ closely follow
the change in gas temperature at the edge. The increase in intensity
thus becomes larger as we go to larger levels, but those intensities
remain most probably unobservable in this case.

\begin{figure}
\includegraphics[width=1\columnwidth]{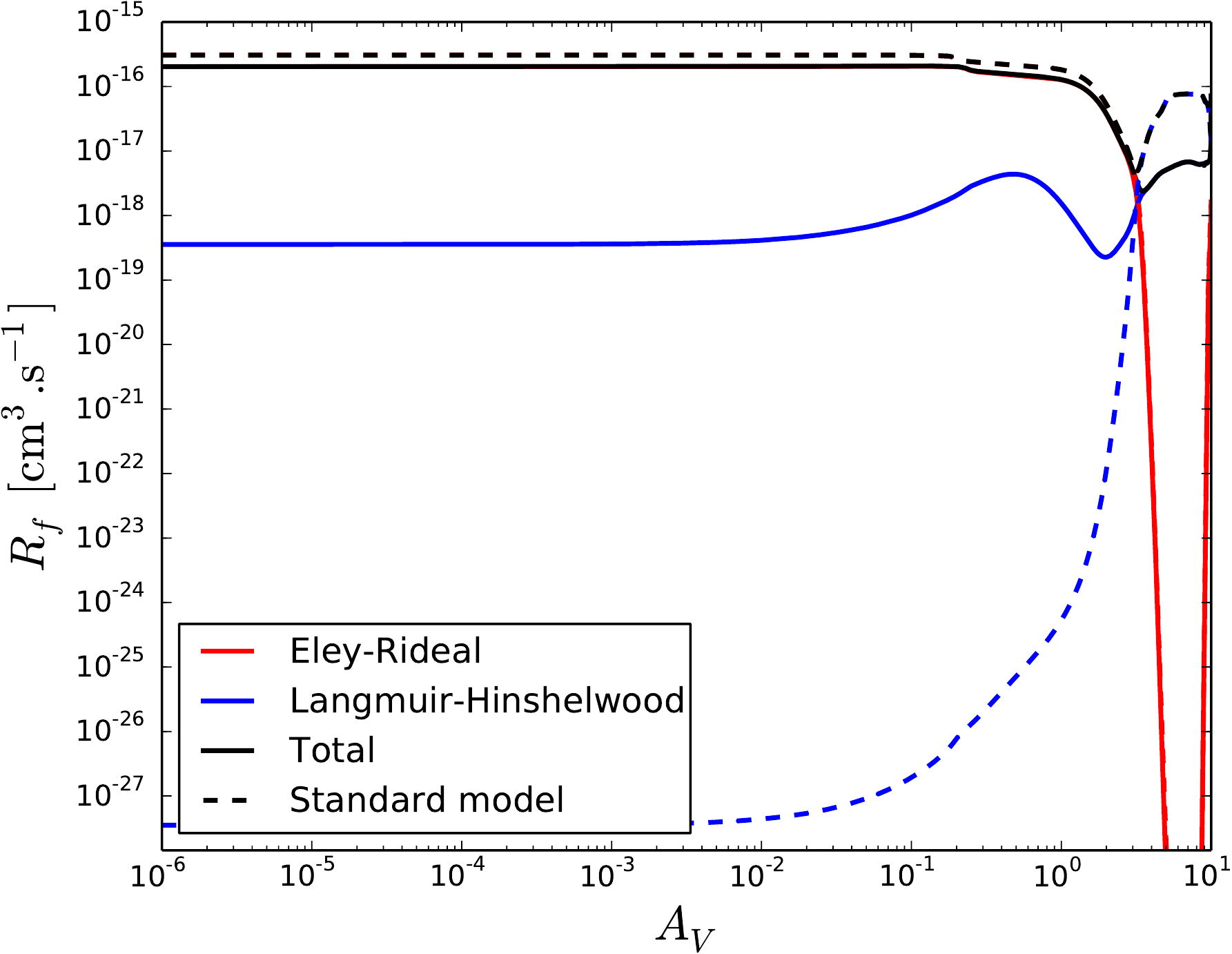}

\caption{Same as Fig.~\ref{fig:n1E3khi10_form} for a cloud with $n_{\mathrm{H}}=10^{3}\,\mathrm{cm}^{-3}$
and $G_{0}=10^{3}$.\label{fig:n1E3khi1000_form}}
\end{figure}

The situation is different for high radiation field PDRs. The increase
in the efficiency of the LH mechanism at the edge is not sufficient
to make it comparable to the ER mechanism and the net result is a
decrease of the total formation rate as the ER efficiency is reduced
(see Fig.~\ref{fig:n1E3khi1000_form}). This results in inverse effects
with the $\mathrm{H}/\mathrm{H}_{2}$ transition slightly shifted
away from the edge, but of smaller amplitude as the formation rate
difference is much smaller.

We present a wider investigation of the effects on observable intensities
in Sect. \ref{sub:Grid}, covering a large domain of cloud conditions.

\subsubsection{Effects on cloud cores}

\begin{figure}
\includegraphics[width=1\columnwidth]{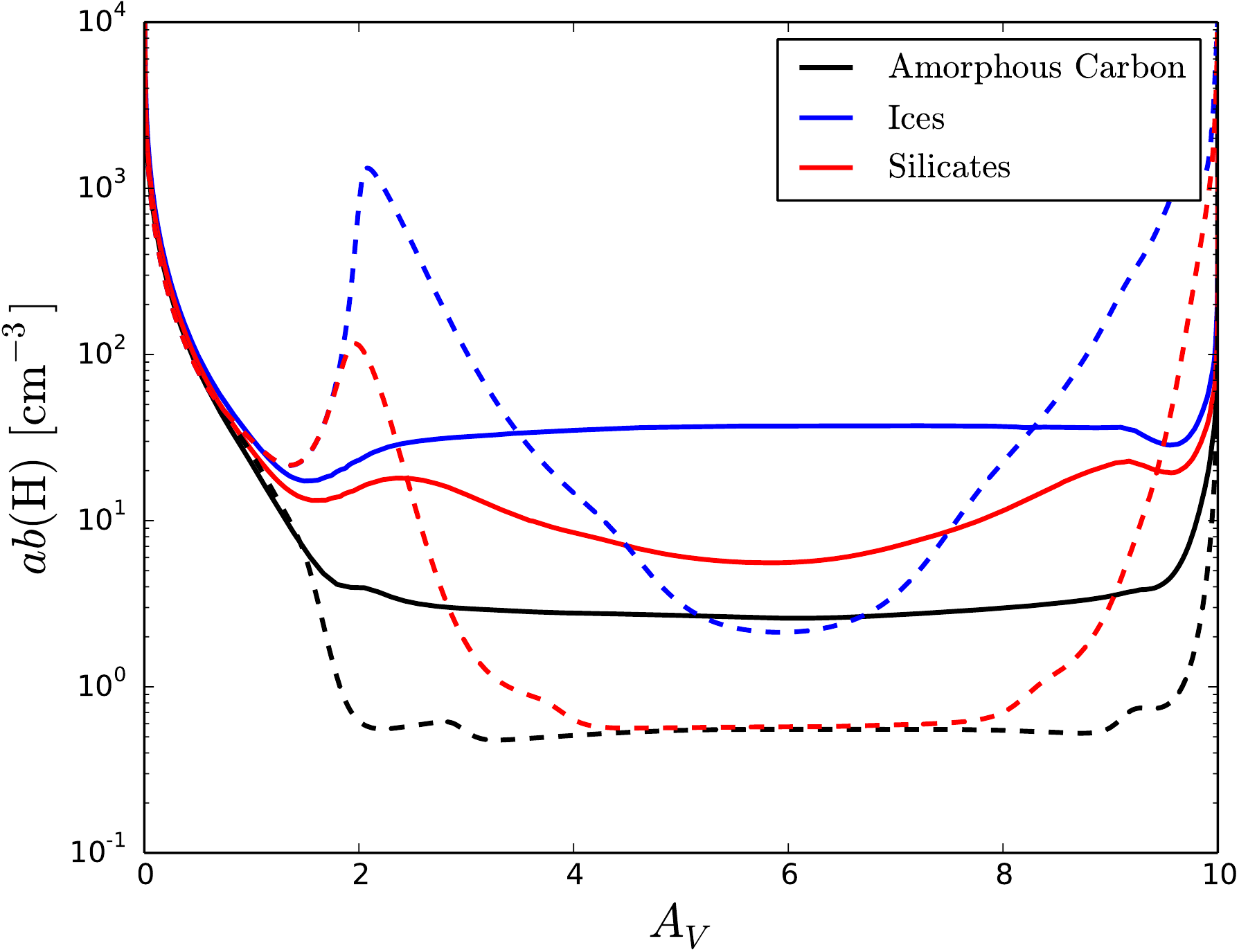}

\caption{Atomic $\mathrm{H}$ abundance in a cloud with $n_{\mathrm{H}}=10^{4}\,\mathrm{cm}^{-3}$
and $G_{0}=10^{2}$. Comparison of rate equations results (dashed
lines) with our full treatment (solid lines).\label{fig:H_Cmp}}
\end{figure}
\begin{figure}
\includegraphics[width=1\columnwidth]{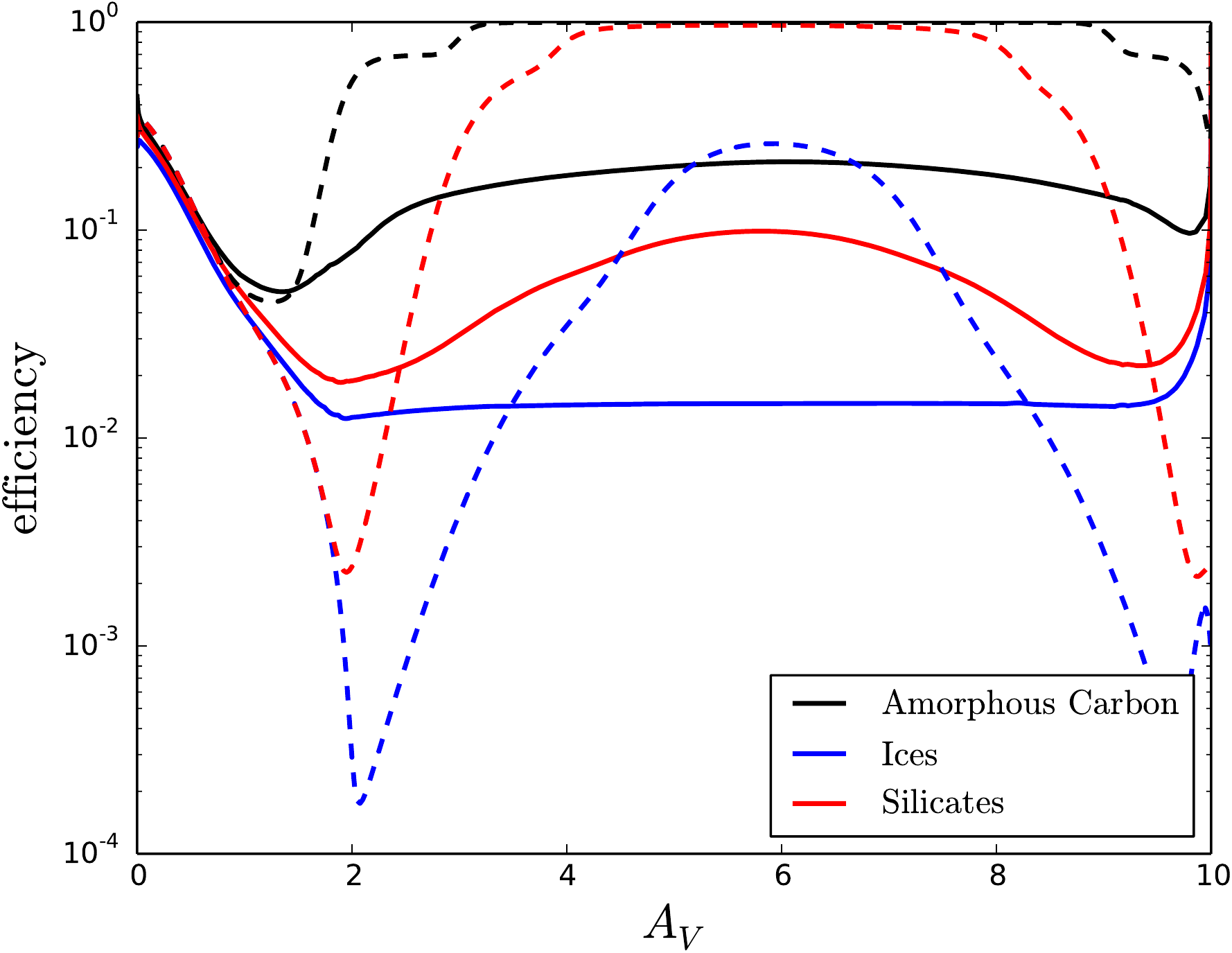}

\caption{$\mathrm{H}_{2}$ formation efficiency. Same case as Fig.~\ref{fig:H_Cmp}.\label{fig:Gr_efficiency}}
\end{figure}

Unexpectedly, temperature fluctuations of small dust grains are found
to have an effect on $\mathrm{H}_{2}$ formation even in cloud cores
where the UV field has been fully extinguished. As an example, Fig.
\ref{fig:H_Cmp} shows the atomic hydrogen abundance in a specific
cloud ($n_{\mathrm{H}}=10^{4}\,\mathrm{cm}^{-3}$, $\chi=100$, total
$A_{V}=10$) and compares the standard models to the coupled models
with full treatment of the fluctuations for the different binding
energy values of Table \ref{tab:H_binding}. The residual atomic $\mathrm{H}$
abundance in the core is found to be systematically higher when taking
the fluctuations into account, which is a direct consequence of a
lower formation efficiency (the efficiencies are shown for the same
models on Fig. \ref{fig:Gr_efficiency}). We recall that the formation
efficiency is defined as the ratio between the formation rate and
half the collision rate between $\mathrm{H}$ atoms and grains (whose
expression is $k_{\mathrm{max}}=\frac{1}{2}\, v_{\mathrm{th}}\left(\mathrm{H}\right)\,\frac{\mathcal{S}_{\mathrm{gr}}\, n_{\mathrm{H}}}{4}\, n(\mathrm{H})$)),
as this rate is the maximum formation rate that would occur if one
molecule was formed every two collisions. We also notice that standard
models tend to show a strong variability of the formation efficiency
(and consequently of the $\mathrm{H}$ abundance) with a gap around
$A_{V}=2$ with very low efficiency, where neither the ER mechanism
(the gas is not warm enough) nor the LH mechanism (the grains are
too warm) are efficient. This variability disappears in the coupled
models as fluctuations make the LH mechanism much less dependent on
the grain equilibrium temperature (see Sect. \ref{sub:Results-LH}),
and the efficiency stays of the order of $10\%$ (for amorphous carbon
surfaces) across the cloud.
\begin{figure}
\includegraphics[width=1\columnwidth]{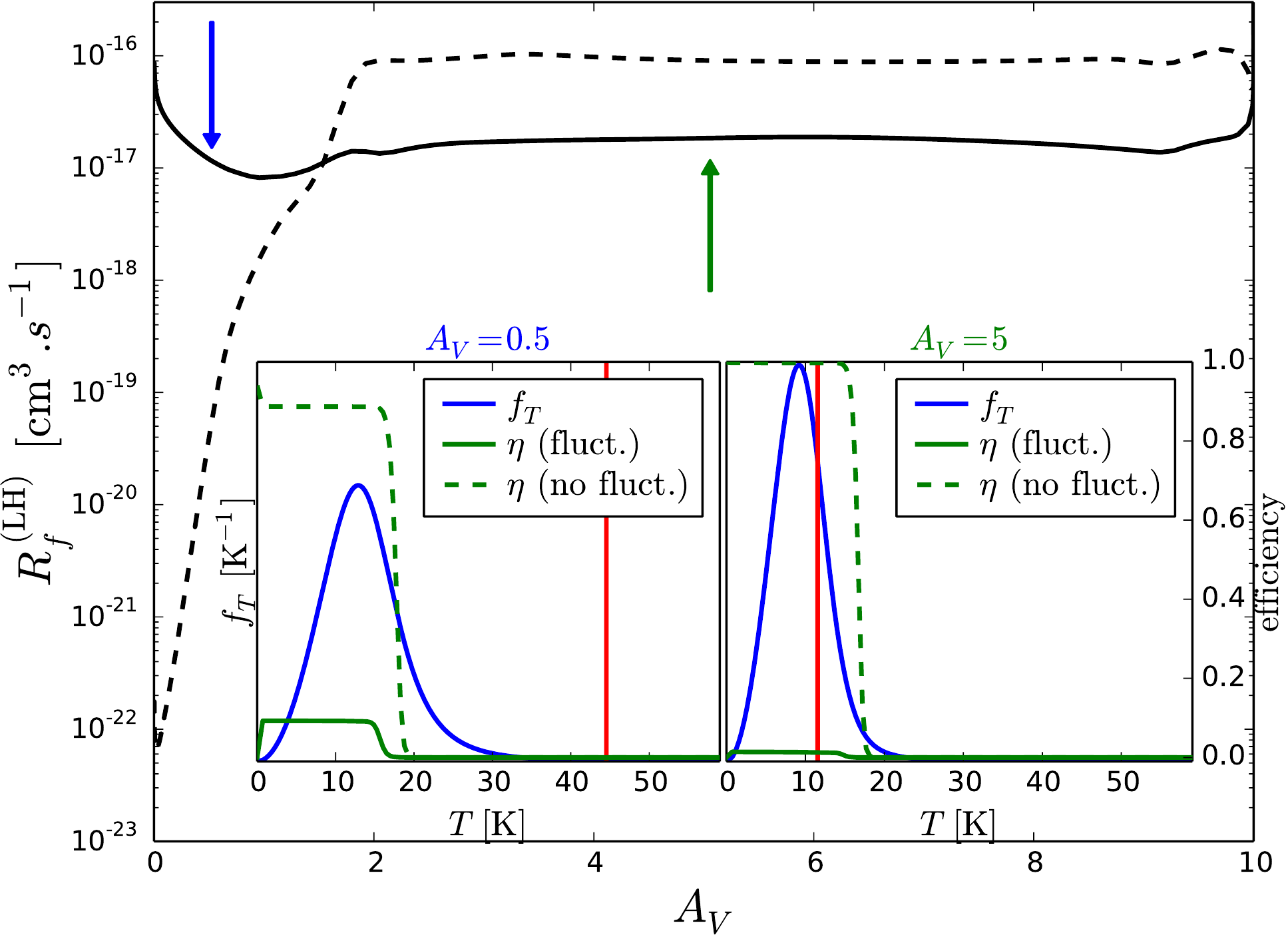}

\caption{Same cloud as Fig.~\ref{fig:H_Cmp}, with formation parameter $R_{f}$
for the LH mechanism, comparing standard (dashed black) and coupled
(solid black) models. The insets show the effect of fluctuations on
a $1\,\mathrm{nm}$ grain at the two positions in the cloud marked
by the arrows. Blue: temperature PDF, red: equilibrium temperature,
dashed green: equilibrium efficiency, and solid green: conditional
efficiency with fluctuations. See the text for more details.\label{fig:Av6_Tdust}}
\end{figure}

The formation efficiency at the center of the cloud is decreased by
almost one order of magnitude when including the fluctuations. This
unexpected effect is detailed on Fig. \ref{fig:Av6_Tdust} for amorphous
carbon surfaces: the formation parameter $R_{f}$ for the LH mechanism
is shown for the standard model (black dashed line) and the coupled
model (black solid line), and two inset plots show the effect of fluctuations
on a $1\,\mathrm{nm}$ grain at two positions ($A_{V}=0.5$ and $A_{V}=5$,
marked, respectively, by the blue and green vertical arrows on the
main plot). Each inset contains a plot similar to Fig. \ref{fig:overplot_fT_efficiency},
displaying the temperature PDF (blue line), the equilibrium temperature
(red line), the equilibrium efficiency curve (dashed green line),
and the effective conditional efficiency curve when taking fluctuations
into account (solid green line). 

Close to the edge of the cloud (at $A_{V}=0.5$), the situation is
similar to the discussion of Fig. \ref{fig:overplot_fT_efficiency}
(see Sect. \ref{sub:Results-LH}). The main effect occurs because
the temperature PDF has an important part inside the efficiency domain,
while the equilibrium temperature is out of this domain. It results
in increased efficiency. The dangers of using the equilibrium temperature
are strikingly emphasized on this figure: when allowing fluctuations,
a very small fraction of high temperature grains is sufficient to
radiate away all the absorbed energy and most of the grains can be
cold, while the grains must all be warm enough to balance the absorbed
energy if we force them all to have a single temperature. 

Deeper in the cloud (at $A_{V}=5$), the situation is different: the
equilibrium temperature now falls close to the actual peak of the
PDF and inside the efficiency domain, so that the actual position
and spread of the PDF does not change the result much. The dominant
effect is now that the conditional efficiency with fluctuations is
strongly reduced compared to the equilibrium efficiency. As discussed
in Sect. \ref{sub:Results-LH}, this reduction comes from the competition
between the fluctuation timescale and the adsorption timescale. At
this optical depth, the grains are mainly heated by IR photons emitted
by the hot dust at the edge, and for $1\,\mathrm{nm}$ grains, the
IR photons are sufficient to bring the grains temporarily out of the
efficiency domain (above $18\,\mathrm{K}$ here) where they lose their
surface population. On the other hand, adsorption of $\mathrm{H}$
atoms are extremely rare, as the gas is almost completely molecular.
As a result, the grain rarely manages to have two atoms on its surface,
and the formation efficiency is thus reduced.

The effect described here concern very small grains, and one might
question the presence of such very small grains and PAHs deep in the
cloud core (the grain population is independent of position in our
model). However, secondary UV photons, which were not considered for
grain heating in these models, could induce a similar effect for much
bigger grains, and further reduce the formation efficiency at high
optical depth inside the cloud. This mechanism tends to increase the
residual atomic $\mathrm{H}$ fraction in the core and could be relevant
in explaining the slightly higher than expected atomic $\mathrm{H}$
abundances deduced from observations of $\mathrm{HI}$ self-absorption
in dark clouds (e.g., \citealt{LiGoldsmith2003}).

No detectable effects on intensities were found in these models but
the inclusion of secondary UV photons may induce stronger effects.

\subsection{Grid of models\label{sub:Grid}}

\begin{figure*}
\includegraphics[width=0.5\textwidth]{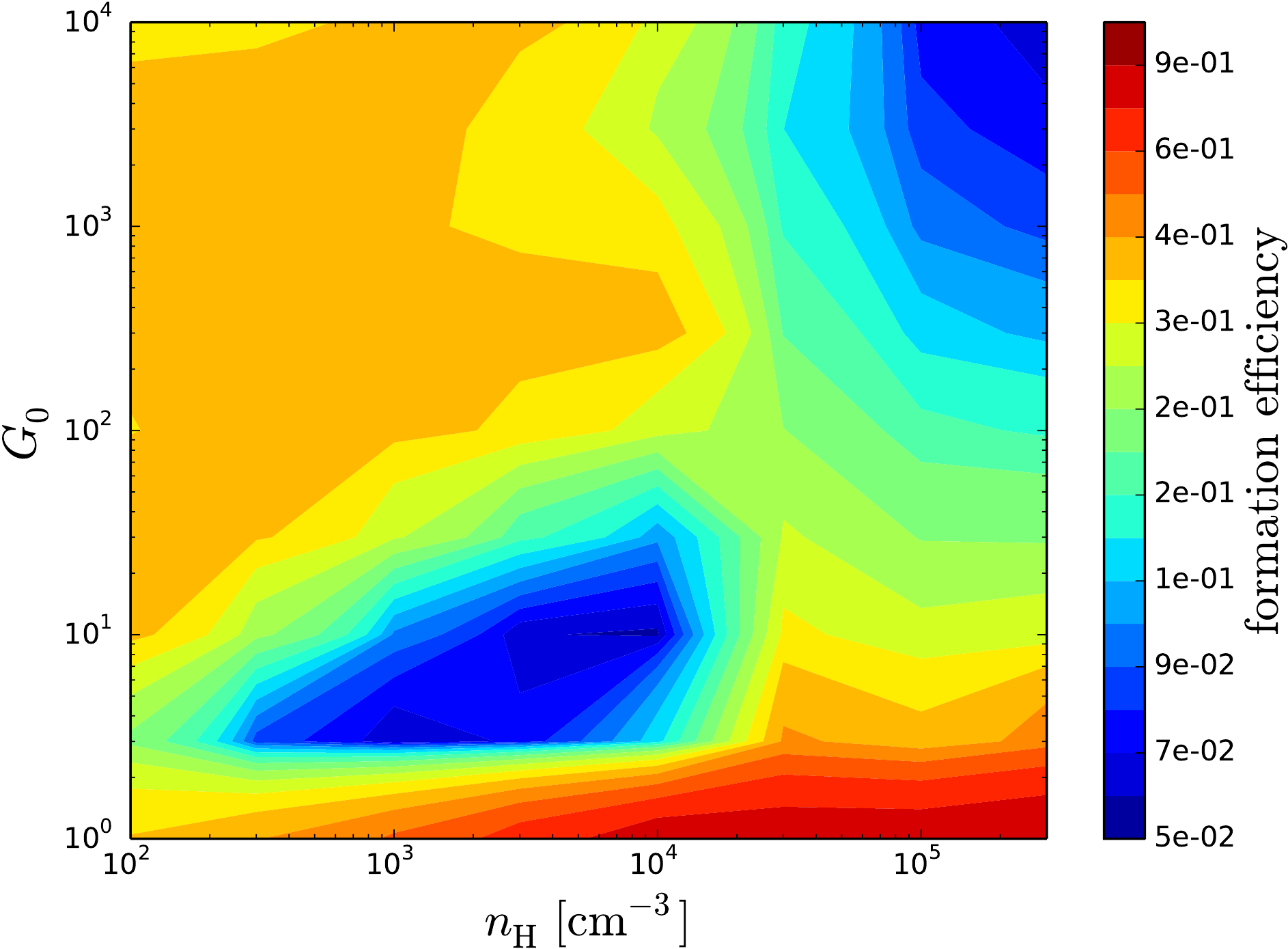}\includegraphics[width=0.5\textwidth]{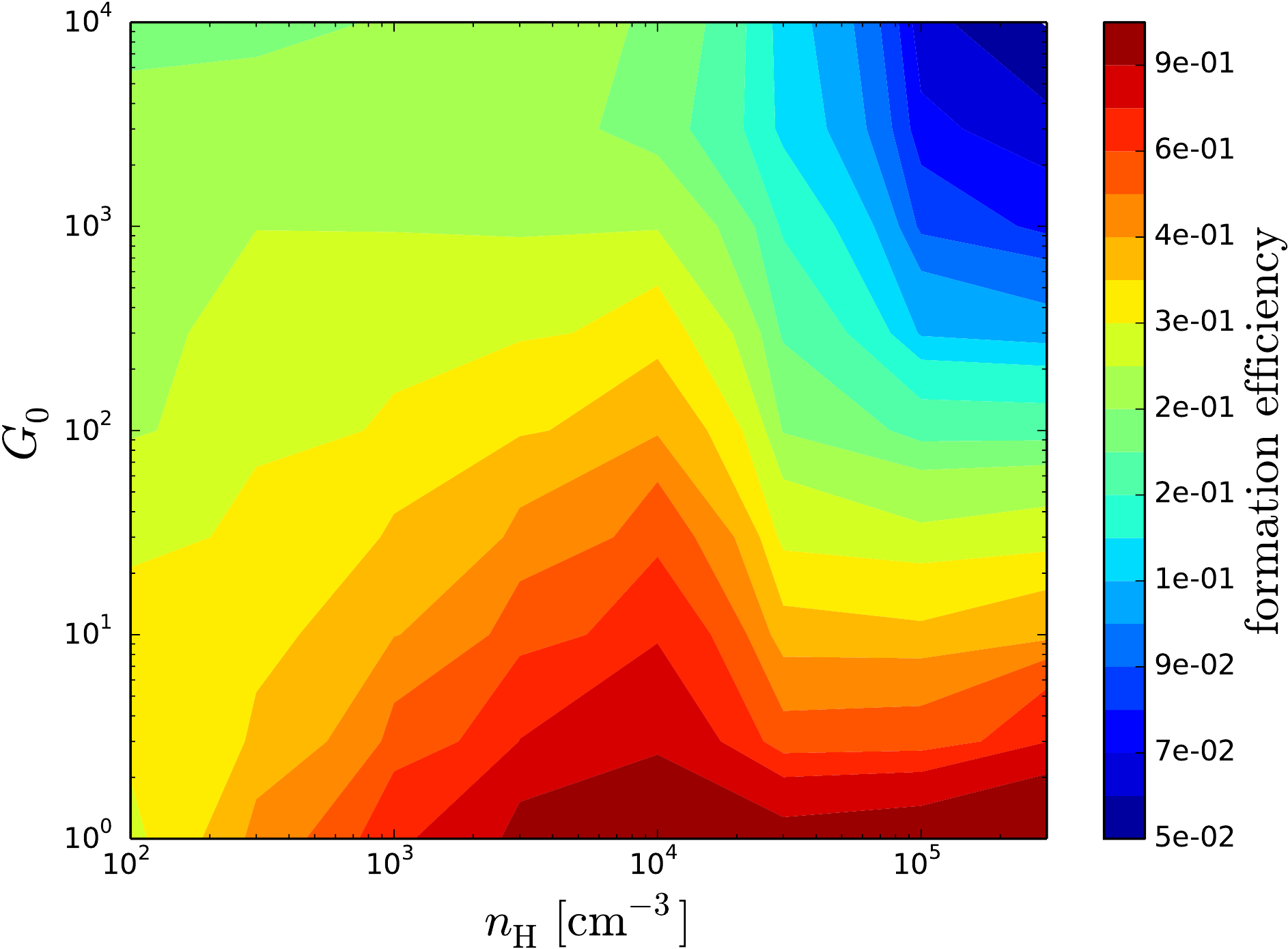}

\caption{$\mathrm{H}_{2}$ formation efficiency at the edge of the cloud as
a function of the gas density $n_{\mathrm{H}}$ and the radiation
field intensity $G_{0}$. The standard models (left) are compared
with the full coupled models (right). \label{fig:maps-eff}}
\end{figure*}

\begin{figure*}
\includegraphics[width=0.5\textwidth]{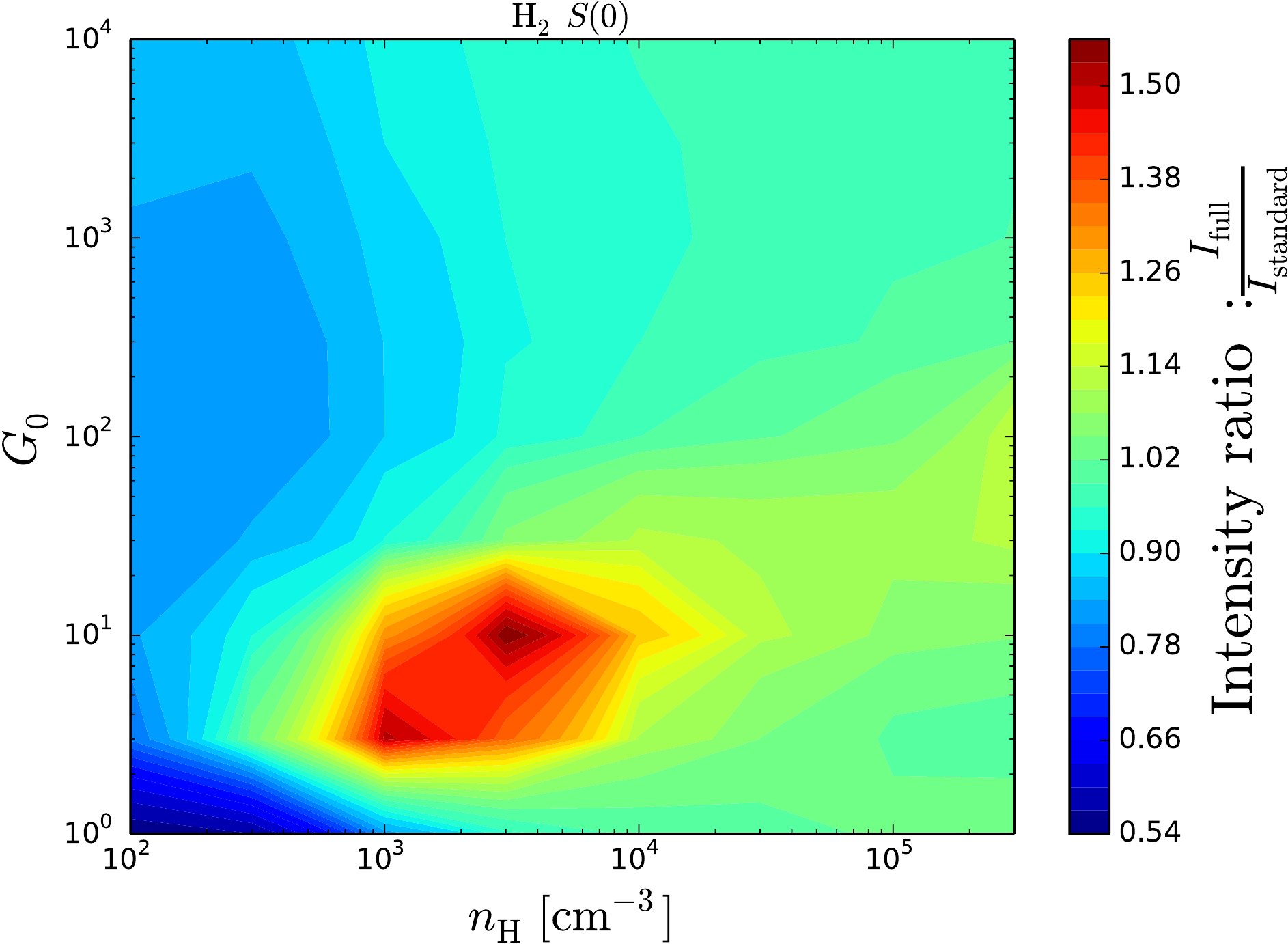}\includegraphics[width=0.5\textwidth]{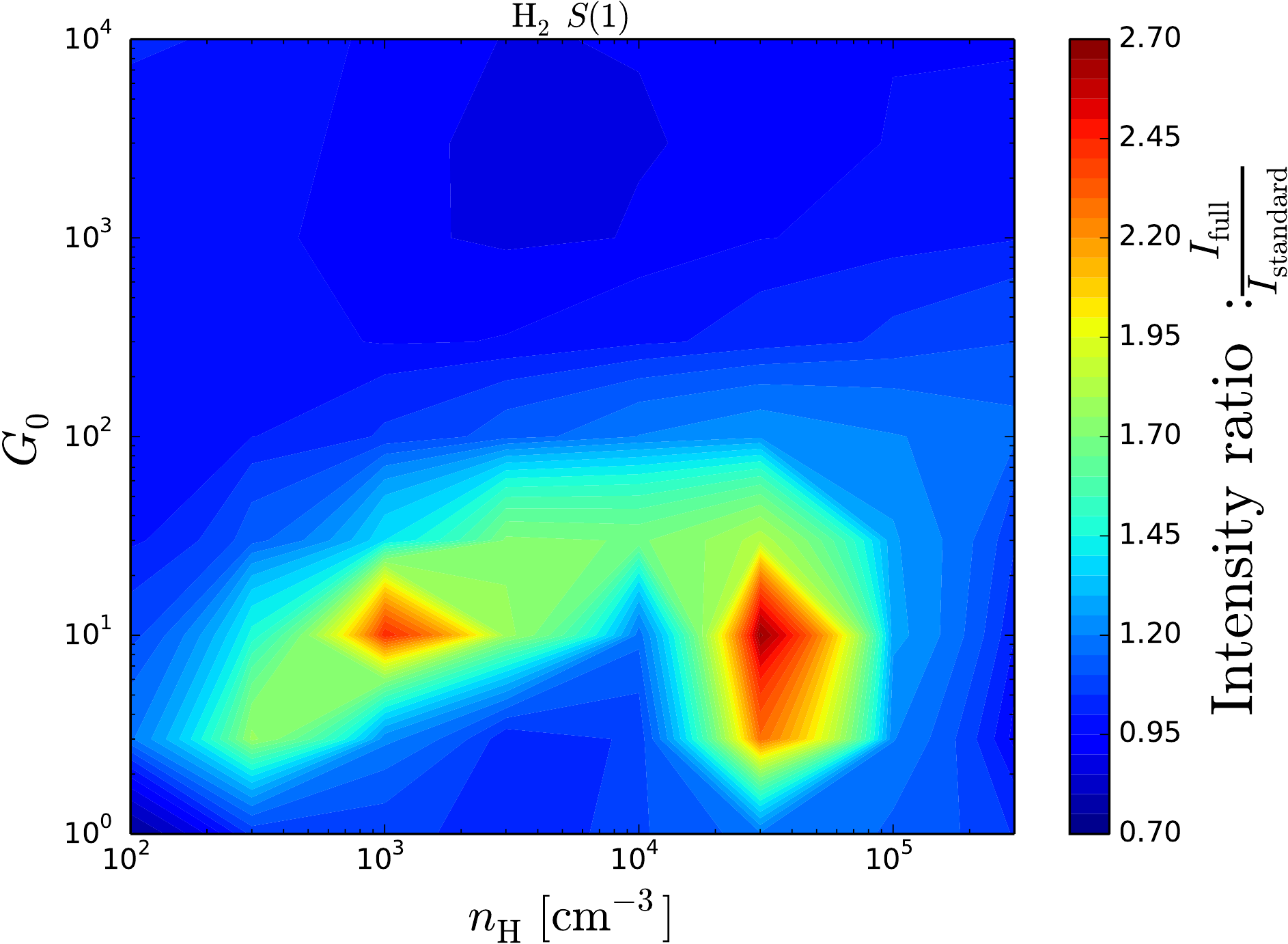}

\includegraphics[width=0.5\textwidth]{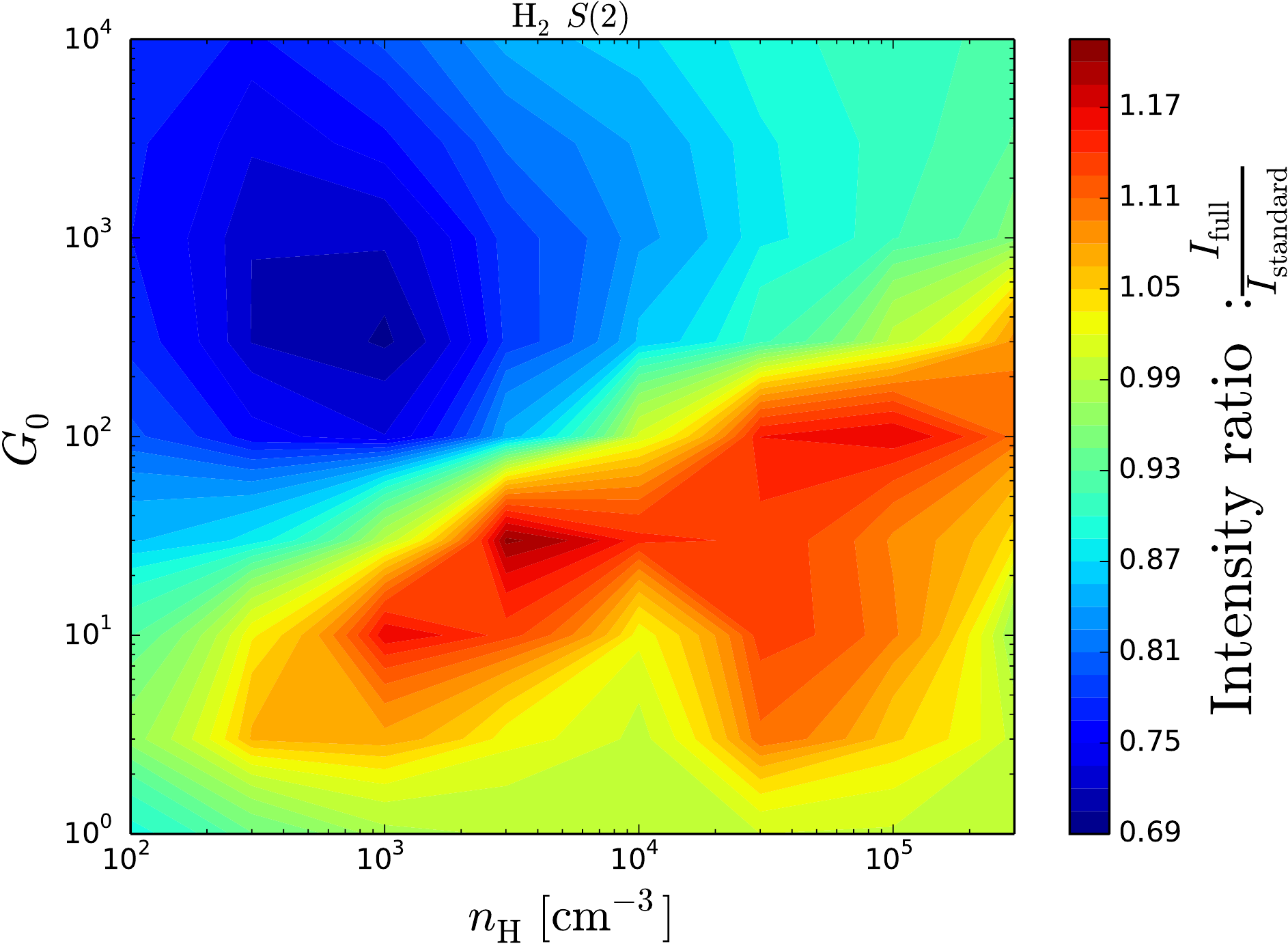}\includegraphics[width=0.5\textwidth]{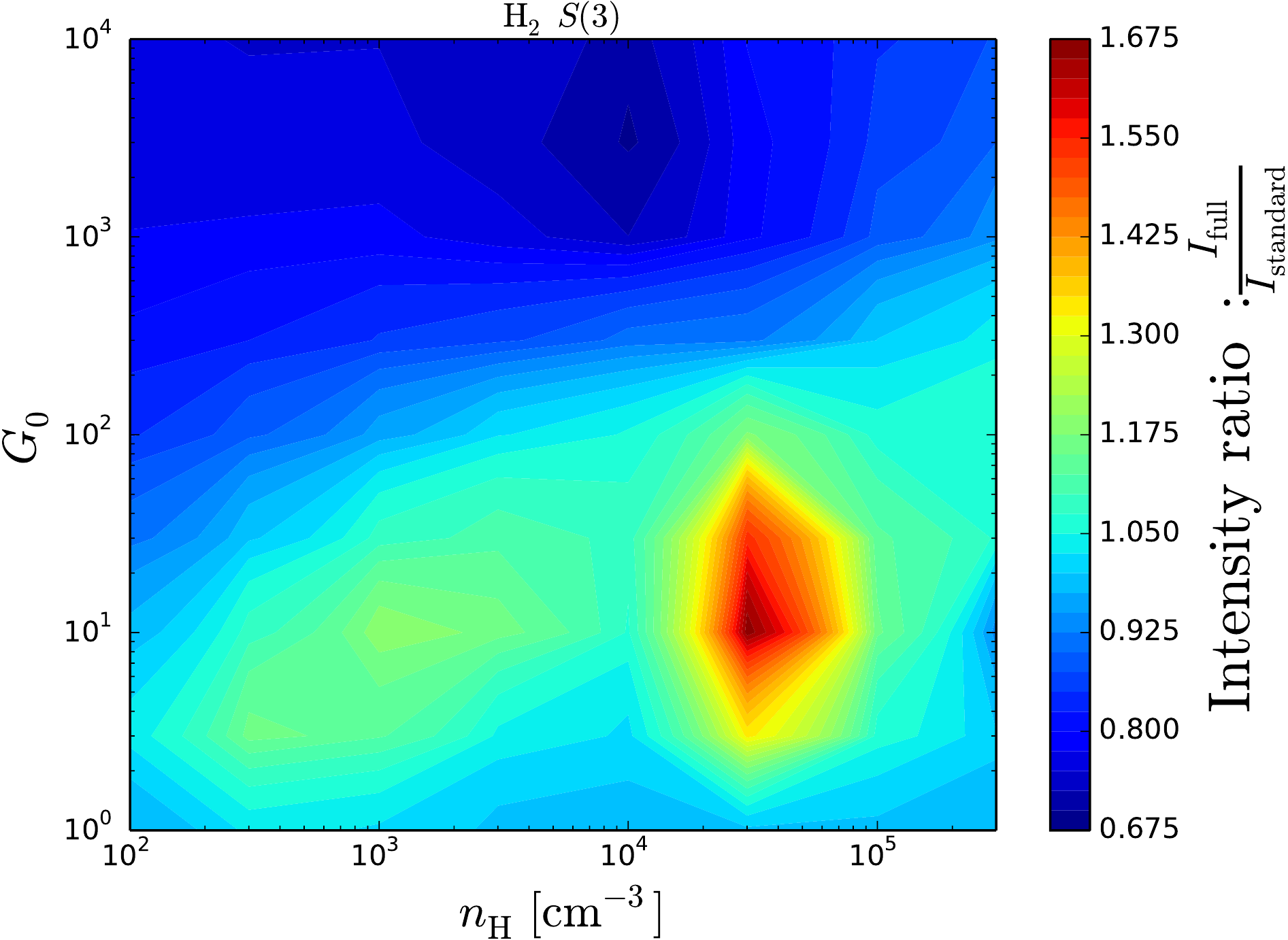}

\caption{Effect on H$_{2}$ rotational lines intensities $S(0)$ to $S(3)$,
shown as a comparison between the full computation result and the
standard PDR result (neglecting dust temperature fluctuations). Parameters
are the gas density $n_{\mathrm{H}}$ and the scaling factor of the
radiation field $G_{0}$.\label{fig:maps-H2}}
\end{figure*}

\begin{figure}
\includegraphics[width=1\columnwidth]{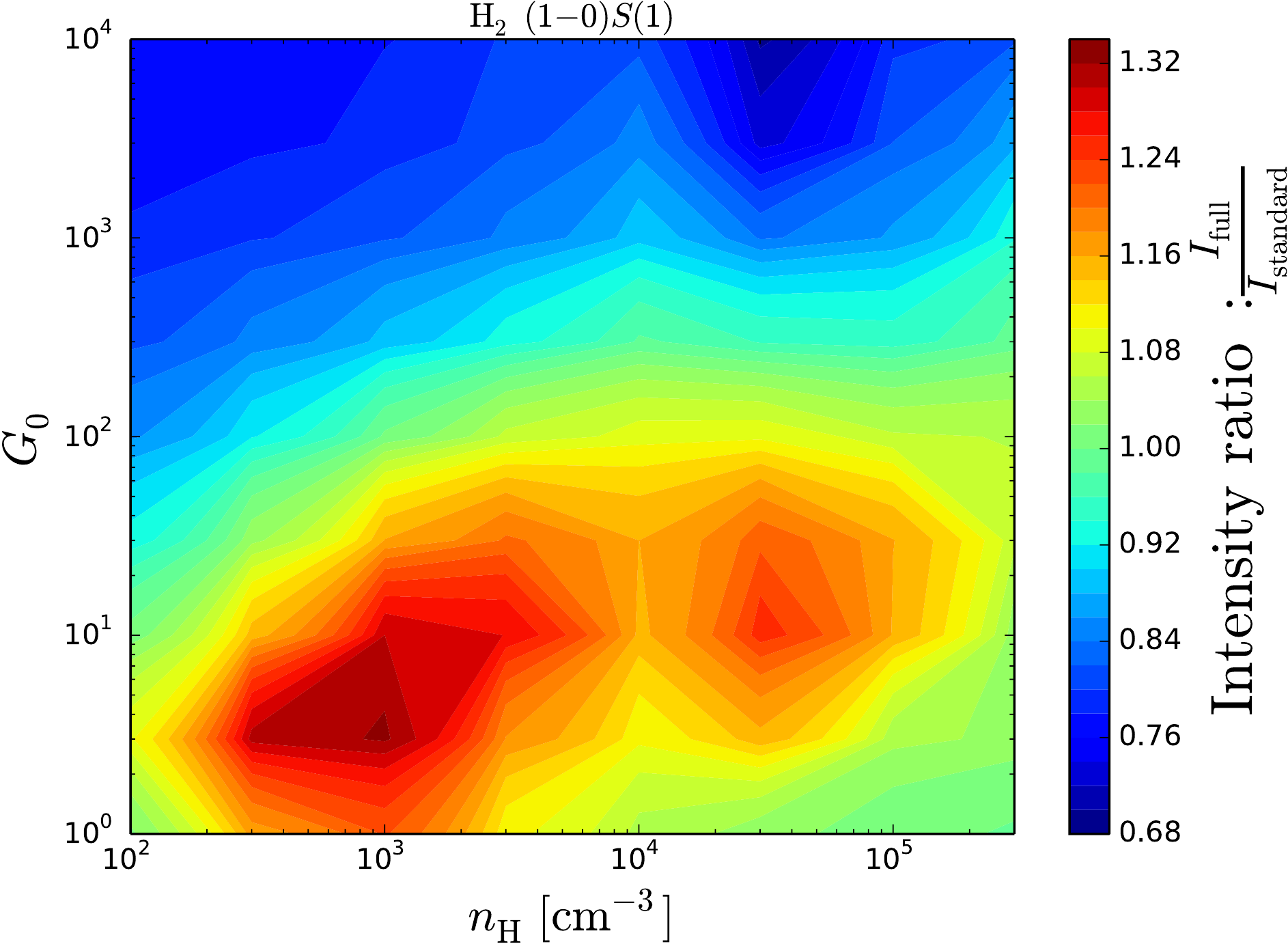}

\caption{Effect on H$_{2}$ ro-vibrational line intensity $(1-0)\, S(1)$,
shown as a comparison between the full computation result and the
standard PDR result (neglecting dust temperature fluctuations). Parameters
are the gas density $n_{\mathrm{H}}$ and the scaling factor of the
radiation field $G_{0}$.\label{fig:map_H2_10S1}}

\end{figure}

A grid of constant density PDR models was run to check the effect
on observable intensities over a wider parameter range. We explore
gas densities from $n_{\mathrm{H}}=10^{2}\mathrm{\, cm}^{-3}$ to
$n_{\mathrm{H}}=3\times10^{5}\,\mathrm{cm}^{-3}$ and scaling factors
on the standard ISRF (\citealp{mathis:83}) from $1$ to $10^{4}$
(as explained previously, the scaling factor only affects the UV part
of the radiation field).

The effects of the fluctuations are mostly found on tracers of the
PDR edges. The formation efficiency at the edge of the cloud is shown
as a color map on Fig. \ref{fig:maps-eff}, which compares the standard
models with the full models including fluctuations. In the upper part
of the map, the ER formation is decreased, and the LH mechanism remains
negligible, resulting in an overall (small) decrease in efficiency.
In the lower part, the LH mechanism is not negligible anymore, and
its strong increase due to fluctuations results in an overall increase
in efficiency. The region around $G_{0}=10$ and $n_{\mathrm{H}}=10^{3}-10^{4}\mathrm{cm}^{-3}$
is strikingly affected. In this region, the ER mechanism stops being
efficient due to the low gas temperature at the edge. In standard
models, the equilibrium temperature of the dust at the edge is still
too high for the LH mechanism to work, giving a very low total formation
efficiency. When taking into account fluctuations, the LH mechanism
works efficiently, assuring a high formation efficiency. The effect
is further amplified as gas heating by $\mathrm{H}_{2}$ formation
is dominant in this region. The efficient LH formation heats up the
gas, thus keeping ER formation efficient as well.

Figure~\ref{fig:maps-H2} shows the modification of the H$_{2}$
lower rotational intensities as the ratio between the result of the
full computation, which includes dust temperature fluctuations, and
the result of a standard PDR model. Higher rotational lines are affected
very similarly to the $S(3)$ line. Figure~\ref{fig:map_H2_10S1}
shows the effect on the ro-vibrationnal line $(1-0)\, S(1\text{)}$
in the same way. As expected from our discussion of Fig. \ref{fig:maps-eff},
we see increased intensities for low radiation fields (roughly $G_{0}<200$),
especially in regions where the gas temperature is highly sensitive
on $\mathrm{H}_{2}$ formation heating and where LH formation can
further trigger an amplification of ER formation. For higher radiation
fields, the intensities are slightly decreased. The effect remains
small, the increase for low radiation field models is at most a factor
of 2.7 (on the $S(1)$ line), and the decrease on the high radiation
field models is at most $30\%$. 

\begin{figure*}
\centering

\includegraphics[width=0.5\textwidth]{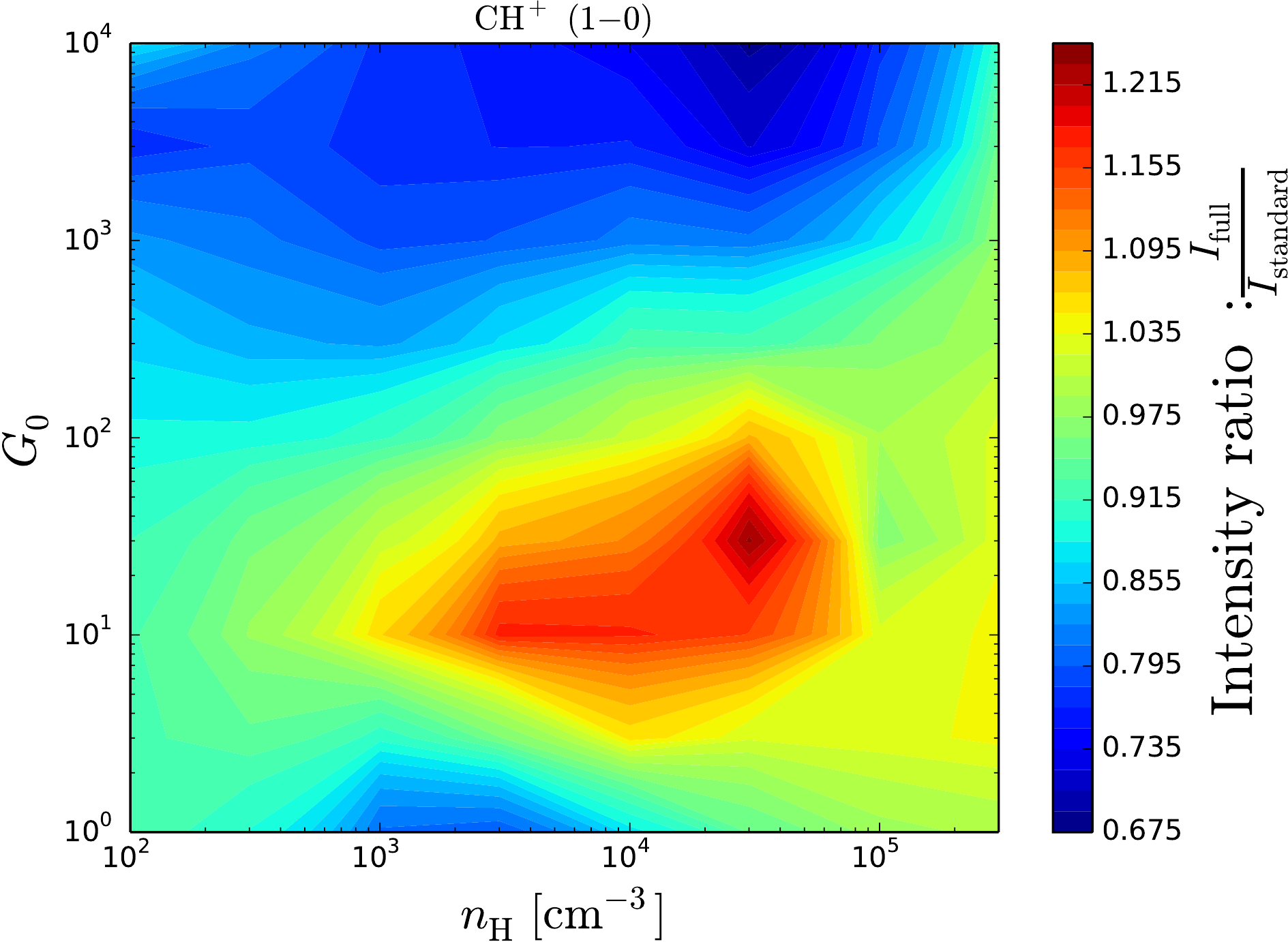}\includegraphics[width=0.5\textwidth]{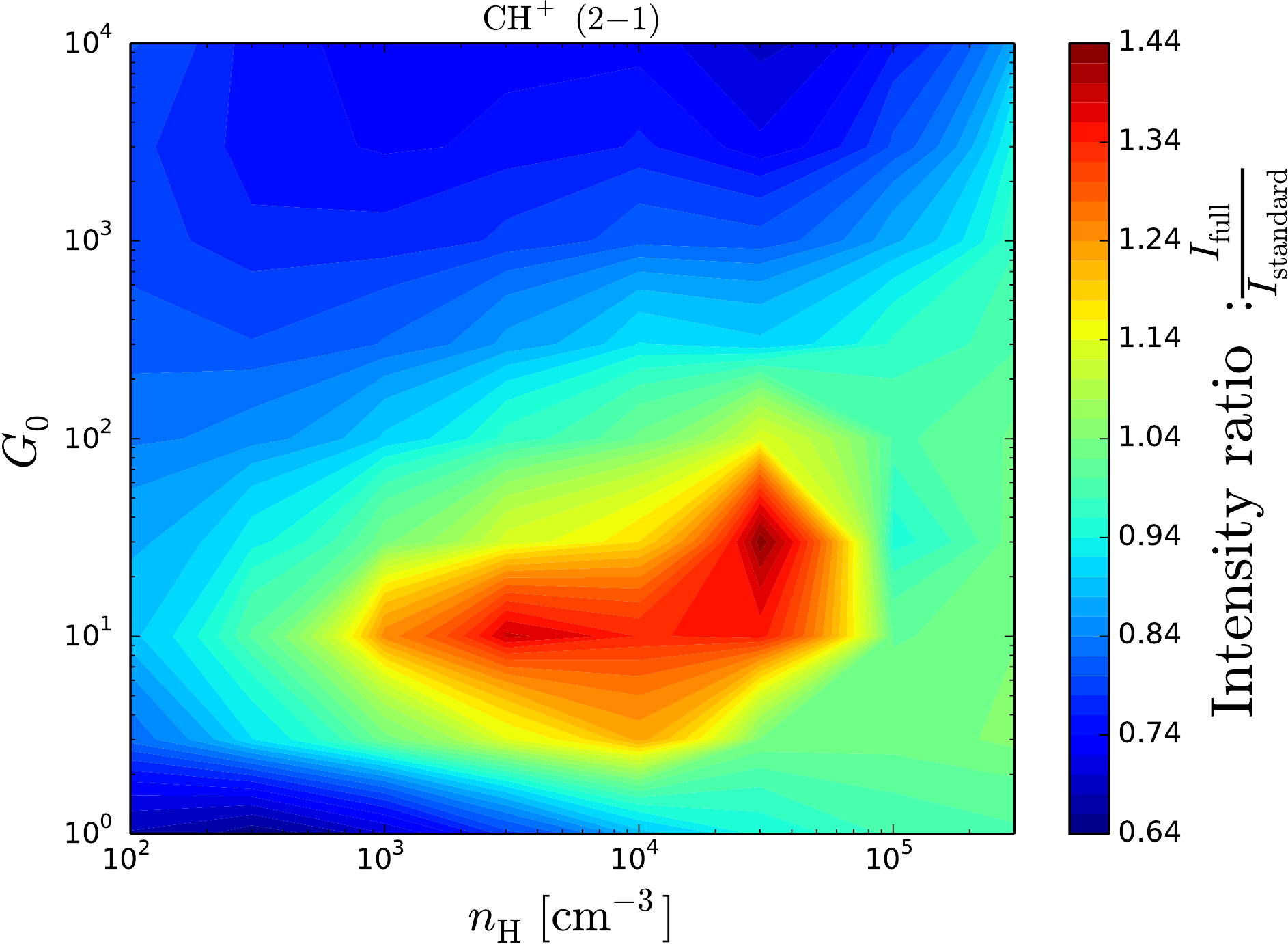}

\includegraphics[width=0.5\textwidth]{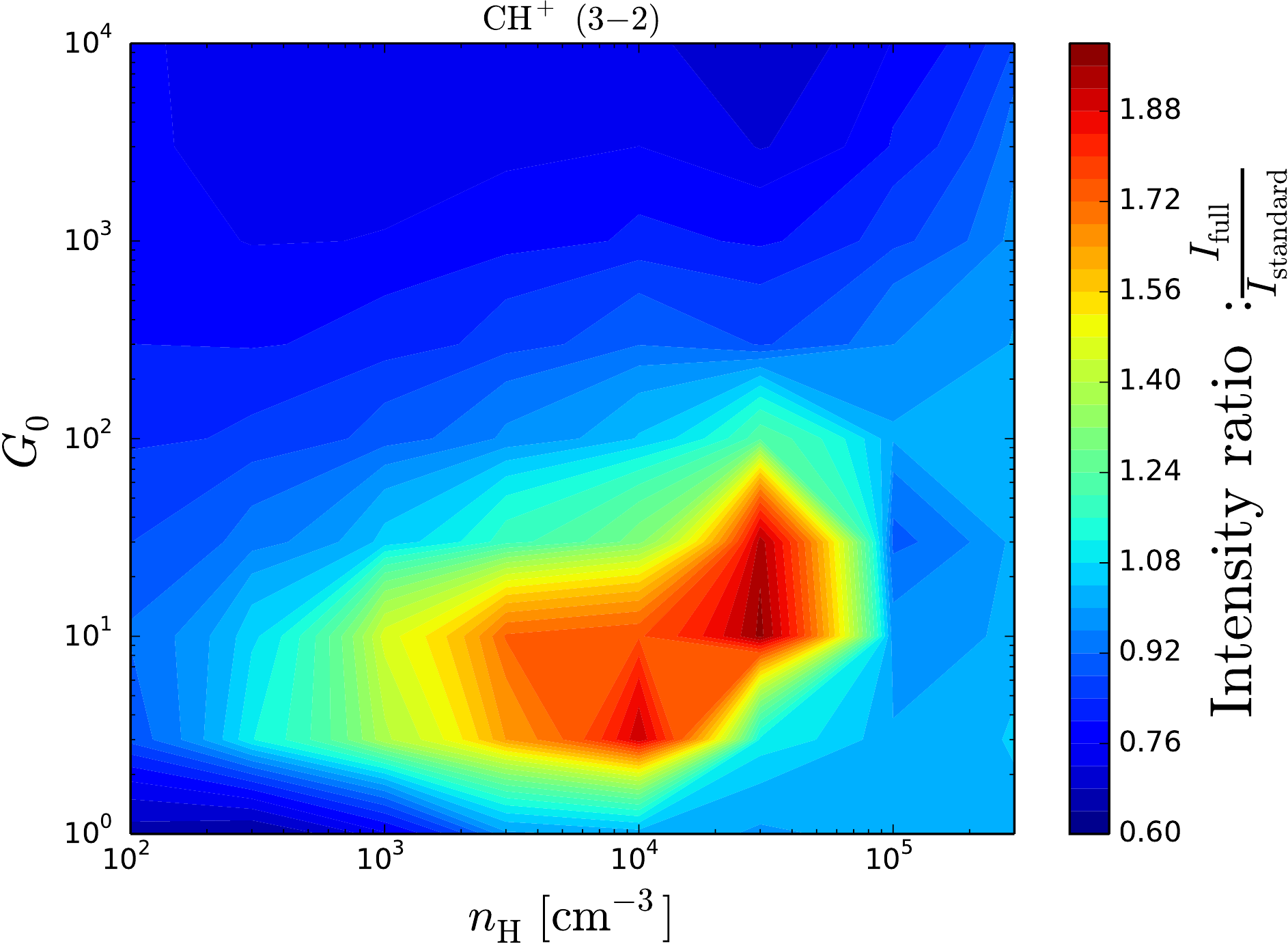}\includegraphics[width=0.5\textwidth]{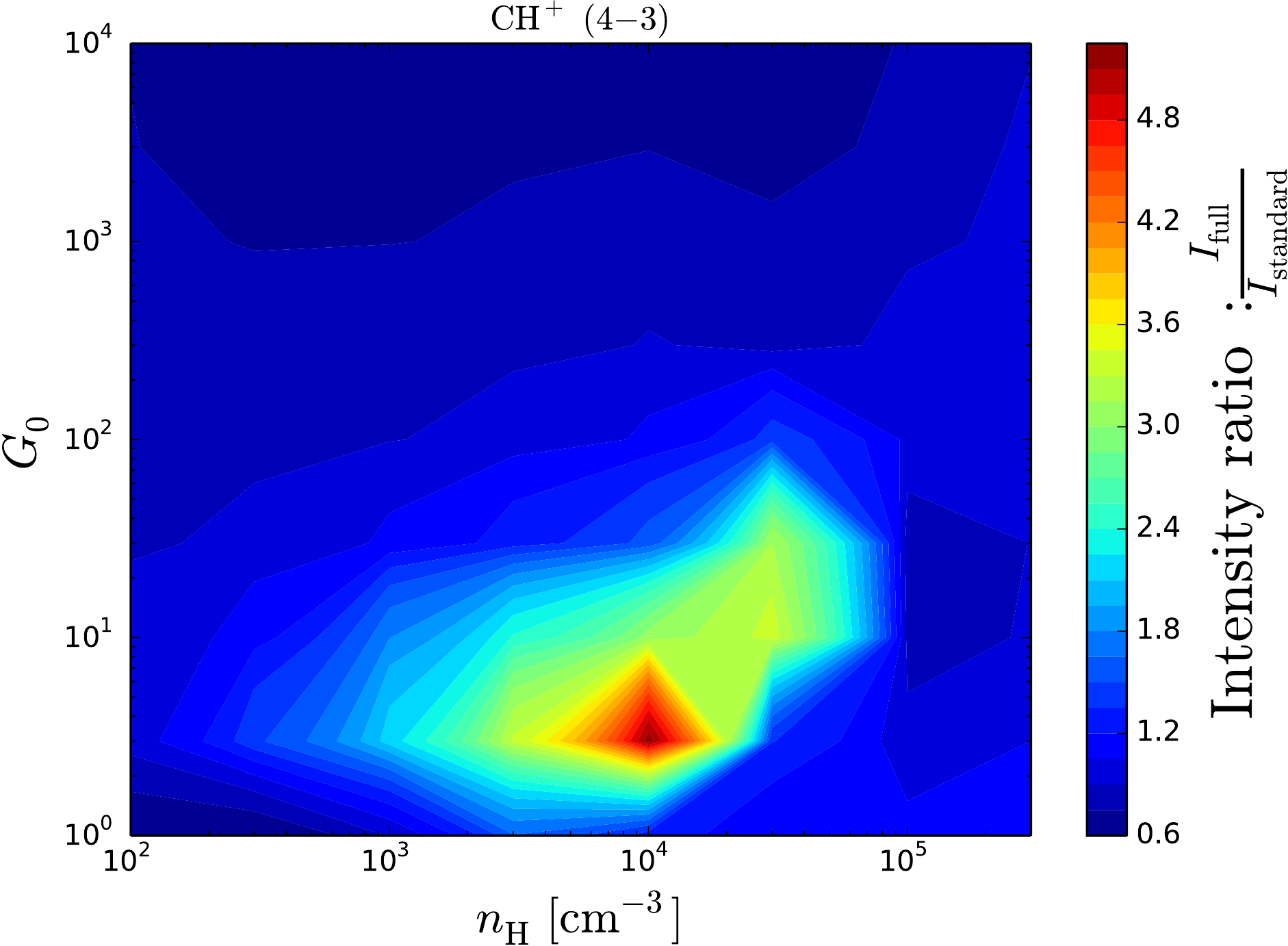}

\caption{Effect on $\mathrm{CH}^{+}$ line intensities, shown as a comparison
between the full computation result and the standard PDR result (neglecting
dust temperature fluctuations). Parameters are the gas density $n_{\mathrm{H}}$
and the scaling factor of the radiation field $G_{0}$.\label{fig:maps-CHplus}}
\end{figure*}

Another species affected by the more efficient formation of $\mathrm{H}_{2}$
at the edge is $\mathrm{CH}^{+}$. It is formed by the reaction of
$\mathrm{C}^{+}+\mathrm{H_{2}}$, which is allowed by the high gas
temperature at the edge of the PDR and by the internal energy of the
$\mathrm{H}_{2}$ molecule, as described in \citet{agundez:2010}
(pumped by UV photons and by its formation process). A higher $\mathrm{H}_{2}$
formation rate at the edge induces a higher $\mathrm{H}_{2}$ fraction
before the transition, which leads to a higher $\mathrm{CH}^{+}$
formation rate. Gas heating by $\mathrm{H}_{2}$ formation can further
enhance its formation. Figure~\ref{fig:maps-CHplus} shows the effect
on $\mathrm{CH}^{+}$ intensities. We again observe increased emission
in the domain where the $\mathrm{H_{2}}$ formation efficiency is
enhanced, due to both higher abundances of $\mathrm{CH}^{+}$ and
higher gas temperatures. The domain of strong enhancement is quite
similar to the one for $\mathrm{H}_{2}$ rotational lines, corresponding
to the domain where gas heating is dominated by $\mathrm{H}_{2}$
formation heating. This temperature effect is visible on all high
excitation lines emitted before the $\mathrm{H/H_{2}}$ transition
(e.g., the most excited lines of $\mathrm{HCO}^{+}$, CS, HCN, ...).

\section{Conclusion and perspectives\label{sec:Conclusion}}

We presented a method to compute $\mathrm{H}_{2}$ formation on dust
grains taking fully the fluctuating temperature of the grain for both
the ER and the LH mechanism into account.

As a side result of this work, we revisited the computation of dust
temperature fluctuations. We confirm that the continuous cooling approximation
(e.g., used in DustEM) is sufficient for an accurate computation of
the mid- and near-infrared spectrum. However, the detailed treatment
of surface processes requires a better estimation of the low temperature
part of the dust temperature PDF, which is done in this paper by including
the discrete nature of photon emission.

The coupled fluctuations problem was solved exactly for both the ER
and the LH mechanisms. A numerically efficient way of computing the
ER formation rate was found. The effect on the ER formation rate is
limited to a factor of $ $$2$, and the effect on full cloud simulations
is limited. The standard rate equation treatment of this ER mechanism
is thus a good treatment for the purpose of complete PDR simulations.

With the exact solution for the LH mechanism being numerically intractable
for big grains, we constructed an approximation based on simple physical
arguments to compute the fluctuation effect on the LH mechanism and
checked for small grains that the approximation yields accurate results.
The effect is strong as fluctuating grains spend a significant part
of their time in the low temperature range, in which formation is
efficient, even when their average temperature is out of the rate
equation efficiency domain. This results in a strong increase in LH
formation efficiency in unshielded environments where the average
dust temperature is high and only fluctuations can allow LH formation.

The LH mechanism is found to become of comparable importance with
the ER mechanism at the edge of PDRs, except for strong radiation
fields. The resulting increase of $\mathrm{H}_{2}$ formation rate
at the edge induces a shift in the $\mathrm{H}/\mathrm{H}_{2}$ transition
toward the edge of the PDR. The resulting effect on observable lines
emitted in this region remains limited and does not reach one order
of magnitude in the strongest cases.

On the contrary, fluctuations of the smallest grains caused by IR
photons reduce the formation efficiency by almost one order of magnitude
in the core of the clouds, as collisions with $\mathrm{H}$ atoms
in a completely molecular gas are much rarer than absorption of IR
photons heating temporarily the grain above 18K. No significant effects
are found on intensities, but including secondary UV photons could
induce stronger effects. 

Overall, the variations induced in observational intensities are comparable
to what can be expected from many other poorly known parameters. The
interpretation of observations with standard PDR codes, thus, need
not use the detailed treatment analyzed here and a simple rate coefficient
approximation can be used most of the time. However, this cannot be
blindly extended to any arbitrary process and must be checked in each
specific case. In particular, the existence of temperature fluctuations
of the smallest grains deep inside the cloud should be taken into
account for other surface chemical processes. Possible consequences
on the abundance of light organic molecules are currently being investigated.
\begin{acknowledgements}
This work was funded by grant ANR-09-BLAN-0231-01 from the French
\textit{Agence Nationale de la Recherche} as part of the SCHISM project.
We thank Laurent Verstraete for fruitful discussions on dust properties,
and Evelyne Roueff for many discussions. We thank our referee for
his useful comments.
\end{acknowledgements}
\appendix

\section{Effect of $\mathrm{H_{2}}$ formation energy on the equilibrium rate
equation calculation\label{app:form_energy}}

We first consider the heating of the grain by the Eley-Rideal mechanism.
Each formation reaction gives $1\,\mathrm{eV}$ to the grain.

The liberation of energy left in the grain by each formation reaction
creates an additional coupling between the surface population and
the grain temperature. The method described in Sect. \ref{sub:Eley-Rideal-mechanism},
which was based on a one-way coupling between the two variables, is
thus not possible here.

To evaluate this effect, we use the standard equilibrium temperature
equation and chemical rate equation with a coupling term representing
the heating of the grain by the formation reaction. We assume that
each formation reaction gives an energy $E_{f_{\mathrm{H}_{2}}}=1\,\mathrm{eV}$
to the grain, which is instantly converted into thermal energy.

For ER formation on chemisorption sites, we thus have the system of
equations:
\begin{multline*}
n_{\mathrm{eq}}(T_{\mathrm{eq}})=N_{\mathrm{s}}\,\frac{s(T_{\mathrm{gas}})}{1+s(T_{\mathrm{gas}})+\frac{N_{\mathrm{s}}\, k_{\mathrm{des}}(T_{\mathrm{eq}})}{k_{\mathrm{coll}}}},\\
\int_{0}^{+\infty}dU\, P_{\mathrm{abs}}(U)+E_{f_{\mathrm{H}_{2}}}\, k_{\mathrm{coll}}\,\frac{n_{\mathrm{eq}}(T_{\mathrm{eq}})}{N}=\int_{0}^{E_{\mathrm{grain}}}dU\, P_{\mathrm{em}}(U,T_{\mathrm{eq}}),
\end{multline*}
which can be easily solved. However, we find that the effect of this
average heating of the grain by formation reactions is completely
negligible in the range of realistic values of parameters, resulting
in a change of the grain temperature by less than $1\%$.

Independently of the detail of the formation mechanism, we can put
an upper limit on the heating term caused by $\mathrm{H}_{2}$ formation
by assuming that all $\mathrm{H}$ atoms hitting the grain are turned
into $\mathrm{H}_{2}$. In this case, the formation rate is simply
$\frac{1}{2}k_{\mathrm{coll}}$, and the heating power is $\frac{1}{2}k_{\mathrm{coll}}E_{f_{H_{2}}}$.
We can then compare this term to the absorbed power coming from photons. 

\begin{figure}
\includegraphics[width=1\columnwidth]{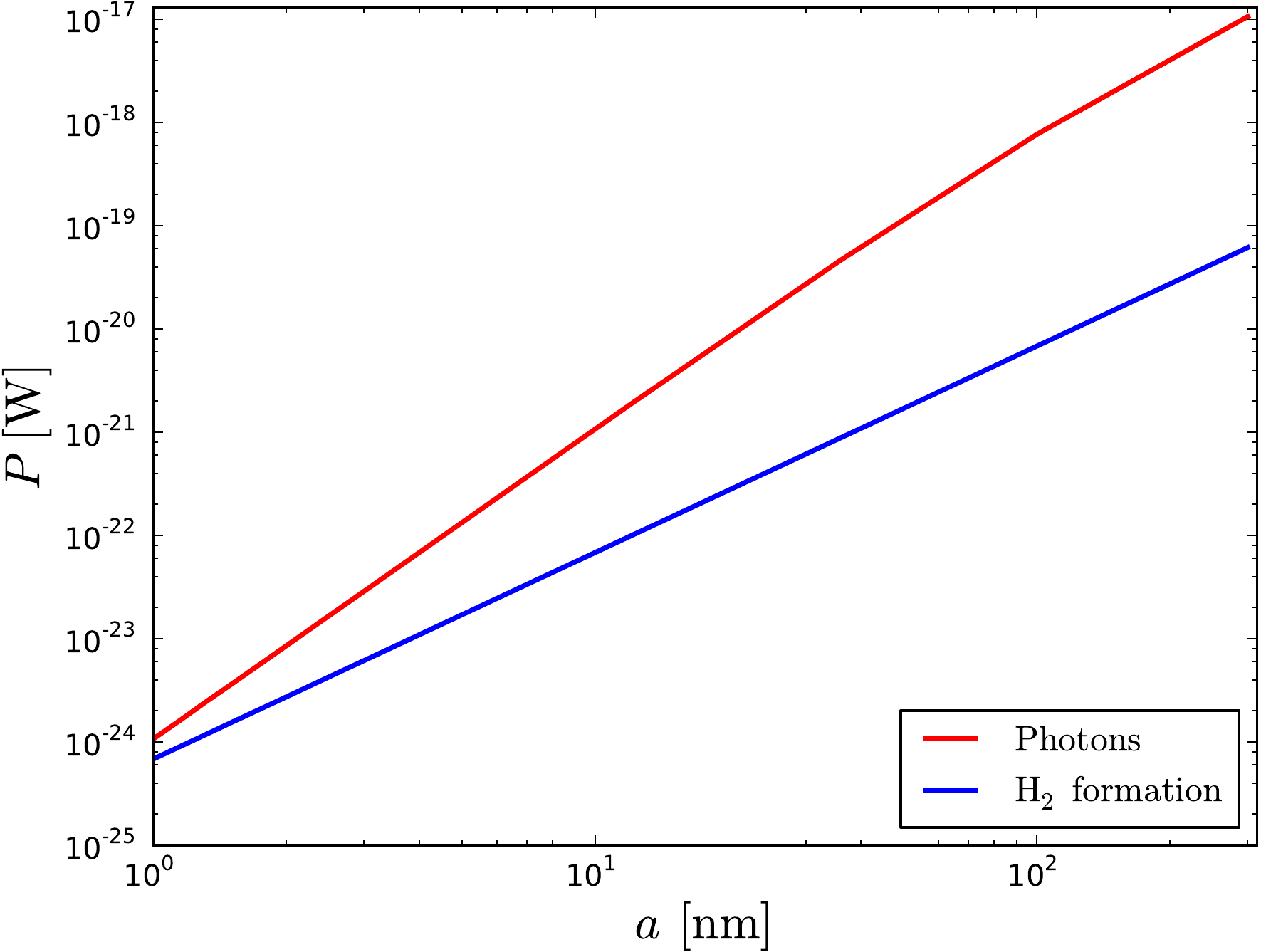}

\caption{Heating power received by a grain as a function of grain size. Power
received from photons in red and $\mathrm{H}_{2}$ formation in blue.\label{fig:Puiss_form}}

\end{figure}

The result is shown here on Fig.~\ref{fig:Puiss_form} in an extreme
case with a low radiation field ($G_{0}=1$) and a high collision
rate (atomic gas with $n=10^{3}\,\mathrm{cm}^{-3}$ and $T=350\,\mathrm{K}$).
Despite assuming a total formation efficiency in a low radiation field,
high collision rate environment, we can see that the heating term
due to $\mathrm{H}_{2}$ formation becomes barely significant only
for the smallest grain sizes. We can thus safely neglect the grain
heating coming from $\mathrm{H}_{2}$ formation (through the ER mechanism,
and all the more for the LH mechanism as the $E_{f_{H_{2}}}$ is probably
much lower as physisorbed atoms are much more weakly bound) in most
usual conditions. Moreover, under a radiation field with $G_{0}=1$,
the gas would not be as warm as we assumed here and the ER efficiency
would be very low.

\section{Eigenvalues of the integral operator for the thermal energy PDF\label{app:Eigenvalues-f_E}}

\textcolor{black}{The operator $\mathcal{L}$ defined in equation
\ref{eq:thermal} can be rewritten as
\[
\mathcal{L}[f](E)=\int_{0}^{+\infty}dE'\, A(E,E')\, f(E')
\]
}

with
\[
A(E,E')=\begin{cases}
{\displaystyle \frac{R_{\mathrm{abs}}(E-E')}{P(E)}} & \mathrm{if}\quad E'<E\\
{\displaystyle \frac{R_{\mathrm{em}}(E'-E,T(E'))}{P(E)}} & \mathrm{if}\quad E'\ge E
\end{cases},
\]

where 
\[
P(E)=\int_{0}^{E}dE'\, R_{\mathrm{em}}(E-E',T(E))\,+\,\int_{E}^{+\infty}dE'\, R_{\mathrm{abs}}(E'-E).
\]

The existence of a solution to our problem is equivalent to the existence
of a positive eigenfunction associated with the eigenvalue 1.

This operator is continuous, compact, and strongly positive. The Krein-Rutman
theorem (see \citealp[Chap. 1]{du:06}) tells us that its spectral
radius is a simple eigenvalue associated with a positive eigenfunction
and that this eigenfunction is the only positive eigenfunction.

We first show that 1 is an eigenvalue of the adjoint operator $\mathcal{L}^{*}[f](E)=\int_{0}^{+\infty}dE'\, A(E',E)\, f(E')$
and that it is associated with a positive eigenfunction. We can easily
find such a positive eigenfunction with the eigenvalue 1 for this
adjoint operator:
\[
\begin{array}{rcl}
\mathcal{L}^{*}[\text{P](E)} & = & {\displaystyle \int_{0}^{+\infty}dE'\, A(E',E)\, P(E')}\\
\, & = & {\displaystyle {\displaystyle \int_{0}^{E}dE'\frac{R_{\mathrm{em}}(E-E',T(E))}{P(E')}P(E')}}\\
\, & + & {\displaystyle {\displaystyle \int_{E}^{+\infty}dE'\frac{R_{\mathrm{abs}}(E'-E)}{P(E')}P(E')}}\\
\, & = & {\displaystyle P(E)}
\end{array},
\]
and P(E) is a positive function. Thus, 1 is the spectral radius of
$\mathcal{L}^{*}$ and hence of $\mathcal{L}$. This proves the existence
and the unicity of the solution to our problem. Moreover, it proves
that all the other eigenvalues $\lambda$ are such that $|\lambda|\le1$
and thus $Re(\lambda)<1$. We can therefore iterate the application
of the operator to an initial function and converge toward the solution,
assuming that the initial function is not orthogonal to the solution.
Starting with a positive initial function is sufficient.

\section{Verification of the approximations\label{sec:Approx_comparison}}

In this appendix, we evaluate the accuracy of our approximations by
comparing their results to those of the exact methods presented in
Sect. \ref{sub:Method}.

\subsection{LH mechanism}

For the LH mechanism, the exact method (cf. Sect. \ref{sub:LH-method})
is only tractable for a limited range of grain sizes. We compare the
approximation (cf. Sect. \ref{sub:LH-Approx}) with it on this range
and verify that the approximation joins the equilibrium result for
bigger grains.

\begin{figure}
\begin{centering}
\includegraphics[width=1\columnwidth]{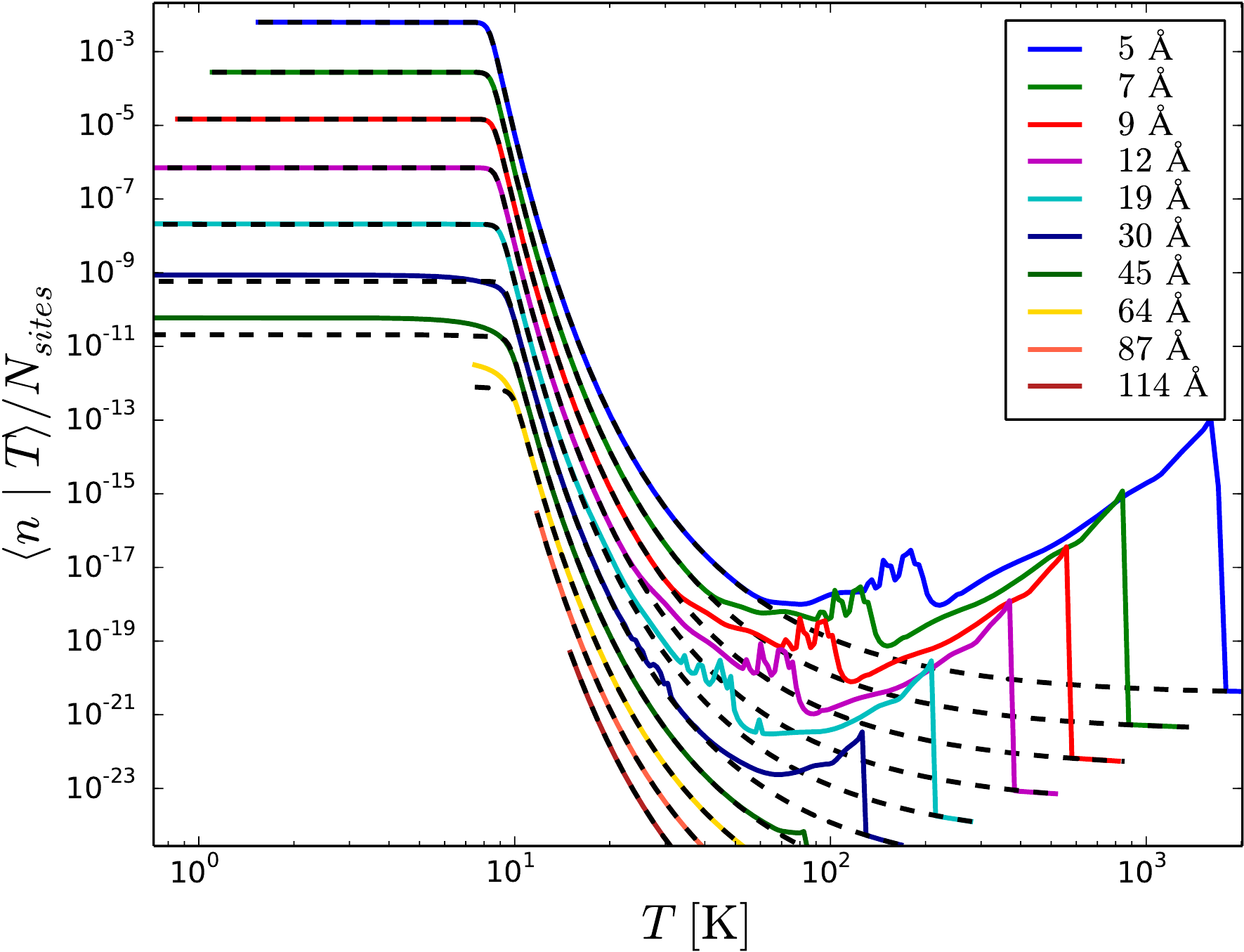}
\par\end{centering}

\caption{Approximated $\left\langle n\left|T\right.\right\rangle $ (dashed
lines) for the LH mechanism compared to the exact results (solid lines).
Model with gas density $n=10^{3}\mathrm{cm}^{-3}$ and $G_{0}=20$
with barrier and binding energies for ices. For clarity, each successive
curve has been shifted by a factor of ten. \label{fig:ApproxLH_n_Es}}
\end{figure}

Figure \ref{fig:ApproxLH_n_Es} shows the resulting curves for $\left\langle \left.n\right|T\right\rangle $
for different grain sizes and compares the approximation to the exact
result. The binding and barrier energies are taken for ice surfaces
in Table \ref{tab:H_binding}. We see a very good match for small
grains in the low temperature domain. Large discrepancies appear at
high temperature but do not matter for the average formation rate
on the grain as the contribution of this regime is negligible. For
bigger grains, the approximation of the low temperature regime starts
to be less accurate. However, we know that total formation rate is
mostly influenced by the smallest grain sizes due to the $-3.5$ exponent
of the size distribution.

\begin{figure}
\begin{centering}
\includegraphics[width=1\columnwidth]{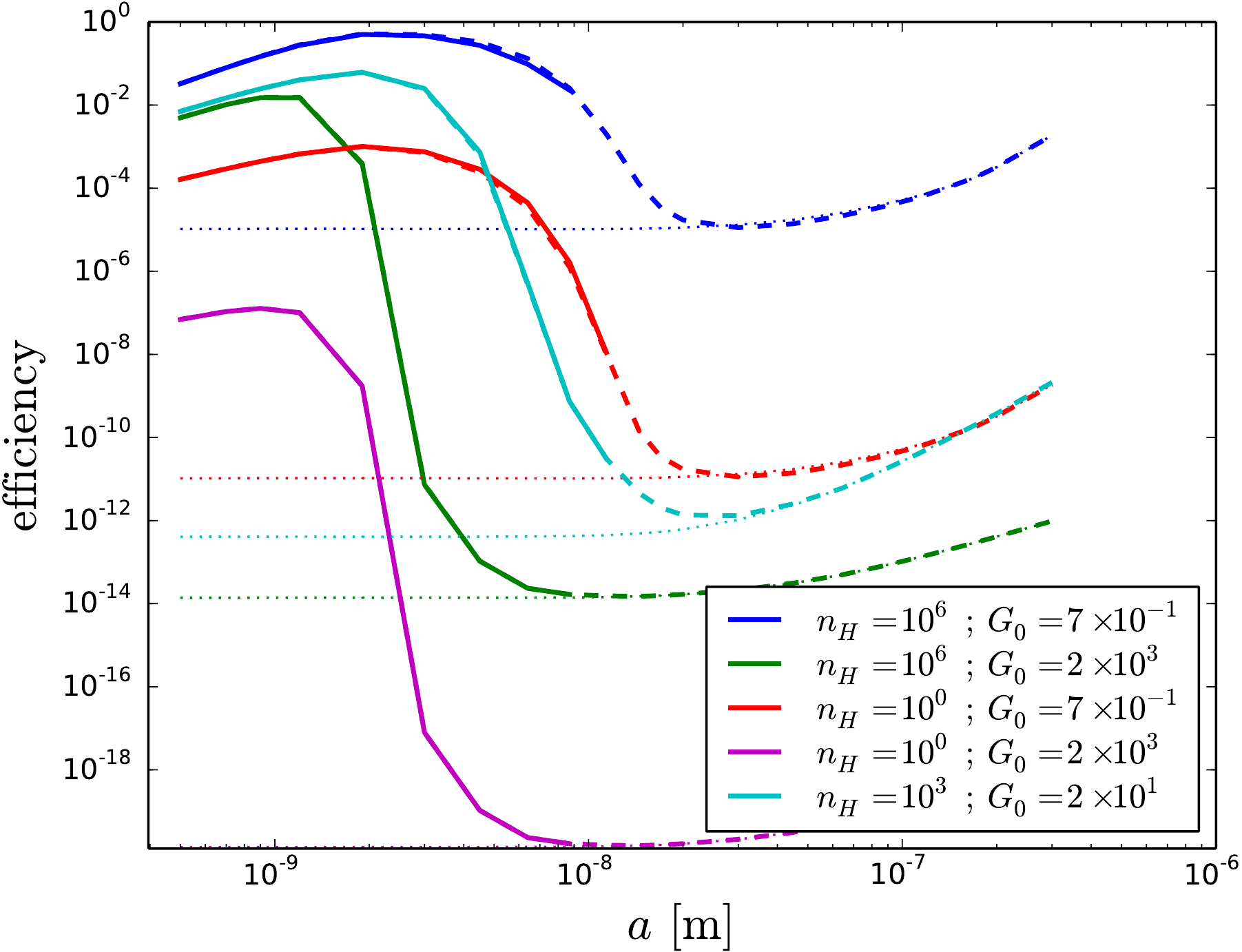}
\par\end{centering}

\caption{Approximated LH formation efficiency for one grain (dashed lines)
compared to the exact results (solid lines). The formation efficiencies
at the constant equilibrium temperatures are shown in dotted lines.
The densities are given in $\mathrm{cm}^{-3}$. The binding and barrier
energies correspond to ices.\label{fig:ApproxLH-rh2vsa_ices}}
\end{figure}

The resulting formation rate per grain is shown on Fig. \ref{fig:ApproxLH-rh2vsa_ices}
as a function of grain size for 5 models spanning the parameter domain.
We again assumed here an ice surface. The match is very good on the
domain on which the exact computation could be performed, and the
approximation smoothly joins with the equilibrium result neglecting
fluctuations for big grains. The relative errors on the final formation
rate integrated over the size distribution are given for each model
in the first part of Table \ref{tab:ApproxLH}. The results are given
for two size distributions. Both are MRN-like power-law distributions
with exponent $-3.5$, but one starts at $5\,\AA$ and the other at
$1\mbox{ nm}$. Both go up to $0.3\,\mu\mathrm{m}$. We find extremely
accurate results for the approximation in most case with a relative
error at most around 5\%.

\begin{figure}
\begin{centering}
\includegraphics[width=1\columnwidth]{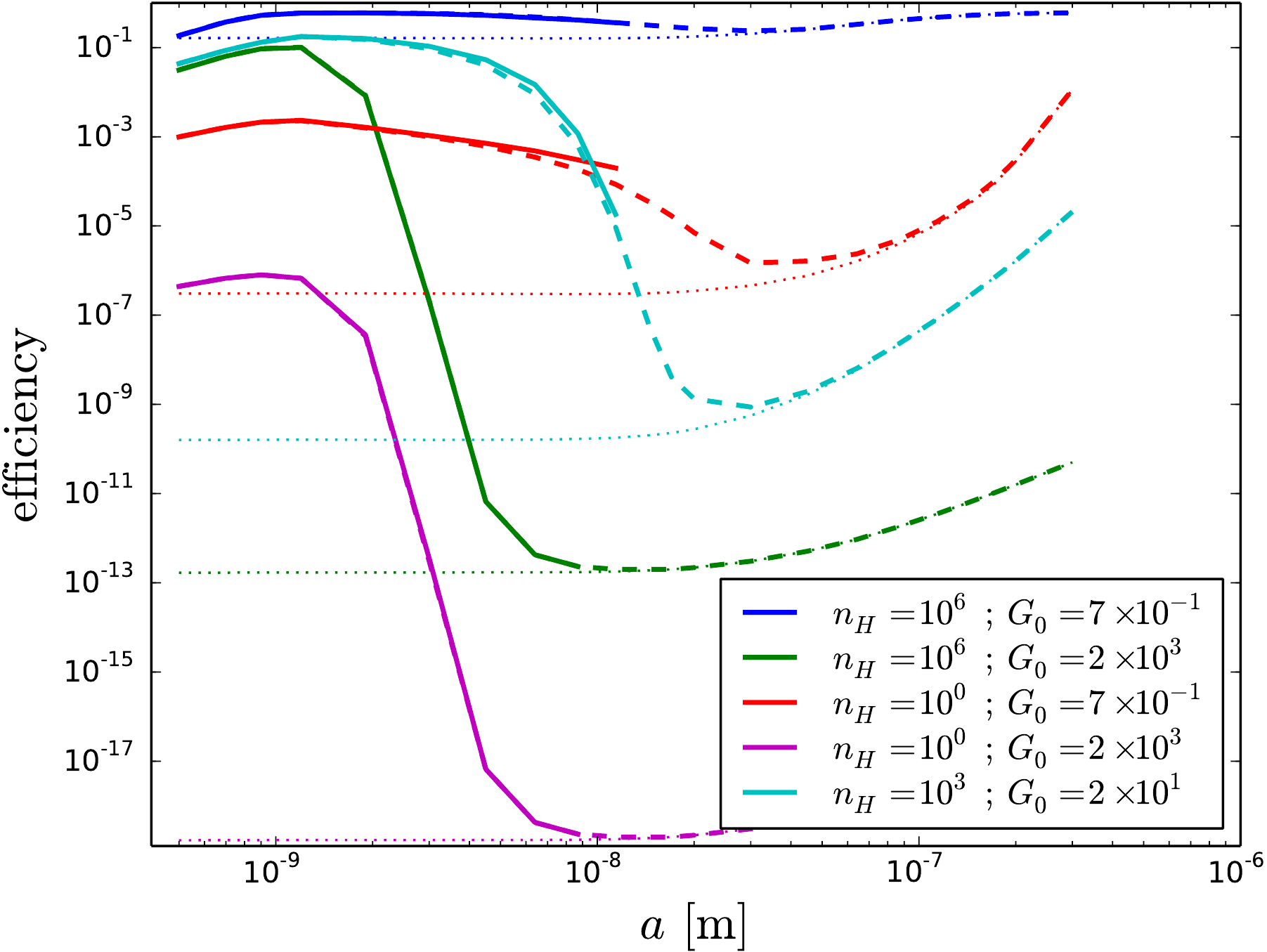}
\par\end{centering}

\caption{Same as Fig. \ref{fig:ApproxLH-rh2vsa_ices} for binding and barrier
energies corresponding to amorphous carbon grains.\label{fig:ApproxLH-rh2vsa_amC}}
\end{figure}

When using the energy values corresponding to an amorphous carbon
surface (see Table \ref{tab:H_binding}), the same levels of accuracy
are achieved. The formation rate as a function of grain size is shown
on Fig. \ref{fig:ApproxLH-rh2vsa_amC}. When integrated over the size
distribution, the formation rate obtained from the approximation is
again within 5\% from exact result (see second part of Table \ref{tab:ApproxLH}).

\begin{table}
\caption{Approximation error on the final formation rate integrated over the
full size distribution for the LH mechanism.\label{tab:ApproxLH}
}

\centering{}\textbf{\small }%
\begin{tabular}{cr@{\extracolsep{0pt}.}lr@{\extracolsep{0pt}.}lr@{\extracolsep{0pt}.}lr@{\extracolsep{0pt}.}l}
\hline 
{\small Surface type :} & \multicolumn{4}{c}{{\small Ices}} & \multicolumn{4}{c}{{\small amC}}\tabularnewline
{\small $a_{\mathrm{min}}=$} & \multicolumn{2}{c}{{\small $0.5\mbox{ nm}$}} & \multicolumn{2}{c}{{\small $1\mbox{ nm}$}} & \multicolumn{2}{c}{{\small $0.5\mbox{ nm}$}} & \multicolumn{2}{c}{{\small $1\mbox{ nm}$}}\tabularnewline
\hline 
{\small Model :} & \multicolumn{2}{c}{} & \multicolumn{2}{c}{} & \multicolumn{2}{c}{} & \multicolumn{2}{c}{}\tabularnewline
{\small $n=1\,,\, G_{0}=0.7$} & {\small 1}&{\small 64\%} & {\small 2}&{\small 10\%} & {\small 2}&{\small 95\%} & {\small 4}&{\small 44\%}\tabularnewline
{\small $n=1\,,\, G_{0}=2\times10^{3}$} & {\small 0}&{\small 53\%} & {\small 0}&{\small 45\%} & {\small 0}&{\small 32\%} & {\small 0}&{\small 02\%}\tabularnewline
{\small $n=1\times10^{6}\,,\, G_{0}=0.7$} & {\small 6}&{\small 09\%} & {\small 6}&{\small 65\%} & {\small 1}&{\small 18\%} & {\small 1}&{\small 13\%}\tabularnewline
{\small $n=1\times10^{6}\,,\, G_{0}=2\times10^{3}$} & {\small 1}&{\small 46\%} & {\small 1}&{\small 82\%} & {\small 1}&{\small 52\%} & {\small 1}&{\small 85\%}\tabularnewline
{\small $n=1\times10^{3}\,,\, G_{0}=2\times10^{1}$} & {\small 0}&{\small 35\%} & {\small 0}&{\small 50\%} & {\small 3}&{\small 68\%} & {\small 4}&{\small 95\%}\tabularnewline
\hline 
\end{tabular}
\end{table}

\subsection{ER mechanism}

\begin{figure}
\begin{centering}
\includegraphics[width=1\columnwidth]{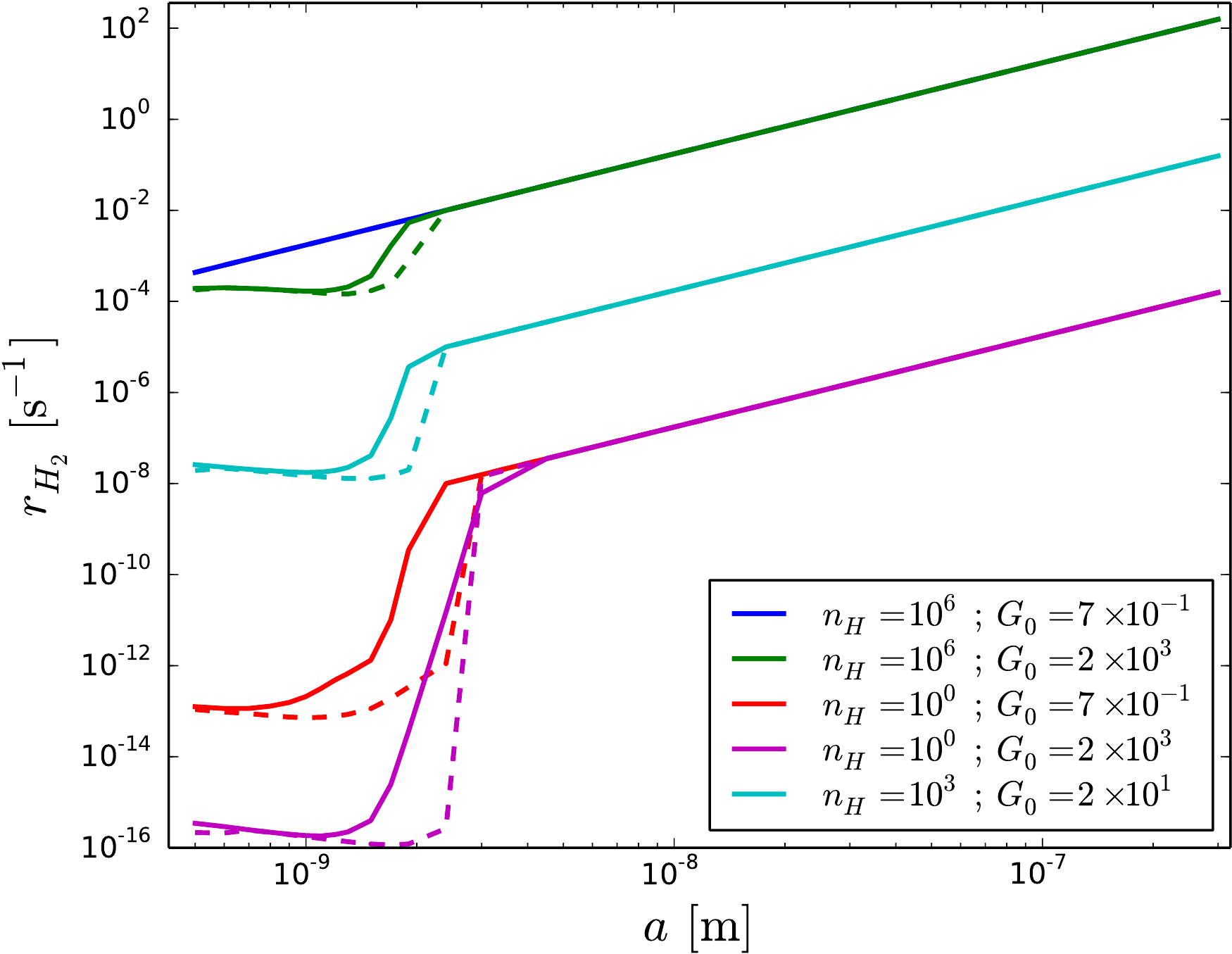}
\par\end{centering}

\caption{Approximated ER formation rate for one grain (dashed lines) compared
to the exact results (solid lines). The densities are given in $\mathrm{cm}^{-3}$.\label{fig:ApproxER-rh2vsa}}
\end{figure}

We now perform the same verification for the ER mechanism. The approximation
of Sect. \ref{sub:ER-Appprox} is compared to the exact method of
Sect. \ref{sub:Eley-Rideal-mechanism}. Figure \ref{fig:ApproxER-rh2vsa}
shows the formation rate as a function of grain size. The qualitative
effect is captured by the approximation, but large discrepancies appear
locally for some sizes, due to a shifted transition from a low efficiency
regime to a high efficiency one. However, the results when integrating
over the dust size distribution remain reasonably accurate, as shown
in Table \ref{tab:ApproxER}. We recall that this approximation is
not used in this paper as the exact method is numerically efficient.

\begin{table}
\caption{Approximation error on the final formation rate integrated over the
full size distribution for the ER mechanism.\label{tab:ApproxER}
}

\centering{}\textbf{\small }%
\begin{tabular}{cr@{\extracolsep{0pt}.}lr@{\extracolsep{0pt}.}l}
{\small $a_{\mathrm{min}}=$} & \multicolumn{2}{c}{{\small $0.5\mbox{ nm}$}} & \multicolumn{2}{c}{{\small $1\mbox{ nm}$}}\tabularnewline
\hline 
{\small Model :} & \multicolumn{2}{c}{} & \multicolumn{2}{c}{}\tabularnewline
{\small $n=1\,,\, G_{0}=0.7$} & 0&0028\% & \multicolumn{2}{c}{0\%}\tabularnewline
{\small $n=1\,,\, G_{0}=2\times10^{3}$} & 8&19\% & 8&73\%\tabularnewline
{\small $n=1\times10^{6}\,,\, G_{0}=0.7$} & 12&08\% & 12&08\%\tabularnewline
{\small $n=1\times10^{6}\,,\, G_{0}=2\times10^{3}$} & 10&22\% & 10&22\%\tabularnewline
{\small $n=1\times10^{3}\,,\, G_{0}=2\times10^{1}$} & 6&26\% & 6&15\%\tabularnewline
\hline 
\end{tabular}
\end{table}

\bibliographystyle{aa}
\bibliography{Transfert}

\end{document}